\documentclass[article]{article}

\usepackage[hmargin={25mm,25mm},vmargin={20mm,25mm}]{geometry}

\usepackage{amsmath,amsfonts,amssymb}
\usepackage{bm}
\usepackage{mathtools}
\usepackage{paralist}
\usepackage{multirow}
\usepackage{hyperref}
\usepackage[affil-it]{authblk}
\newcommand{\email}[1]{\href{mailto:#1}{#1}}

\newcommand{\MAT}[1]{\boldsymbol{#1}}
\newcommand{\VEC}[1]{{\sf #1}}
\newcommand{\UVEC}[1]{\underline{#1}}

\newcommand{\elem}[0]{T} 
\newcommand{\face}[0]{F} 

\DeclareMathOperator{\CARD}{card}
\DeclareMathOperator{\SPAN}{span}
\newcommand{\card}[1]{\CARD(#1)}
\newcommand{\cardc}[1]{\CARD#1}

\newcommand{\Th}{\mathcal{\elem}_h}
\newcommand{\Fh}{\mathcal{\face}_h}
\newcommand{\Mh}{\mathcal{M}_h}
\newcommand{\FT}{\mathcal{\face}_\elem}
\newcommand{\normal}{\boldsymbol{n}}
\newcommand{\nTF}{\boldsymbol{n}_{\elem \face}}
\newcommand{\npT}{\boldsymbol{n}_{\partial \elem}}

\newcommand{\fVec}{\boldsymbol{f}}
\newcommand{\gVec}{\boldsymbol{g}}
\newcommand{\gvb}[0]{\bm{\ell}}
\newcommand{\fvb}[0]{\bm{f}}

\newcommand{\uVec}{\boldsymbol{u}}
\newcommand{\vVec}{\boldsymbol{v}}
\newcommand{\uTF}{\underline{\boldsymbol{u}}_\elem}
\newcommand{\uTFn}{\underline{\boldsymbol{u}}_{\elem'}}
\newcommand{\uT}{\boldsymbol{u}_\elem}
\newcommand{\uTn}{\boldsymbol{u}_{\elem'}}
\newcommand{\uF}{\boldsymbol{u}_\face}
\newcommand{\upT}{\boldsymbol{u}_{\partial \elem}}
\newcommand{\vT}{\boldsymbol{v}_\elem}
\newcommand{\vpT}{\boldsymbol{v}_{\partial \elem}}
\newcommand{\vTF}{\underline{\boldsymbol{v}}_\elem}
\newcommand{\vF}{\boldsymbol{v}_\face}
\newcommand{\xVec}{\boldsymbol{x}}

\newcommand{\eg}{e.g., }
\newcommand{\ea}{et al.}
\newcommand{\ie}{i.e., }

\newcommand{\pT}{{p}_\elem}
\newcommand{\pF}{{p}_\face}
\newcommand{\pTF}{\underline{p}_\elem}
\newcommand{\ppT}{p_{\partial \elem}}
\newcommand{\qT}{{q}_\elem}
\newcommand{\qpT}{{q}_{\partial \elem}}
\newcommand{\qF}{{q}_\face}
\newcommand{\qTF}{\underline{q}_\elem}

\newcommand{\rTp}{{\rho}_{\elem'}}
\newcommand{\rT}{{\rho}_\elem}
\newcommand{\rTn}{{\rho}_{\elem'}}
\newcommand{\rF}{{\rho}_\face}
\newcommand{\rTF}{\underline{\rho}_\elem}
\newcommand{\rTFn}{\underline{\rho}_{\elem'}}
\newcommand{\rpT}{\rho_{\partial \elem}}
\newcommand{\zT}{{z}_\elem}
\newcommand{\zpT}{{z}_{\partial \elem}}
\newcommand{\zF}{{z}_\face}
\newcommand{\zTF}{\underline{z}_\elem}

\newcommand{\wT}{\boldsymbol{w}_\elem}

\newcommand{\intT}[0]{\int_{\elem}}
\newcommand{\intpT}[0]{\int_{\partial\elem}}
\newcommand{\intpTN}{\int_{\partial\elem^\neu}}
\newcommand{\intpTD}{\int_{\partial\elem^\dir}}
\newcommand{\intpTND}{\int_{\partial\elem^{\dir,\neu}}}
\newcommand{\pTbullet}{\partial \elem^\bullet}
\newcommand{\intF}[0]{\int_{\face}}

\newcommand{\eulern}{\mathrm{e}}
\newcommand{\rbrackets}[1]{\left(#1\right)}
\newcommand{\iVec}{\boldsymbol{i}}
\newcommand{\jVec}{\boldsymbol{j}}

\newcommand{\Real}{\mathbb{R}}
\newcommand{\Reynolds}{\mathrm{Re}}
\newcommand{\Atwood}{\mathrm{At}}
\newcommand{\tF}{t_{\rm F}}

\newcommand{\Poly}[1]{\mathcal{P}^{#1}}
\newcommand{\Polyd}[2]{\mathbb{P}_{#1}^{#2}}
\newcommand{\dir}{{\rm D}}
\newcommand{\neu}{{\rm N}}
\newcommand{\In}{{\rm In}}
\newcommand{\internal}{{\rm i}}

\newcommand{\GhT}[0]{{\mathcal{G}}_\elem }
\newcommand{\GhV}[0]{{\bm{\mathfrak{g}}}_\elem }

\newcommand{\sd}{sd}

\newcommand{\pBC}{{\rm !bc}}

\newcommand{\lm}{{\rm slm}}
\newcommand{\ilm}{{\rm silm}}
\newcommand{\ulm}{{\rm mnt}}
\newcommand{\uilm}{{\rm imnt}}
\newcommand{\mas}{{\rm sms}}
\newcommand{\umas}{{\rm mss}}
\newcommand{\vol}{{\rm vol}}

\newcommand\C[1]{#1}

\begin{document}

\title{A HHO formulation for variable density incompressible flows where the density is purely advected}

\author[1]{Lorenzo Botti} 
\author[2]{Francesco Carlo Massa} 
\affil[1]{Department of Engineering and Applied Sciences, University of Bergamo, Italy, \email{lorenzo.botti@unibg.it}}
\affil[2]{Department of Engineering and Applied Sciences, University of Bergamo, Italy, \email{francescocarlo.massa@unibg.it}}

\maketitle

\begin{abstract}
    We propose a Hybrid High-Order (HHO) formulation of the incompressible Navier--Stokes equations with variable density
    that provides exact conservation of volume and, accordingly, pure advection of the density variable.
    The spatial discretization relies on hybrid velocity-density-pressure spaces and the
    temporal discretization is based on Explicit Singly Diagonal Implicit Runge-Kutta (ESDIRK) methods.
    The formulation possesses some attractive features that can be fruitfully exploited for the 
    simulation of mixtures of immiscible incompressible fluids, namely:
    conservation of volume enforced cell-by-cell up to machine precision, pressure-robustness, 
    ability to preserve density bounds at low-order,
    robustness in the convection dominated regime, 
    weak imposition of boundary conditions,
    implicit high-order accurate time stepping,
    reduced memory footprint thanks to static condensation,
    possibility to exploit inherited $p$-multilevel solution strategies to improve the performance of iterative solvers.
    After addressing stability at the discrete level, numerical validation is performed showcasing spatial and temporal convergence rates.
    To conclude, we tackle the Rayleigh-Taylor instability at different Atwood and Reynolds numbers focusing on mesh independence capabilities.
\end{abstract}

\section{Introduction} \label{sec:Intro}
In recent years, the introduction of Hybrid High-Order (HHO) and Hybridizable Discontinuous Galerkin (HDG) methods 
has provided new lifeblood to promote the development of high-order accurate computational modelling tools for incompressible fluid flows.
In contrast to Discontinuous Galerkin (DG) methods, HHO and HDG methods are based 
on degrees of freedom that are broken polynomials on both the mesh and its skeleton (that is, the set collecting interelement boundaries).
Relevant features of hybrid schemes are:
\begin{inparaenum}[i)]
\item local (cell-by-cell) conservation of physical quantities,
\item increased convergence rates in the diffusion-dominated regime 
\item availability of upwind stabilizations to improve the robustness in the convection-dominated regime
\item resiliency to mesh distortion and grading,
\item implicit time integration with reduced memory footprint of the Jacobian matrix.
\end{inparaenum}

In this work we extend the HHO formulations of \cite{BottiDiPietroMassa25}-\cite{BottiMassa22}, devoted to constant density incompressible flows, 
to cope with mixtures of immiscible incompressible fluids with variable density. 
The present research effort relies on a hybrid discretization of all the variables of the problem, including velocity, pressure and density.
The choice of a hybrid pressure, combined with an appropriate selection of local polynomial spaces, 
allows to reconstruct a $\boldsymbol{H}(\operatorname{div})$-conforming velocity approximation starting from non-conforming polynomial spaces.
In particular, polynomial functions in the pressure space, with support on faces and cells, 
enforce, respectively, the continuity of the normal velocity components 
on the mesh skeleton and the incompressibility constraint within each mesh cell.
The $\boldsymbol{H}(\operatorname{div})$-conforming solenoidal velocity field is used to good advantage 
in the HHO discretization of the mass conservation equation, see \eg \C{Di Pietro \ea~}\cite{Di-Pietro.Droniou.ea:15},
therein employed as the advective velocity for the material time derivative of the density.
Pressure-robustness implies that irrotational body forces only affect the pressure field
and, as a consequence, the velocity and density errors are insensitive to the accuracy of the pressure field representation. 
Robustness in the convection-dominated regime is achieved by introducing skew-symmetric forms with 
upwinding stabilizations for the transport of mass and momentum, and suitable time derivative stabilizations for the density variable.

In order to treat time-dependent problems, we rely 
on high-order accurate Explicit Singly Diagonal Implicit Runge-Kutta (ESDIRK) time integration schemes 
especially conceived for Differential Algebraic Equations (DAEs) of differential index equal to two.  
Efficient resolution of the algebraic problems resulting from the space-time discretization 
relies on $p$-multigrid preconditioners in the spirit of \C{Botti \ea~}\cite{BottiDiPietroHHOpMG2021}, 
designed to work in tandem with Schur complement solution strategies. 
Thanks to static condensations, the blocked structure, the dimension of the blocks and, ultimately, 
the sparsity of the global matrix resulting from HHO discretizations is dictated by face unknowns.
Interestingly, since the dimension of $d{-}1$-variate polynomial spaces grows
slower that the dimension of $d$-variate polynomial spaces when increasing $k$, 
both the memory footprint of the global matrix
and the cost of matrix-vector products involved in iterative solution strategies benefit from static condensation.

Several methods for variable density incompressible flows have been proposes in literature.
A DG formulation based on exact Riemann solvers for the
artificial compressibility perturbation of the problem was proposed by \C{Bassi \ea~}\cite{Bassi.Massa.ea:18}
and assessed in \cite{BassiVDAss22}.
A Volume Of Fluid (VOF) DG formulation for two-phase flows was considered by \cite{LANDET2020},
interestingly, the transport equation for the density is modified introducing 
a slope-limited $H$-div conforming velocity field serving as advection velocity.
An entropy–stable DG approximation was devised by \C{Manzanero \ea~}\cite{Manzanero.Rubio.ea:20} while a post-processed
semi-implicit DG scheme was introduced by \cite{LiQiuYang22}. 
A Discrete Duality Finite Volume (DDFV)
scheme was proposed by \C{Goudon \ea~}\cite{GoudonKrell14}, while
\C{Fu~}\cite{FU2020} addressed the Cahn-Hilliard phase-field model 
with a HDG formulation based on divergence-free velocity spaces. Recently, \C{Nordstr\"{o}m \ea~}\cite{NORDSTROM25}
proposed an energy stable formulation for multi-phase incompressible flows and \C{Dauphin \ea~}\cite{dauphin2026} 
analysed a low-order hybrid method for the variable-density incompressible Navier-Stokes equations.
Earlier pioneering works focused on Galerkin discretizations with projection methods, consider for example \C{Guermond \ea~}\cite{Guermond2000}-\cite{Guermond2009}.

The paper is organized as follows.
In Sections~\ref{sec:INS} and~\ref{sec:HHOESDIRK} we respectively introduce the model problem and present its discrete formulation.
$\boldsymbol{H}(\operatorname{div})$-conformity and pressure robustness are addressed in Section~\ref{sec:hdiv}.
In Section~\ref{sec:fluxForm} we seek to recast the HHO discretizations of the mass conservation and linear momentum equations, 
first presented in skew-symmetric form, in the, so called, flux formulation.
Numerical validation of spatial and temporal convergence rates is performed in Section~\ref{test:kova} and Section~\ref{test:GQ} considering 
exact and manufactured solutions.
A realistic flow configuration is tackled in Section~\ref{sec:RTI} where we consider a 
low Atwood number Rayleigh-Taylor instability at $\Reynolds{=}1~000$ and $\Reynolds{=}5~000$ and a
high Atwood number Rayleigh-Taylor instability at $\Reynolds{=}1~000$.
Finally, some conclusions are drawn in Section~\ref{sec:conclusion}.

\section{Continuous setting} \label{sec:INS}
Given a polygonal or polyhedral domain $\Omega\subset\Real^d$, $d\in\{2,3\}$, with boundary $\partial \Omega$, 
the initial divergence-free velocity field $\uVec_0: \Omega \rightarrow \Real^d$, the initial density field $\rho_0: \Omega \rightarrow \Real^{+}$ 
and a finite time $\tF$, the variable density incompressible Navier--Stokes problem 
consists in finding the velocity field $\uVec: \Omega \times \lbrack 0,\tF\rbrack \rightarrow \Real^d$, 
the density field $\rho: \Omega  \times (0,\tF\rbrack \rightarrow \Real^{+}$,
and the pressure field $p: \Omega  \times (0,\tF\rbrack \rightarrow \Real$,
such that $\uVec(\cdot,0) = \uVec_0$, $\rho(\cdot,0) = \rho_0$ and
\begin{subequations}
  \label{nstokesProb}
  \begin{alignat}{2}
    \frac{\partial (\rho \uVec)}{\partial t} + \nabla \cdot \left[{(\rho \uVec \otimes \uVec) -\mu \nabla \uVec }\right] + \nabla p &= \boldsymbol{f}
    &\qquad& \text{in $\Omega \times (0,\tF\rbrack$}, \label{stokesProb:momentum}\\ 
    \frac{\partial \rho}{\partial t} + \nabla \cdot ( \rho \uVec) &= 0 
    &\qquad& \text{in $\Omega \times (0,\tF\rbrack$}, \label{stokesProb:mass}\\ 
    \nabla \cdot \uVec &= 0
    &\qquad& \text{in $\Omega\times (0,\tF\rbrack$}, \label{stokesProb:volume}\\
    \uVec &= \boldsymbol{g}_\dir
    &\qquad& \text{on $\partial\Omega_\dir\times (0,\tF\rbrack$}, \\
    p \normal - \mu (\nabla \uVec)\normal  &= \boldsymbol{g}_\neu
    &\qquad& \text{on $\partial\Omega_\neu\times (0,\tF\rbrack$}, \\
    \rho &= g_{\In}  
    &\qquad& \text{on $\partial \Omega_{\In} \times(0,t_F\rbrack$  }.
  \end{alignat}
\end{subequations}
where $\normal$ denotes the unit vector normal to $\partial\Omega$ pointing out of $\Omega$, $\mu > 0$ is the (constant) viscosity,
$\boldsymbol{g}_\dir$ and $\boldsymbol{g}_\neu$ denote, respectively, 
the prescribed velocity on the Dirichlet boundary $\partial\Omega_\dir\subset\partial\Omega$ and 
the prescribed traction on the Neumann boundary $\partial\Omega_\neu\coloneqq\partial\Omega\setminus\partial\Omega_\dir$,
while $\boldsymbol{f}:\Omega\to\Real^d$ is a given body force.
Since, by virtue of the incompressibility constraint, we consider the Laplace operator in place of the divergence of the symmetric gradient, 
tractions do not account for tangential variations of the normal velocity component.
The density is prescribed on inflow boundaries, defined as follows 
$$
\partial \Omega_{\In} \coloneqq \left\{\xVec \in \partial \Omega \; \text{such that} \; 
\left\{
\begin{array}{lcl} 
 \boldsymbol{g}_\dir \cdot \normal < 0 & \text{if} & \xVec \in \partial \Omega_\dir, \\ 
               \uVec \cdot \normal < 0 & \text{if} & \xVec \in \partial \Omega_\neu
\end{array} \right. \right\}.
$$
Equations \eqref{stokesProb:momentum},  \eqref{stokesProb:mass} and \eqref{stokesProb:volume} are the linear momentum, 
the mass conservation and the volume conservation equations, respectively.
We remark that conservation of volume will be exactly enforced at the discrete level, ensuring that the density is a purely advected quantity.
       
\section{ESDIRK-HHO discretization} \label{sec:HHOESDIRK}
\subsection{Mesh}
\label{sec:meshSet}
The HHO formulation proposed in this work is capable of dealing with two and three dimensional flow problems. 
Nevertheless, since numerical test cases focus on two space dimensions, discrete settings are provided for the 2D case focusing on simplicial meshes.
We consider meshes of the domain $\Omega$ corresponding to couples $\Mh\coloneq(\Th,\Fh)$, where $\Th$ 
is a finite collection of triangular elements such that $h\coloneq\max_{\elem \in \Th}h_T>0$ with $h_\elem$ denoting the diameter of $\elem$, 
while $\Fh$ is a finite collection of line segments, with $h_\face$ denoting the length of $\face$. 
It is assumed henceforth that the mesh $\Mh$ is shape- and contact-regular, as detailed in \cite[Definition 1.4]{Di-Pietro.Droniou:20} and that its trace on $\partial \Omega$ is compatible with the partition $\partial \Omega = \partial \Omega_\dir \sqcup \partial \Omega_\neu$.
We respectively denote by $\Fh^\dir$ and $\Fh^\neu$ the sets of Dirichlet and Neumann faces, 
whereas the set of internal faces is $\Fh^\internal \coloneq \Fh\setminus\big(\Fh^\dir\cup\Fh^\neu\big)$.
For each mesh cell $\elem \in \Th$, the faces contained in the cell boundary $\partial T$ are collected in the set 
$\FT$ and we additionally let $\pTbullet \coloneqq \partial T \cap \partial \Omega_\bullet$ for $\bullet \in \{ \dir, \neu \}$.
For all $T\in\Th$ and all $F\in\FT$, $\normal_{TF}$ denotes the unit vector normal to $F$ 
pointing out of $T$ and $\npT:\partial T \to \Real^d$ is such that $\npT|_F \coloneqq \normal_{TF}$ for all $F\in\FT$.

\subsection{Discrete spaces}
Hybrid High-Order methods hinge on local polynomial spaces on mesh cells and faces.
For given integers $\ell\ge 0$ and $n\ge 1$, we denote by $\Polyd{n}{\ell}$ the space of $n$-variate polynomials of total degree $\le\ell$ (in short, of degree $\ell$).
For $X$ mesh cell or face, we denote by $\Poly{\ell}(X)$ the space spanned by the restriction to $X$ of functions in $\mathbb{P}_d^\ell$.

The global discrete spaces for the velocity, density and pressure unknowns are respectively defined as follows:
\[
\underline{\boldsymbol{V}}_h^{k}\coloneqq \left\{ \underline{\vVec}_h = \big( (\vVec_T)_{T\in\Th} , (\vVec_F)_{F \in \Fh} \big) :
\begin{array}{l}
  \text{${\vVec}_T \in \Poly{k+1}(T)^d$ for all $T\in\Th$},\\ \text{$\vVec_F  \in \Poly{k}(F)^d$ for all $F \in \Fh \setminus \Fh^\neu$,} \\ \text{$\vVec_F  \in \Poly{k+1}(F)^d$ for all $F \in \Fh^{\neu}$
  }
\end{array}
\right\},
\]
\[
\underline{Z}_h^{k}\coloneqq\left\{
\underline{z}_h=\big((z_T)_{T\in\Th}, (z_F)_{F\in\Fh}\big)\,:\,
\text{$z_T\in\Poly{k}(T)$ for all $T\in\Th$}, \; \text{$z_F\in\Poly{k}(F)$ for all $F\in\Fh$}
\right\},
\]
\[
\underline{Q}_h^{k+1}\coloneqq\left\{
\underline{q}_h=\big((q_T)_{T\in\Th}, (q_F)_{F\in\Fh}\big)\,:\,
\text{$q_T\in\Poly{k}(T)$ for all $T\in\Th$}, \; \text{$q_F\in\Poly{k+1}(F)$ for all $F\in\Fh$}
\right\}.
\]
The restriction of these spaces and their elements to a mesh cell $T \in \Th$ are denoted replacing the subscript $h$ by $T$ and are obtained collecting the components attached to $T$ and its boundary.
Given $\vTF \in \underline{\boldsymbol{V}}_T^{k}$, we will also denote by $\vVec_{\partial T}$ the function such that $\vVec_{\partial T}|_F = \vVec_F$ for all $F \in \FT$.
Given $\underline{q}_T \in  \underline{Q}_T^{k+1}$ and $\underline{z}_T \in  \underline{Z}_T^{k}$, the notation $q_{\partial T}$ and $z_{\partial T}$ are defined similarly.

\subsection{Gradient reconstruction operators}
For each mesh cell $\elem \in \Th$,
the tensor and vector gradient reconstruction operators
$\GhT^k : \underline{\boldsymbol{V}}_T^{k} \to \Poly{k}(\elem)^{d \times d}$
and $\GhV^{k+1} : \underline{Q}_T^{k+1} \to \Poly{k+1}(\elem)^{d}$
are respectively defined as follows:
For all $(\vTF, \qTF) \in \underline{\boldsymbol{V}}_T^{k} \times \underline{Q}_T^{k+1}$,
\begin{gather*} 
  \intT \GhT^k \vTF : \bm{\tau}
  = \intT \nabla \vT : \bm{\tau}
  - \intpT (\vT - \vpT) \otimes \npT :  \bm{\tau}
  \qquad \forall \bm{\tau} \in \Poly{k}(\elem)^{d \times d},
  \\ 
  \intT \GhV^{k+1} {\qTF} \cdot {\vVec}   
  = - \intT \qT \;  \nabla \cdot {\vVec}
  + \intpT \qpT {\vVec} \cdot \npT
  \qquad \forall {\vVec} \in \Poly{k+1}(\elem)^{d}.
\end{gather*}

\subsection {Semi-discrete HHO formulation} \label{sec:SpaceDiscr}
We first present the semi-discrete formulation of the variable density problem in \eqref{nstokesProb} omitting
the terms that are responsible of the weak imposition of boundary conditions. 
Those will be introduced thereafter for the sake of readability.

Given $(\uTF, \rTF, \pTF)\in \underline{\boldsymbol{V}}_T^{k}\times \underline{Z}_T^{k}\times \underline{Q}_T^{k+1}$, the 
(local) spatial discretizations
$\sd^{\lm,\pi^p}_{T,\pBC} ((\uTF,\rTF,\pTF);\cdot):\underline{\boldsymbol{V}}_T^{k}\to\Real$ of the steady momentum equation,
$ \sd^\mas_{T,\pBC}(\uTF,\rTF;\cdot):\underline{Z}^{k}_T\to\Real$ of the steady mass conservation equation and
$ \sd^\vol_{T,\pBC}(\uTF;\cdot):\underline{Q}^{k+1}_T\to\Real$ of the volume conservation equation
are such that,
for all $\vTF\in\underline{\boldsymbol{V}}_T^{k}$, all $\zTF\in\underline{Z}_T^{k}$ and all $\qTF\in\underline{Q}_T^{k+1}$,
\begin{alignat}{1}
\sd^{\lm,\pi^p}_{T,\pBC} \bigl((\uTF,\rTF,\pTF);\vTF\bigr) =  
   &  - \frac{1}{2}\int_T  \rho_T (\bm{\pi}_{T}^p\uT \otimes \uT) : \nabla \bm{\pi}_{T}^k\vT 
    + \frac{1}{2} \int_T  \rho_T (\bm{\pi}_{T}^k\vT \otimes \uT) : \nabla \bm{\pi}_{T}^p\uT \label{lmConvHHO1}\\
   &  + \frac{1}{2} \intpT \rpT \bigl(\uT \cdot \npT \bigr) 
      \left(\bm{\pi}_{\partial T}^p \upT \cdot \bm{\pi}_{T}^k\vT - \bm{\pi}_{\partial T}^k \vpT \cdot \bm{\pi}_{T}^p\uT \right) \label{lmConvHHO2}\\
   &  + \frac{1}{2} \intpT \rpT \bigl|\uT \cdot \npT \bigr| 
      \left( \bm{\pi}_{T}^p\uT -\bm{\pi}_{\partial T}^p \upT \right) \cdot \left(\bm{\pi}_{T}^k\vT - \bm{\pi}_{\partial T}^k \vpT \right) \label{lmConvHHO3}\\
   & +  \intT \mu \, {\GhT^k} \uTF : {\GhT^k} \vTF   
    + \intpT \frac{\mu}{h_F} \, \bm{\pi}_{\partial T}^k(\uT - \upT) \cdot \bm{\pi}_{\partial T}^k(\vT - \vpT) \label{lmDiffHHO}\\
    & + \intT \GhV^{k+1} \pTF \cdot \vT-\intT \fVec \cdot \vT, \label{lmPresHHO} \\
\sd^\mas_{T,\pBC}\bigl((\uTF,\rTF);\zTF\bigr)  = &  
  - \frac{1}{2} \int_T \rho_T \uT \cdot \nabla z_T + \frac{1}{2} \int_T z_T \uT \cdot \nabla \rho_T \label{massHHO} \\
   & + \frac{1}{2} \intpT \bigl(\uT \cdot \nTF \bigr) (\rpT \zT - \rT \zpT )  \label{massHHO2} \\
   & + \frac{1}{2} \intpT \bigl|\uT \cdot \nTF \bigr| (\rT - \rpT) (\zT - \zpT)   \label{massHHO3} \\
\sd^\vol_{T,\pBC}(\uTF;\qTF)  = &
   \intT \GhV^{k+1} \qTF \cdot  \uT, \label{volHHO}
\end{alignat}
where 
$\bm{\pi}_{\partial T}^\ell$ is the $L^2$-orthogonal projector onto the broken polynomial space $\Poly{\ell}(\partial T)^d$ defined as follows
$$\Poly{\ell}(\partial T)^d \coloneqq \left\{ \vVec \in L^2(\partial T)^d \;:\; \text{$\vVec|_F \in \Poly{\ell}(F)^d$ for all $F \in \FT$} \right\},$$ and 
{$\bm{\pi}_{T}^\ell$ is the $L^2$-orthogonal projector onto $\Poly{\ell}(T)^d$}.

Both in the momentum and the mass conservation equation, advection terms are treated with skew-symmetric formulations 
see \eqref{lmConvHHO1}-\eqref{lmConvHHO2} and \eqref{massHHO}-\eqref{massHHO2}, respectively,
plus stabilization bilinear forms, see \eqref{lmConvHHO3} and \eqref{massHHO3}, respectively.
We remark that, in the convective term discretization of \eqref{lmConvHHO1}-\eqref{lmConvHHO2}-\eqref{lmConvHHO3}, we consider
$L^2$-projections of degree $k$ for test functions and $L^2$-projections of degree $p$, with $p{=}k$ or $p{=}k{+}1$, for the advected velocity unknowns,
leading to two different variants of the method. Note that the advective velocity is not projected.
In the case $p{=}k{+}1$, since $\uT \in \Poly{k+1}(T)$, it holds $\bm{\pi}_{T}^p\uT = \uT$, 
furthermore, since $\upT \in \Poly{k}(\partial T^{\internal,\dir}) \subset \Poly{k+1}(\partial T^{\internal,\dir})$ and $\upT \in \Poly{k+1}(\partial T^{\neu})$, 
it holds $\bm{\pi}_{\partial T}^p \upT = \upT$, thus, contrary to the case $p{=}k$, the formulations is not skew-symmetric.
The choice $p{=}k$, proposed in \cite[Remark 3.7]{Beirao-da-Veiga.Di-Pietro.ea:25}, 
leads to a provably stable formulation, see Sec.~\ref{sec:stab}, while the choice $p{=}k{+}1$, 
proposed in \cite{BottiDiPietroMassa25}, provides improved accuracy in the convection-dominated regime.
The diffusion term stabilization of \eqref{lmDiffHHO}, also involving $L^2$-projection operators,
was first proposed by \cite{Lehrenfeld:10}, see also \cite{Cockburn.Di-Pietro.ea:16}.
Note that the HHO discretization of the mass conservation equation in \eqref{massHHO} only relies on the cell velocity $\uT$ 
which, as result of the HHO discretization of the volume conservation in \eqref{volHHO}, 
is $\boldsymbol{H}(\operatorname{div})$-conforming and divergence-free, see Sec.~\ref{sec:hdiv} for details.

Let's now take into account Dirichlet and Neumann boundary conditions.
Given $(\uTF, \rTF, \pTF)\in \underline{\boldsymbol{V}}_T^{k}\times \underline{Z}_T^{k}\times \underline{Q}_T^{k+1}$, the (local) spatial discretizations
$\sd^{\lm,\pi^p}_{T} ((\uTF,\rTF,\pTF);\cdot):\underline{\boldsymbol{V}}_T^{k}\to\Real$ of the steady momentum equation,
$ \sd^\mas_{T}(\uTF,\rTF;\cdot):\underline{Z}^{k}_T\to\Real$ of the steady mass conservation equation and
$ \sd^\vol_{T}(\uTF;\cdot):\underline{Q}^{k+1}_T\to\Real$ of the volume conservation equation 
are such that,
for all $\vTF\in\underline{\boldsymbol{V}}_T^{k}$, all $\zTF\in\underline{Z}_T^{k}$ and all $\qTF\in\underline{Q}_T^{k+1}$,
\begin{subequations}
\label{HHObc}
\begin{alignat}{1}
  \sd^{\lm,\pi^p}_{T} \bigl((\uTF,\rTF,\pTF); \vTF\bigr) &\coloneqq \sd^{\lm,\pi^p}_{T,\pBC} ((\uTF,\pTF); \vTF)  
   - \frac{1}{2} \intpTND \rpT \bigl( \uT {\cdot} \npT \bigr)  \bigl(\bm{\pi}_{\partial T}^p \upT {\cdot} {\bm{\pi}_{\partial T}^k}\vpT \bigr) \nonumber \\
  & + \intpTN \left( \bigl( \upT {\cdot} \npT \bigr)^{\oplus} \rpT + \bigl( \upT {\cdot} \npT \bigr)^{\ominus} g_\In \right) \bigl( {\bm{\pi}_{\partial T}^p}\upT {\cdot} {\bm{\pi}_{\partial T}^k}\vpT \bigr) \nonumber \\
  & + \intpTD \left(\bigl( \bm{g}_\dir \cdot \npT \bigr)^{\oplus} \rpT + \bigl( \bm{g}_\dir \cdot \npT \bigr)^{\ominus} g_\In \right) \; (\bm{g}_\dir \cdot \bm{\pi}^k_{\partial T}\vpT) \nonumber \\
  &  + \intpTD  \bigl( (\upT - \bm{g}_\dir) \otimes \npT \bigr) : \left( \nu \GhT \vTF + \frac{\nu}{h_\face} \vpT \otimes \npT \right) \nonumber \\
  &  - \intpTD \nu \GhT \uTF : (\vpT \otimes \npT )
   - \intpTN \ppT (\vpT \cdot \npT) + \intpTN \bm{g}_\neu \cdot \vpT , \label{lmHHObc} \\
  \sd^\mas_{T}\bigl((\uTF,\rTF);\zTF\bigr)  & \coloneqq \sd^\mas_{T,\pBC}(\uTF,\rTF;\zTF)  
  - \frac{1}{2} \intpTND \bigl( \uT \cdot \npT \bigr)  \bigl(\rpT \zpT \bigr) \nonumber\\
  & + \intpTN \left( \bigl( \upT \cdot \npT \bigr)^{\oplus} \rpT +
  	\bigl( \upT \cdot \npT \bigr)^{\ominus}  g_\In \right) \zpT  \nonumber \\
  &+ \intpTD \left( \bigl( \gVec_\dir \cdot \npT \bigr)^{\oplus} \rpT 
  	+ \bigl( \gVec_\dir \cdot \npT \bigr)^{\ominus}  g_\In \right) \zpT, \label{massHHObc} \\
  \sd^{\vol}_{T} (\uTF;\qTF)
  &\coloneqq \sd^{\vol}_{T,\pBC} (\uTF;\qTF)
  - \intpTN (\upT \cdot \npT) \, \qpT
  - \intpTD (\bm{g}_\dir \cdot \npT) \, \qpT, \nonumber 
\end{alignat}
\end{subequations}
where $\alpha^{\oplus} \coloneqq \frac{1}{2} (\alpha + |\alpha|)$ and $\alpha^{\ominus} \coloneqq \frac{1}{2} (\alpha - |\alpha|)$.

Given $(\uTF, \rTF, \pTF)\in \underline{\boldsymbol{V}}_T^{k}\times \underline{Z}_T^{k}\times \underline{Q}_T^{k+1}$,
the local spatial discretization of the unsteady momentum and mass conservation are such that,
for all $\vTF\in\underline{\boldsymbol{V}}_T^{k}$ and all $\zTF\in\underline{Z}_T^{k}$ 
\begin{alignat}{1}
&\sd_{T}^{\ulm,\pi^p}\bigl((\uTF,\rTF,\pTF); \vTF\bigr) \coloneqq  
  \int_T \left( \rT \frac{\partial \uT}{\partial t} + \frac{\uT}{2} \frac{\partial \rT }{\partial t} \right) \cdot \vT
 + \sd^{\lm,\pi^p}_{T}\bigl((\uTF,\rTF,\pTF);\vTF\bigr), \label{eq:uLmSD} \\
&\sd_{T}^\umas\bigl((\uTF,\rTF); \zTF\bigr) \coloneqq  
 \int_T \frac{\partial \rho_T}{\partial t}  z_T 
 + \frac{h_\elem}{(k+1)^2} \intpT \left( \frac{\partial \rT}{\partial t}- \frac{\partial \rpT}{\partial t}\right)  (\zT - \zpT) 
  + \sd^\mas_{T}\bigl((\uTF,\rTF);\zTF\bigr). \label{eq:uMasSD}
\end{alignat}
We remark that the time derivative in the momentum equation has been modified compatibly with the skew-symmetric treatment 
of the convective term, as first proposed by \cite{Guermond2000}, see Sec.~\ref{sec:qdmFlux} for details.
We further remark that the second term of \eqref{eq:uMasSD} is a time derivative stabilization 
designed to guarantee the robustness of the HHO discretization of the mass conservation equation
whenever the normal velocity component vanishes on a mesh face, see Sec.~\ref{sec:massFlux} for additional details.

The global residuals of the spatial discretization
$r^{\ulm,\pi^p}_{h}((\underline{\boldsymbol{u}}_h,\underline{\rho}_h,\underline{p}_h);\cdot):\underline{\boldsymbol{V}}^{k}_h\to\Real$,
$r^\umas_{h}((\underline{\boldsymbol{u}}_h,\underline{\rho}_h);\cdot):\underline{Z}^{k}_h\to\Real$, and
$r^\vol_{h}(\underline{\boldsymbol{u}}_h;\cdot):\underline{Q}^{k+1}_h\to\Real$ are obtained by cell-by-cell assembly of the local discretizations, \textit{i.e.}:
For all $\underline{\boldsymbol{v}}_h\in\underline{\boldsymbol{V}}_h^{k}$, all $\underline{z}_h \in \underline{Z}_h^k$ and all $\underline{q}_h\in\underline{Q}_h^{k+1}$,
\begin{subequations}
  \label{eq:HHOdiscrGlobalResHDIV}
\begin{alignat}{1}
  r_{h}^{\ulm,\pi^p}\bigl((\underline{\boldsymbol{u}}_h,\underline{\rho}_h,\underline{p}_h);\underline{\boldsymbol{v}}_h\bigr)
  &\coloneqq \sum_{T\in\Th}  \sd^{\ulm,\pi^p}_{T}\bigl((\uTF,\rTF,\pTF);\vTF\bigr), \\
  r^\umas_{h}\bigl((\underline{\boldsymbol{u}}_h,\underline{\rho}_h);\underline{z}_h\bigr) &\coloneqq \sum_{T\in\Th} \sd^\umas_{T}\bigl((\uTF,\rTF);\zTF\bigr), \\
  r_{h}^\vol(\underline{\boldsymbol{u}}_h;\underline{q}_h) &\coloneqq \sum_{T\in\Th} \sd^\vol_{T}(\uTF;\qTF). 
\end{alignat}
\end{subequations}
The global problem reads: Find $(\underline{\boldsymbol{u}}_h,\underline{\rho}_h,\underline{p}_h)\in\underline{\boldsymbol{V}}_h^{k}\times\underline{Z}_h^k\times\underline{Q}^{k+1}_h$ such that
\begin{equation}\label{eq:hhoGlobalProblem}
  \begin{alignedat}{2}
    r_{h}^{\ulm,\pi^p}\bigl((\underline{\boldsymbol{u}}_h,\underline{\rho}_h,\underline{p}_h);\underline{\boldsymbol{v}}_h\bigr) &= 0 &\qquad& \forall 
\underline{\boldsymbol{v}}_h\in\underline{\boldsymbol{V}}_h^{k},
    \\
    r_{h}^\umas\bigl((\underline{\boldsymbol{u}}_h,\underline{\rho}_h);\underline{z}_h\bigr) & = 0 &\quad& \forall \underline{z}_h \in \underline{Z}_h^k, \\
    r_{h}^\vol(\underline{\boldsymbol{u}}_h;\underline{q}_h) &= 0 &\qquad& \forall \underline{q}_h\in\underline{Q}_h^{k+1}.
  \end{alignedat}
\end{equation}

\subsection{H(div)-conformity and independence from irrotational body forces} \label{sec:hdiv}
Let's focus on the HHO discretization of the volume conservation equation.
The requirement $r_{h}^\vol(\underline{\boldsymbol{u}}_h;\underline{q}_h) = 0$
for each $\underline{\boldsymbol{u}}_h \in \underline{\boldsymbol{V}}_h$ and for all $\underline{q}_h\in\underline{Q}^{k+1}_h$
reads 
$$
\begin{aligned} 
0 =  
  - \sum_{T \in \Th} \intT \qT \;  \nabla \cdot \uT 
  + \sum_{T \in \Th} \left(  \intpT \qpT \uT \cdot \nTF  
  - \intpTN \qpT \upT \cdot \npT   - \intpTD \qpT \bm{g}_\dir \cdot \npT \right).
\end{aligned}
$$
Among all the $\underline{q}_h\in\underline{Q}^{k+1}_h$, we consider the following relevant choices
\begin{enumerate}
\item
$\underline{q}_h = \left( (q_T)_{T\in\Th},(0)_{F\in \Fh} \right)$, accordingly, $\forall T \in \Th$ and $\forall \qT \in \Poly{k}(T)$, we get 
$
\intT \qT \; \nabla \cdot \uT = 0, \; 
$
\ie $\nabla \cdot \uT = 0, \forall \xVec \in T$, since $\nabla \cdot \uT \in \Poly{k}(T)$;
\item
$\underline{q}_h = \left( (1)_{T\in\Th},(0)_{F\in \Fh} \right)$, accordingly, after integration by parts, $\forall T \in \Th$,  we get 
$
\intpT  \uT \cdot \nTF = 0;
$
\item
$\underline{q}_h = \left( (0)_{T\in\Th},(q_F)_{F\in \Fh} \right)$, accordingly, $\forall \qF \in \Poly{k+1}(F)$, we get
\begin{equation}
\label{eq:HDivTraces}
\begin{aligned}
& \forall F \in \Fh^\dir,      &\intF \qF (\uT - \bm{g}_\dir) {\cdot} \nTF = 0,  & \; \text{\ie} \; \uT  {\cdot} \nTF {=} \pi_F^{k+1}(\bm{g}_D {\cdot} \nTF), \\
& \forall F \in \Fh^\neu,      &\intF \qF (\uT - \uF) {\cdot} \nTF = 0,          & \; \text{\ie} \; \uT {\cdot} \nTF  {=} \uF {\cdot} \nTF, \\
& \forall F \in \Fh^\internal, &\intF \qF \; (\uT - \uTn) {\cdot} \nTF = 0,      & \; \text{\ie} \; \uT {\cdot} \nTF  {=} \uTn{\cdot} \nTF,  \\
\end{aligned}
\end{equation}
where the equalities involving the normal velocity component hold true $\forall \xVec \in F$.
Note that a Neumann face is such that $\uF \in \Poly{k+1}(F)$.
Note that an internal face is such that $F=\partial T \cap \partial T'$, with $T,T'\in\Th$, $T\neq T'$ and $\nTF = - \normal_{T'F}$.
\end{enumerate}
The above constraints ensure that the velocity field is pointwise divergence-free and $\boldsymbol{H}(\operatorname{div})$-conforming, 
and allow to infer that conservation of volume is enforced cell-by-cell.

Let's consider the Helmholtz decomposition of the body force appearing in the momentum equation: $\bm{f} = \gvb + \nabla \psi$, 
where $\gvb$ and $\nabla \psi$ are the divergence free and irrotational components, respectively.
Focusing on the HHO discretization of the momentum equation, $\forall \vT \in \Poly{k+1}(\elem)^{d}$, we are able to infer
\begin{equation}
\label{eq:bodyF}
\begin{aligned}
\intT \fvb \cdot \vT &= \intT \gvb \cdot \vT + \intT (\nabla \psi) \cdot \vT 
              = \intT \gvb \cdot \vT - \intT \psi \; (\nabla \cdot \vT) + \intpT \psi \; \vT \cdot \npT \\
              &= \intT \gvb \cdot \vT - \intT \pi^k_T \psi \; (\nabla \cdot \vT) + \intpT \pi^{k+1}_{\partial T} \psi \; \vT \cdot \npT 
              = \intT \gvb \cdot \vT + \intT \GhV^{k+1} (\underline{I}_{{Q},T} \psi) \cdot \vT, \quad 
\end{aligned}
\end{equation}
where, for all $q \in H^1(\Omega)$,   
$\underline{I}_{\underline{Q},T} q \coloneqq \bigl( \pi_T^{k} q, \pi_{\partial T}^{k+1} q \bigr)$,
is the local interpolation operator onto $\underline{Q}_T$.
Note that integration is employed on the last step of the first line.
Moreover, since $\vT \in \Poly{k+1}(T)$, $\nabla \cdot \vT \in \Poly{k}(T)$, and $\nTF$ is a constant vector over $F$,
$L^2$-projectors can be introduced on the second line.

Introducing \eqref{eq:bodyF} in $\sd^{\lm,\pi^p}_{T,\pBC}$, see in particular \eqref{lmPresHHO}, it becomes
clear that the irrotational part of the body force can be combined with the pressure gradient discretization
to obtain an equivalent reformulation of the momentum equation spatial dicretization.
The procedure required to achieve the desired reformulation can be summarized as follows: replace
\begin{enumerate}
\item 
$\intT \GhV^{k+1} \pTF \cdot \vT$ with $\intT \GhV^{k+1}(\pTF - \underline{I}_{{Q},T} \psi) \cdot \vT$ in \eqref{lmPresHHO},
\item
$\intT \fvb \cdot \vT$ with $\intT \gvb \cdot \vT$ in \eqref{lmPresHHO},
\item 
-$\intpTN \ppT \vpT \cdot \npT$ with $-\intpTN (\ppT - \pi_{\partial T}^{k+1} \psi) \vpT \cdot \npT$ in \eqref{lmHHObc},
\item
$\intpTN \bm{g}_\neu \cdot \vpT $ with $\intpTN (\bm{g}_\neu - \pi_{\partial T}^{k+1} \psi \npT) \cdot \vpT$ in \eqref{lmHHObc}.
\end{enumerate}
Note that the last two occurrences in the list above are required to make Neumann boundary conditions consistent 
with the reformulation and can be interpreted as adding and subtracting $\intpTN \pi_F^{k+1} \psi \, \vpT \cdot \npT$ 
to $\sd^{\lm,\pi^p}_{T}$.
Interestingly, the refactored HHO discretization, 
can be rewritten in terms of the modified pressure unknown 
$\widetilde{\underline{p}}_h \coloneqq \bigl( (\pT - \pi_T^k\psi)_{T \in \Th}, (\pF - \pi_F^{k+1}\psi)_{F \in \Fh} \bigr)$,
being otherwise indistinguishable from the original version.
This procedure allows to conclude that only the pressure field is affected by the irrotational part of body forces.

\subsection{Flux formulations} \label{sec:fluxForm}
\subsubsection{Flux formulation of the mass conservation equation} \label{sec:massFlux}
By means of the following integration by parts formula
\begin{equation} \label{eq:byPartsMass}
\int_T z_T \uT \cdot \nabla \rho_T = \intpT \bigl(\uT \cdot \nTF \bigr) \rT \zT - \int_T \rho_T \uT \cdot \nabla z_T,
\end{equation}
it is straightforward to obtain the following equivalent reformulation of the steady mass conservation equation HHO formulation introduced in Sec.~\ref{sec:SpaceDiscr}
\begin{alignat}{1}  
\sd^\mas_{T,\pBC}\bigl((\uTF,\rTF);\zTF\bigr)= & 
 - \int_T \rho_T \uT \cdot \nabla \zT + \frac{1}{2} \intpT \bigl( \uT \cdot \npT \bigr)  \bigl(\rpT \zpT \bigr)\,\nonumber \\
 &    + \intpT \left( \frac{1}{2} \bigl( \uT \cdot \npT \bigr) (\rT + \rpT) + \frac{1}{2} \bigl| \uT \cdot \npT \bigr|(\rT - \rpT) \right) (\zT - \zpT) 
    \label{eq:HHOsmasref}.
\end{alignat}
Introducing \eqref{eq:HHOsmasref} in the definition of $\sd^\mas_{T}$, see \eqref{massHHObc}, 
we finally get the following reformulation of the global residual of the mass conservation equation 
defined in \eqref{eq:HHOdiscrGlobalResHDIV}:
for all $\underline{z}_h \in\underline{Z}_h^{k}$,
$$
\begin{aligned}
&r^\umas_{h}\bigl((\underline{\boldsymbol{u}}_h,\underline{\rho}_h);\underline{z}_h\bigr) 
 = \sum_{T \in \Th} \left(\int_T \frac{\partial \rho_T}{\partial t}  z_T - \int_T \rho_T \uT \cdot \nabla \zT + \intpT \phi^\umas_{\partial T}(\uT{,}\rTF) (\zT {-} \zpT) \right)\\ 
  &  {+} \sum_{T \in \Th} \left( \intpTN \left( \bigl( \upT \cdot \npT \bigr)^{\oplus} \rpT {+}
  	\bigl( \upT \cdot \npT \bigr)^{\ominus}  g_\In \right) \zpT  
   {+} \intpTD \left( \bigl( \gVec_\dir \cdot \npT \bigr)^{\oplus} \rpT 
  	{+} \bigl( \gVec_\dir \cdot \npT \bigr)^{\ominus}  g_\In \right) \zpT \right) \\
  & + \sum_{T \in \Th} \left(\frac{1}{2} \intpT \bigl( \uT \cdot \npT \bigr)  \bigl(\rpT \zpT \bigr) - 
                             \frac{1}{2} \intpTND \bigl( \uT \cdot \npT \bigr) \bigl(\rpT \zpT \bigr) \right).\\
\end{aligned}
$$
Notice that, 
since $\rpT$ and $\zpT$ and are single valued over internal faces and $\uT$ is $H$(div)-conforming, it holds
\begin{equation}\label{eq:massBT}
\sum_{T \in \Th} \intpT \bigl( \uT \cdot \npT \bigr)  \bigl(\rpT \zpT \bigr) {=} \sum_{T \in \Th}\intpTND \bigl( \uT \cdot \npT \bigr)  \bigl(\rpT \zpT \bigr),
\end{equation}
and, accordingly, the terms on the last line of $r^\umas_{h}$ cancel.
In the spatial discretization residual above, the \emph{numerical flux} reads
$\phi^\umas_{\partial T}(\uT{,}\rTF) \coloneqq \pi^k_{\partial T}\bigl( \phi^{\rm{td}}_{\partial T}(\uT{,}\rTF) + \phi^{up}_{\partial T}(\uT{,}\rTF) \bigr)$, with
$$
\begin{aligned}
\phi_{\partial T}^{\rm{td}}(\rTF) &\coloneqq \widetilde{h}_T \left(\frac{\partial \rT}{\partial t}- \frac{\partial \rpT}{\partial t}\right), 
\; \rm{where} \; \widetilde{h}_T = \frac{h_T}{(k+1)^2};\\
\phi_{\partial T}^{\rm{up}}(\uT{,}\rTF) &\coloneqq \frac{1}{2} \left(\uT \cdot \npT \right) (\rT + \rpT) + \frac{1}{2}\bigl| \uT \cdot \npT \bigr|(\rT - \rpT)  
=  \left\{ \begin{tabular}{ll}
$\left(\uT \cdot \npT \right)  \rT$ & if $\uT \cdot \npT \geq 0$,\\
$\left(\uT \cdot \npT \right) \rpT$ & if $\uT \cdot \npT < 0$.\\
\end{tabular} \right. 
\end{aligned}
$$ 
We remark that, since $\zT - \zpT \in \Poly{k}(\partial \elem)$, we introduced the $L^2$-projection operator in the above definition of $\phi^\umas_{\partial T}$.

Let's consider the requirement $r_{h}^\umas\bigl((\underline{\boldsymbol{u}}_h,\underline{\rho}_h);\underline{z}_h\bigr) = 0$
for each $(\underline{\boldsymbol{u}}_h,\underline{\rho}_h) \in \underline{\boldsymbol{V}}_h \times \underline{Z}_h^k$ and for all $\underline{z}_h \in \underline{Z}_h^k$
focusing on the following relevant choices of the test function
\begin{enumerate}
\item $\underline{z}_h = \left( (\zT)_{T\in\Th},(0)_{F\in \Fh} \right)$, accordingly, $\forall T \in \Th$ and $\forall \zT \in \Poly{k}(T)$,
we obtain the following local mass conservation statement
$$\int_T \frac{\partial \rho_T}{\partial t}  z_T - \int_T \rho_T \uT \cdot \nabla \zT + \intpT \phi^\umas_{\partial T}(\uT{,}\rTF) \zT = 0;$$
\item $\underline{z}_h = \left( (0)_{T\in\Th},(\zF)_{F\in \Fh} \right)$, accordingly, we enforce the following continuity requirements for the numerical fluxes 
\[
\begin{aligned}
& \forall F \in \Fh^\dir,      &\pi^k_F \left(\bigl(\gVec_\dir {\cdot} \nTF \bigr)^{\oplus} \rpT {+} \bigl(\gVec_\dir {\cdot} \nTF \bigr)^{\ominus} g_\In \right)  = \pi^k_F \left(\phi^\umas_{\partial T}(\uT{,}\rTF) \right),          &  \\
& \forall F \in \Fh^\neu,      &\pi^k_F \left(\bigl(\uT {\cdot} \nTF \bigr)^{\oplus} \rpT {+} \bigl(\uT {\cdot} \nTF \bigr)^{\ominus} g_\In \right)  = \pi^k_F \left( \phi^\umas_{\partial T}(\uT{,}\rTF) \right),          &  \\
& \forall F \in \Fh^\internal, &\; \pi^k_F \left(\phi^\umas_{\partial T}(\uT{,}\rTF)\right) + \pi^k_F \left(\phi^\umas_{\partial T'}(\uTn{,}\rTFn)\right) = 0.      & 
\end{aligned}
\]
The above conditions can be rewritten to emphasize the upwind nature of the numerical flux 
\[
\begin{aligned}
& \forall F \in \Fh^\dir,      & \pi^k_{F}\left( \widetilde{h}_T\frac{\partial \rF}{\partial t} {+} \left| \uT {\cdot} \nTF \right| \, \rF \right) =& \pi^k_{F}\left( \widetilde{h}_T\frac{\partial \rT}{\partial t} {+} \mathrm{upw}(\gVec_\dir , \rT,  g_\In)  \right), & \\
& \forall F \in \Fh^\neu,      & \pi^k_{F}\left( \widetilde{h}_T\frac{\partial \rF}{\partial t} {+} \left| \uT {\cdot} \nTF \right| \, \rF  \right) =& \pi^k_{F}\left( \widetilde{h}_T\frac{\partial \rT}{\partial t} {+} \mathrm{upw}(\uT, \rT, g_\In)  \right), & \\
& \forall F \in \Fh^\internal, &\pi^k_{F}\left( (\widetilde{h}_T {+} \widetilde{h}_{T'}) \frac{\partial \rF}{\partial t} {+} \left| \uT {\cdot} \nTF \right| {\rF} \right)  
 =& \pi^k_{F}\left({\widetilde{h}_T}\frac{\partial \rT}{\partial t} {+} \widetilde{h}_{T'}\frac{\partial \rTn}{\partial t} {+} \mathrm{upw}( \uT, \rT, \rTp) \right), &\\
\end{aligned}
\]
where $$\mathrm{upw}(\bm{u}, a , b)  \coloneqq \left\{
\begin{tabular}{ll}
$\left| \bm{u} \cdot \nTF \right| a$, & \rm{if} \; $\bm{u} \cdot \nTF \geq 0$,\\
$\left| \bm{u} \cdot \nTF \right| b$, & \rm{if} \; $\bm{u} \cdot \nTF < 0$.\\
\end{tabular}
\right.$$
\end{enumerate}
Note that an internal face is such that $F=\partial T \cap \partial T'$, with $T,T'\in\Th$, $T\neq T'$ and $\nTF = - \normal_{T'F}$.
We further remark that $\phi_{\partial T}^{\rm{up}}(\uT{,}\rTF) = 0$ if the normal velocity component vanishes on a face $F \in \FT$.
In this event, thanks to the additional time derivative stabilization provided by $\phi_{\partial T}^{\rm{td}}(\rTF)$, 
the face density is still well defined.
A similar strategy was proposed to stabilize the linear momentum equations in
the limit of vanishing viscosity, see \cite{Beirao-da-Veiga.Di-Pietro.ea:25}.
Note that, disregarding the time derivative stabilisation, \ie~$\phi_{\partial T}^{\rm{td}}(\rTF) = 0$, $\forall F \in \Fh^\internal$, we simply get
\begin{equation} \label{eq:massFluxNoTDS}
\pi^k_{F} (\uT \cdot \nTF \; \rF)  = \left\{
\begin{tabular}{ll}
$ \pi^k_{F} (\uT \cdot \nTF \; \rT)$, & \rm{if} \; $\bm{u} \cdot \nTF \geq 0$,\\
$ \pi^k_{F} (\uT \cdot \nTF \; \rTp)$, & \rm{if} \; $\bm{u} \cdot \nTF < 0$. \\ 
\end{tabular} \right.
\end{equation}

\subsubsection{Tentative flux formulation of the linear momentum equation} \label{sec:qdmFlux}
In this section show that, by enforcing the HHO discretization of the mass conservation equation, 
it is possible to reformulate the skew-symmetric (plus stabilization) HHO discretization of the momentum equation
as a HHO discretization of the momentum equation in conservative form.
Thus, we investigate the possibility to derive a flux formulation for the conservative form of inviscid momentum equation.
We refer to \cite[Ch. \ 4.2.3]{Di-Pietro.Droniou:20} for a flux formulation of the viscous terms.

Using integration by parts formula \eqref{eq:byPartsMass}, reworked to express the last term based on the others two terms,
it is straightforward to obtain the following equivalent reformulation of the steady mass conservation equation HHO formulation introduced in Sec.~\ref{sec:SpaceDiscr}
\begin{alignat}{1}  
\sd^\mas_{T,\pBC}\bigl((\uTF,\rTF);\zTF\bigr)= & 
 \int_T \zT \uT \cdot \nabla \rT - \frac{1}{2} \intpT \bigl( \uT \cdot \npT \bigr)  (\rT - \rpT) ( \zT + \zpT )\,\nonumber \\
 &    + \frac{1}{2} \intpT \bigl| \uT \cdot \npT \bigr|(\rT - \rpT) (\zT - \zpT) - \frac{1}{2} \intpT \bigl( \uT \cdot \npT \bigr)  \rpT \zpT.
    \label{eq:HHOsmasrefbis}.
\end{alignat}
Introducing \eqref{eq:HHOsmasrefbis} in the definition of $\sd^\mas_{T}$, see \eqref{massHHObc}, 
and summing over all mesh elements we get
$$
\begin{aligned}
&\sum_{T \in \Th} \sd^\mas_{T}\bigl((\uTF,\rTF);\zTF\bigr)=  
 \sum_{T \in \Th} \left(\intpTN \bigl( \upT \cdot \npT \bigr)^{\ominus}  (g_\In - \rpT) \zpT 
 + \intpTD \bigl( \gVec_\dir \cdot \npT \bigr)^{\ominus}  (g_\In - \rpT) \zpT \right) \nonumber \\
& {+} \sum_{T \in \Th}\left( \int_T \zT \uT \cdot \nabla \rT {-} \frac{1}{2} \intpT \bigl( \uT \cdot \npT \bigr)  (\rT - \rpT) ( \zT + \zpT )
 {+} \frac{1}{2} \intpT \bigl| \uT \cdot \npT \bigr|(\rT - \rpT) (\zT - \zpT)\right). \nonumber 
\end{aligned}
$$
Notice that, in order to recast the boundary terms and get to the first line, we used \eqref{eq:massBT} and
we rewrote the normal velocity component as follows $\uT \cdot \npT =  (\uT \cdot \npT)^\oplus + (\uT \cdot \npT)^\ominus$. 
Let's consider once again the requirement 
$$r_{h}^\umas\bigl((\underline{\boldsymbol{u}}_h,\underline{\rho}_h);\underline{z}_h\bigr) = \sum_{T \in \Th}\int_T \frac{\partial \rho_T}{\partial t}  z_T + \sum_{T \in \Th} \sd^\mas_{T}\bigl((\uTF,\rTF);\zTF\bigr)= 0$$
for each $(\underline{\boldsymbol{u}}_h,\underline{\rho}_h) \in \underline{\boldsymbol{V}}_h \times \underline{Z}_h^k$ and for all $\underline{z}_h \in \underline{Z}_h^k$.
Notice that we here disregard the time derivative stabilization for the sake of simplicity.
Choosing $\underline{z}_h = \left( (\zT)_{T\in\Th},(0)_{F\in \Fh} \right)$
we obtain the following local mass conservation statement: $\forall T \in \Th$ and $\forall \zT \in \Poly{k}(T)$,
\begin{equation} 
 \label{eq:localMassCons}
 \int_T \frac{\partial \rho_T}{\partial t}  z_T +  \int_T \zT \uT \cdot \nabla \rT {-} \frac{1}{2} \intpT \bigl( \uT \cdot \npT \bigr)  (\rT - \rpT)  \zT 
 {+} \frac{1}{2} \intpT \bigl| \uT \cdot \npT \bigr|(\rT - \rpT) \zT = 0. 
\end{equation}

Let's now introduce the local skew-symmetric (plus stabilization) 
convective term HHO function $c_T: \underline{Z}_T^{k} \times \Poly{k}(T) \times [\underline{V}_T^{k}]^2 \to \Real$, such that,
for all $(\rTF, \wT, \uTF, \vTF)\in \underline{Z}_T^{k} \times \Poly{k}(T) \times [\underline{V}_T^{k}]^2$
$$
\begin{aligned}
\bm{c}_T&(\rTF, \wT, \uTF, \vTF) \coloneqq 
     - \frac{1}{2}\intT  \rT (\uT \otimes \wT) : \nabla \vT 
    + \frac{1}{2} \intT  \rT (\vT \otimes \wT) : \nabla \uT \\
   &  + \frac{1}{2} \intpT \rpT \bigl(\wT \cdot \npT \bigr) 
      \left( \upT \cdot \vT -  \vpT \cdot \uT \right) 
     + \frac{1}{2} \intpT \rpT \bigl|\wT \cdot \npT \bigr| 
      \left( \uT - \upT \right) \cdot \left(\vT -  \vpT \right). \\
\end{aligned}
$$
The HHO formulation of \eqref{lmConvHHO1}-\eqref{lmConvHHO2}-\eqref{lmConvHHO3} can be recovered with
$\bm{c}_T \bigl(\rTF, \uT, \underline{\bm{\pi}}^p \uTF , \underline{\bm{\pi}}^k \vTF \bigr)$,
where $ \underline{\bm{\pi}}^p \uTF = (\bm{\pi}_{T}^p \uT, \bm{\pi}_{\partial T}^p \upT)$ and
$\underline{\bm{\pi}}^k \vTF = (\bm{\pi}_{T}^k \vT, \bm{\pi}_{\partial T}^k \vpT)$.
We rely on the following identities
$$
\begin{aligned}
\intpT \rT \wT {\cdot} \npT \; \uT {\cdot} \vT &= \intpT \bigl( \rT (\wT \otimes \uT) \vT \bigr) \cdot \npT 
                                               = \intT \nabla \cdot \bigl( \rT (\wT \otimes \uT) \vT \bigr)  \\
                                               &= \intT \bigl( \nabla \cdot (\rT \uT \otimes \wT) \bigr)\cdot \vT + \int \rT \uT \otimes \wT : \nabla \vT \\  
                                               &= \intT  \rT  \vT \otimes \wT : \nabla \uT{+}  \intT  \bigl(\nabla \cdot (\rT\wT) \bigr) \uT \cdot \vT 
                                                                                                       {+}  \intT \rT \uT \otimes \wT : \nabla \vT,
\end{aligned}
$$
to obtain the useful integration by parts formula 
$$
\intT  \rT  \vT \otimes \wT : \nabla \uT = \intpT \rT \wT \cdot \npT \; \uT \cdot \vT - \intT  \nabla \cdot (\rT \wT) \; \uT \cdot \vT - \intT \rT \uT \otimes \wT : \nabla \vT,
$$
which, introduced in $\bm{c}_T$, yields
$$
\begin{aligned}
\bm{c}_T&(\rTF, \wT, \uTF, \vTF) = 
     - \intT  \rT (\uT \otimes \wT) : \nabla \vT 
    - \frac{1}{2} \intT  \nabla \cdot (\rT  \wT) \; \uT \cdot \vT 
     + \frac{1}{2} \intpT \rT \wT \cdot \npT \; \uT \cdot \vT\\
   &  + \frac{1}{2} \intpT \rpT \bigl(\wT \cdot \npT \bigr) 
      \left( \upT \cdot \vT -  \vpT \cdot \uT \right) 
    + \frac{1}{2} \intpT \rpT \bigl|\wT \cdot \npT \bigr| 
      \left( \uT - \upT \right) \cdot \left(\vT -  \vpT \right). \\
\end{aligned}
$$
We further reformulate $\bm{c}_T$ adding and subtracting $\frac{1}{2} \intpT \rpT \wT \cdot \npT \; \uT \cdot \vT$ and
$\frac{1}{2} \intpT \rpT (\wT \cdot \npT) \upT \cdot \vpT$.
After some trivial algebra the convective term HHO formulation can be rewritten as follows
$$
\bm{c}_T(\rTF, \wT, \uTF, \vTF) 
 = \widetilde{\bm{c}}_T(\rTF, \wT, \uTF, \vTF) - \frac{1}{2} \widehat{\bm{c}}_T(\rTF, \wT, \uT, \vT) 
$$
where $\widetilde{\bm{c}}_T: \underline{Z}_T^{k} \times \Poly{k}(T) \times [\underline{V}_T^{k}]^2 \to \Real$, is such that
for all $(\rTF, \wT, \uTF, \vTF)\in \underline{Z}_T^{k} \times \Poly{k}(T) \times [\underline{V}_T^{k}]^2$ 
$$
\begin{aligned}
\widetilde{\bm{c}}_T(\rTF, \wT, \uTF, \vTF) \coloneqq & 
     - \intT  \rT (\uT \otimes \wT) : \nabla \vT  + \frac{1}{2} \intpT \rpT \bigl(\wT \cdot \npT \bigr)   \upT \cdot  \vpT \\ 
   &  + \frac{1}{2} \intpT \rpT \Bigl( (\wT \cdot \npT ) ( \uT + \upT) + \bigl|\wT \cdot \npT \bigr| ( \uT - \upT ) \Bigr)\cdot (\vT -  \vpT),  \\
\end{aligned}
$$
and $\widehat{\bm{c}}_T(\rTF, \wT, \uT, \vT): \underline{Z}_T^{k} \times [\Poly{k}(T)]^3 \to \Real$ is such that, 
for all $(\rTF, \wT, \uT, \vT) \in \underline{Z}_T^{k} \times [\Poly{k}(T)]^3$
$$
\begin{aligned}
 \widehat{\bm{c}}_T(\rTF, \wT, \uT, \vT)  \coloneqq & \intT  \nabla \cdot (\rT \wT) \; \uT \cdot \vT - \intpT (\rT - \rpT) \wT \cdot \npT \; \uT \cdot \vT. \\
\end{aligned}
$$
We further rework $\widehat{\bm{c}}_T$
adding and subtracting
$
\frac{1}{2} \intpT \bigl( \wT \cdot \npT + \bigl| \wT \cdot \npT \bigr|\bigr) (\rT - \rpT) \uT \cdot \vT
$
to obtain
$$
\begin{aligned}
 \widehat{\bm{c}}_T&(\rTF, \wT, \uT, \vT)  = \intT  \nabla \cdot (\rT \wT) \; \uT \cdot \vT 
- \frac{1}{2} \intpT (\rT - \rpT) \wT \cdot \npT \; \uT \cdot \vT \\
 &+ \frac{1}{2} \intpT (\rT - \rpT) \bigl|\wT \cdot \npT \bigr|\; \uT \cdot \vT - 
\frac{1}{2} \intpT \bigl( \wT \cdot \npT + \bigl| \wT \cdot \npT \bigr|\bigr) (\rT - \rpT) \uT \cdot \vT. \\
\end{aligned}
$$

The local HHO discretization of inertial forces reads 
$$
\begin{aligned}
&\underbrace{\int_T \left( \rT \frac{\partial \uT}{\partial t} + \frac{\uT}{2} \frac{\partial \rT }{\partial t} \right) \cdot \vT + \bm{c}_T(\rTF, \uT, \underline{\bm{\pi}}^p\uTF, \underline{\bm{\pi}}^k\vTF)}_{\text{local HHO inertial forces in skew-symmetric (plus stabilization) form}} \\
& \qquad = \int_T \left( \rT \frac{\partial \uT}{\partial t} + \frac{\uT}{2} \frac{\partial \rT }{\partial t} \right) \cdot \vT +
\widetilde{\bm{c}}_T(\rTF, \uT, \underline{\bm{\pi}}^p\uTF, \underline{\bm{\pi}}^k\vTF) - \frac{1}{2} \widehat{\bm{c}}_T(\rTF, \uT, \bm{\pi}^p_T\uT, \bm{\pi}^k_T \vT) \\ 
& \qquad = \underbrace{\int_T \frac{\partial (\rT \uT)}{\partial t} \cdot \vT + \widetilde{\bm{c}}_T(\rTF, \uT, \underline{\bm{\pi}}^p\uTF, \underline{\bm{\pi}}^k\vTF)}_{\text{local HHO inertial forces in conservative form}} - \frac{1}{2} \underbrace{\left(\int_T \frac{\partial \rT }{\partial t} \uT \cdot \vT + \widehat{\bm{c}}_T(\rTF, \uT, \bm{\pi}^p_T\uT, \bm{\pi}^k_T \vT)\right)}_{\text{related to the local mass conservation statement in \eqref{eq:localMassCons}}}.
\end{aligned}
$$
indeed, since $\uT$ is divergence free, we are able to infer
\begin{equation} \label{eq:localMassAlmost}
\begin{aligned}
 &\int_T \frac{\partial \rT }{\partial t} \uT \cdot \vT + 
     \widehat{\bm{c}}_T(\rTF, \uT, \bm{\pi}^p_T\uT, \bm{\pi}^k_T \vT) \\
  & = \int_T \frac{\partial \rT }{\partial t} \uT \cdot \vT 
   + \intT  \nabla \rT \cdot \uT \; \bm{\pi}^p_T\uT \cdot \bm{\pi}^k_T\vT 
- \frac{1}{2} \intpT (\rT - \rpT) \uT \cdot \npT \; \bm{\pi}^p_T\uT \cdot \bm{\pi}^k_T\vT  \\
 & + \frac{1}{2} \intpT (\rT - \rpT) \bigl|\uT \cdot \npT \bigr|\; \bm{\pi}^p_T\uT \cdot \bm{\pi}^k_T\vT - 
\frac{1}{2} \intpT \bigl( \uT \cdot \npT + \bigl| \uT \cdot \npT \bigr|\bigr) (\rT - \rpT) \bm{\pi}^p_T\uT \cdot \bm{\pi}^k_T\vT. 
\end{aligned}
\end{equation}
Note that, 
since $$\left\{
\begin{tabular}{ll}
$ \pi^k_{\partial T} (\uT \cdot \nTF \; (\rT - \rpT )) = 0 $, & \rm{if} \; $\bm{u} \cdot \nTF \geq 0$, according to \eqref{eq:massFluxNoTDS}, \\
$ \uT \cdot \npT + \bigl| \uT \cdot \npT \bigr| = 0 $, & \rm{if} \; $\bm{u} \cdot \nTF < 0$;  \\
\end{tabular}\right.
$$
it holds
\begin{equation} \label{eq:nullTermMass}
\intpT \bigl( \uT \cdot \npT + \bigl| \uT \cdot \npT \bigr|\bigr)  (\rT - \rpT) \; \zT = 0, \; \forall \zT \in \Poly{k}(T). 
\end{equation}
According to \eqref{eq:nullTermMass} and forgetting about the polynomial degree discrepancies, 
note that, in \eqref{eq:localMassCons}, $\zT \in \Poly{k}(T)$, while, in \eqref{eq:localMassAlmost}, 
$\uT \cdot \vT \in \Poly{2k + 2}(T)$ and  $\bm{\pi}^p_T\uT \cdot \bm{\pi}^k_T \vT \in \Poly{p+k}(T)$,
the formulation in \eqref{eq:localMassAlmost} falls back to the local mass conservation statement in \eqref{eq:localMassCons}.
This implies that the local HHO inertial forces formulation in skew-symmetric (plus stabilization) form can be replaced with 
the local HHO inertial forces in conservative form in order to derive a flux formulation of the momentum equation.
We remark that the use of \eqref{eq:massFluxNoTDS} to obtain \eqref{eq:nullTermMass} 
is consistent with the fact that \eqref{eq:localMassCons} was obtained disregarding the time derivative stabilization.
We further remark that \eqref{eq:nullTermMass} and \eqref{eq:localMassCons}, with $\zT$ replaced by $\bm{\pi}^p_T\uT \cdot \bm{\pi}^k_T \vT$, 
holds true at the lowest polynomial degree, indeed $\bm{\pi}^p_T\uT \cdot \bm{\pi}^k_T\vT \in \Poly{k}(T)$ if $p{=}k{=}0$.

In what follows, we tackle the flux formulation of 
momentum equation in conservative form, focusing on the inviscid limit ($\mu = 0$), considering homogeneous forcing term ($\bm{f} = \bm{0}$) and
Dirichlet boundary conditions, \ie $\partial \Omega_{\dir} = \partial \Omega$.
Given $(\uTF, \rTF, \pTF)\in \underline{\boldsymbol{V}}_T^{k}\times \underline{Z}_T^{k}\times \underline{Q}_T^{k+1}$,
the local spatial discretization of the steady inviscid linear momentum in conservative form 
$\widetilde{\sd}_{T,\pBC}^{\ilm,\pi^p}\bigl((\uTF,\rTF,\pTF); \cdot \bigr):\underline{\boldsymbol{V}}_T^{k}\to\Real$ is such that,
for all $\vTF\in\underline{\boldsymbol{V}}_T^{k}$ 
$$
\begin{aligned}
&\widetilde{\sd}_{T}^{\ilm,\pi^p}\bigl((\uTF,\rTF,\pTF); \vTF \bigr) =  \widetilde{\bm{c}}_T(\rTF, \uT, \underline{\bm{\pi}}^p\uTF, \underline{\bm{\pi}}^k\vTF) + \intT \GhV^{k+1} \pTF \cdot \vT \\
   &- \frac{1}{2} \intpTD \rpT \bigl( \uT {\cdot} \npT \bigr)  \bigl(\bm{\pi}_{\partial T}^p \upT {\cdot} {\bm{\pi}_{\partial T}^k}\vpT \bigr) 
   + \intpTD \left(\bigl( \bm{g}_\dir \cdot \npT \bigr)^{\oplus} \rpT + \bigl( \bm{g}_\dir \cdot \npT \bigr)^{\ominus} g_\In \right) \; (\bm{g}_\dir \cdot \bm{\pi}^k_{\partial T}\vpT) \nonumber 
\end{aligned}
$$
Note that, as compared with skew-symmetric spatial discretization $\sd_{T,\pBC}^{\lm,\pi^p}$, introduced in Sec.~\ref{sec:SpaceDiscr},
the HHO formulation of \eqref{lmConvHHO1}-\eqref{lmConvHHO2}-\eqref{lmConvHHO3} has been replaced by
$\widetilde{\bm{c}}_T(\rTF, \uT, \underline{\bm{\pi}}^p\uTF, \underline{\bm{\pi}}^k\vTF)$, the viscous terms of \eqref{lmDiffHHO}
have been disregarded, and Dirichlet boundary conditions have been accounted for as in \eqref{lmHHObc}.
The global residual of the unsteady inviscid linear momentum HHO discretization can be obtained 
adding the unsteady term and summing over all mesh elements
$$
\begin{aligned}
 &\widetilde{r}_{h}^{\uilm,\pi^p}\bigl((\underline{\boldsymbol{u}}_h,\underline{\rho}_h,\underline{p}_h);\underline{\boldsymbol{v}}_h\bigr)  = \sum_{T \in \Th}\int_T \frac{\partial (\rT \uT)}{\partial t} \cdot \vT  + \sum_{T \in \Th}\widetilde{sd}^{\ilm,\pi^p}_{T} \bigl((\uTF,\rTF,\pTF); \vTF\bigr) \\
& = \sum_{T \in \Th} \left( \int_T \frac{\partial (\rT \uT)}{\partial t} \cdot \vT {-} \intT  \rT (\bm{\pi}^p_T\uT \otimes \uT) : \nabla \bm{\pi}^k_T\vT {-} \intT \pT \;  \nabla \cdot \vT \right)
   {+} \sum_{T \in \Th}  \intpT \ppT \nTF \cdot ( \vT {-} \vpT)   \\
   &  + \sum_{T \in \Th}\frac{1}{2} \intpT \rpT \Bigl( (\uT \cdot \npT ) ( \bm{\pi}^p_T\uT + \bm{\pi}_{\partial T}^p\upT) + \bigl|\uT \cdot \npT \bigr| ( \bm{\pi}^p_T\uT - \bm{\pi}_{\partial T}^p\upT ) \Bigr)\cdot (\bm{\pi}^k_T \vT -  \vpT) \\
  & + \sum_{T \in \Th}\intpTD \left(\bigl( \bm{g}_\dir \cdot \npT \bigr)^{\oplus} \rpT + \bigl( \bm{g}_\dir \cdot \npT \bigr)^{\ominus} g_\In \right) \; (\bm{g}_\dir \cdot \vpT) + \sum_{T \in \Th} \intpTD  \ppT \npT \cdot \vpT.
\end{aligned}
$$
Note that, to obtain the residual formulation above, we relied upon
$$
\begin{aligned}
- \sum_{T \in \Th} \intpT \rpT \bigl(\uT \cdot \npT \bigr) \bm{\pi}_{\partial T}^p\upT \cdot  \vpT + \sum_{T \in \Th}\intpTD \rpT \bigl(\uT \cdot \npT \bigr) \bm{\pi}_{\partial T}^p\upT \cdot  \vpT &= 0, \\
 - \sum_{T \in \Th}  \intpT \ppT \vpT \cdot \nTF  + \sum_{T \in \Th}  \intpTD \ppT \vpT \cdot \nTF &= 0,
\end{aligned}
$$
which can be easily verified since $\rpT$, $\ppT$, $\upT$ and $\vpT$ are single valued over internal faces and $\uT$ is $H$(div)-conforming, 
Moreover, since $\forall F \in \Fh^{\internal,\dir}$, $\vF \in \Poly{k}(F)$, we introduced $\bm{\pi}_{\partial T}^k\vpT = \vpT$ and we
we expanded the discrete pressure gradient operator according to its definition.

Let's consider requirement 
$\widetilde{r}_{h}^{\uilm,\pi^p}\bigl((\underline{\boldsymbol{u}}_h,\underline{\rho}_h,\underline{p}_h);\underline{\boldsymbol{v}}_h\bigr)= 0$
for each $(\underline{\boldsymbol{u}}_h,\underline{\rho}_h) \in \underline{\boldsymbol{V}}_h \times \underline{Z}_h^k$ 
and for all $\underline{\bm{v}}_h \in \underline{\boldsymbol{V}}_h^k$.
Choosing $\underline{\bm{v}}_h = \left( (\vT)_{T\in\Th},(0)_{F\in \Fh} \right)$
we obtain the following local momentum conservation statement: $\forall T \in \Th$ and $\forall \vT \in \Poly{k+1}(T)^d$,
\begin{equation} \label{eq:momTcons}
\int_T \frac{\partial (\rT \uT)}{\partial t} {\cdot} \vT  {-} \intT  \rT (\bm{\pi}^p_T\uT \otimes \uT) {:} \nabla \bm{\pi}^k_T\vT {-} \intT \pT \; \nabla \cdot \vT {+} 
\intpT \bm{\psi}^{\rm{up}}_{\partial T} {\cdot} \bm{\pi}^k_T \vT + \intpT \ppT \npT {\cdot} (\vT {-} \bm{\pi}_{T}^k\vT)  {=} 0,
\end{equation}
where, disregarding the paradigm disruption induced by the last term, the (inviscid) momentum \emph{numerical flux} reads
$$
\bm{\psi}^{\rm{up}}_{\partial T}(\uTF{,}\rpT) \coloneqq \bm{\pi}^k_{\partial T}\left( \frac{1}{2}\rpT \Bigl( (\uT \cdot \npT ) ( \bm{\pi}^p_T\uT + \bm{\pi}_{\partial T}^p\upT) + \bigl|\uT \cdot \npT \bigr| ( \bm{\pi}^p_T\uT - \bm{\pi}_{\partial T}^p\upT ) \Bigr) + \ppT \npT \right).$$
Choosing $\underline{\bm{v}}_h = \left( (0)_{T\in\Th},(\vF)_{F\in \Fh} \right)$, we enforce the following continuity requirements on the numerical fluxes 
\[
\begin{aligned}
& \forall F \in \Fh^\dir,      &\bm{\pi}^k_F \left(\Bigl(\bigl(\gVec_\dir \cdot \nTF \bigr)^{\oplus} \rpT {+} \bigl(\gVec_\dir \cdot \nTF \bigr)^{\ominus} g_\In \Bigr) \gVec_\dir + \ppT \nTF \right)  
= \bm{\pi}^k_F \left(\bm{\psi}^{\rm{up}}_{\partial T}(\uTF{,}\rpT) \right),          &  \\
& \forall F \in \Fh^\internal, &\; \bm{\pi}^k_F \left(\bm{\psi}^{\rm{up}}_{\partial T}(\uTF{,}\rpT)\right) + \bm{\pi}^k_F \left(\bm{\psi}^{\rm{up}}_{\partial T'}(\uTFn{,}\rpT)\right) = 0.      & 
\end{aligned}
\]
We remark that, thanks to the $L^2$-projection of the cell test function $\bm{\pi}^k_T \vT$, $\vT \in \Poly{k+1}(T)$,
which we introduced in all the contributions pertaining to the HHO discretization of the convective term, 
the nonlinear convective flux can be treated as a proper numerical flux.
As opposite, due to the last term in \eqref{eq:momTcons}, the normal hydrostatic stresses contributions do not qualify as numerical fluxes.
Note that, as discussed in Sec.~\ref{sec:hdiv}, choosing $\vT \in \Poly{k+1}(T)$ is crucial for achieving $H$(div) conformity.

\subsection{Stability} \label{sec:stab}
In this section we tackle the stability of the HHO formulation in case of homogeneous Dirichlet boundary conditions and body force, 
\ie $\partial \Omega_{\dir} = \partial \Omega$, and $\boldsymbol{g}_{\dir} = \boldsymbol{f} = 0$. 
In order to show that, in the aforementioned setting,  
the HHO spatial discretizations are dissipative,
we start by choosing $\vTF = \uTF$ in the steady linear momentum HHO discretization 
and $\zTF = \rTF$ in the steady mass HHO discretization, to get
\begin{alignat}{1}
&\sd^{\lm,\pi^k}_{T} \bigl((\uTF,\rTF,\pTF);\uTF\bigr) =  
   \frac{1}{2} \intpT \rpT \bigl|\uT \cdot \npT \bigr| 
      \left| \bm{\pi}_{T}^k\uT -\bm{\pi}_{\partial T}^k \upT \right|^2 \nonumber \\
   &\qquad +  \intT \mu \, {\GhT^k} \uTF : {\GhT^k} \uTF   
    + \intpT \frac{\mu}{h_F} \, \left| \bm{\pi}_{\partial T}^k(\uT - \upT) \right|^2 
    + \intT \GhV^{k+1} \pTF \cdot \uT + \intpTD \frac{\mu}{h_F} \, \left| \upT \right|^2,
\label{eq:stabLM} \\
&\sd^\mas_{T}\bigl((\uTF,\rTF);\rTF\bigr)  =  
   \frac{1}{2} \intpT \bigl|\uT \cdot \nTF \bigr| \bigl( \rT - \rpT \bigr)^2.
\label{eq:stabMas} 
\end{alignat}
We remark that, for the momentum equation spatial discretization, we consider $\sd^{\lm,\pi^p}_{T}$ with $p{=}k$.
In order to recast the HHO spatial discretizations as we did above, the following considerations are in order.
The two terms appearing in lines \eqref{lmConvHHO1}-\eqref{lmConvHHO2}, and the two terms in lines \eqref{massHHO}-\eqref{massHHO2},
are additive inverses if $\vTF = \uTF$ and if $\zTF = \rTF$, respectively.
In case of homogeneous Dirichlet, according to \eqref{eq:HDivTraces}, we get $\uT \cdot \npT = \pi_F^{k+1}(\bm{g}_D {\cdot} \nTF)  = 0$.
Accordingly, in \eqref{massHHObc}, all the terms involving integrals over Dirichlet boundaries can be disregarded.
Following the same reasoning, most of the terms in \eqref{lmHHObc} can be omitted, notice in particular that the terms 
involving gradient reconstruction operators are skew-symmetric. 
As a consequence, the sole contribution arising from the weak imposition of Dirichlet BCs is the last term in \eqref{eq:stabLM}.

The pressure-velocity coupling in \eqref{eq:stabLM} does not alter the energy balance, indeed we are able to infer
\begin{equation}
\label{eq:pvStab}
  \sum_{T \in \Th} \intT \GhV^{k+1} \pTF \cdot \uT = 
  - \sum_{T \in \Th}  \intT \pT \;  \nabla \cdot \uT 
   + \sum_{T \in \Th}  \intpT \ppT \uT \cdot \nTF  
   = \sum_{T \in \Th}  \intpTD \ppT \uT \cdot \nTF 
   = 0,
\end{equation}
where, in the first step, we introduced the definition of the gradient reconstruction operator, and in the second step, 
we used the fact that the element velocity $\uT$ is divergence-free and H-div conforming, see Sec.~\ref{sec:hdiv},
together with the single-valuedness of $\ppT$ on internal faces.
The non-dissipation property follows, since, in case of homogeneous Dirichlet BCs, $\uT \cdot \nTF = \pi_F^{k+1}(\bm{g}_D {\cdot} \nTF) = 0$.
Accordingly, we remark that all the terms on the right hand side of \eqref{eq:stabLM}-\eqref{eq:stabMas} but $\intT \mu \, {\GhT^k} \uTF : {\GhT^k} \uTF$ are non-negative.
As demonstrated in \cite[Ch. \ 5.1.6]{Di-Pietro.Droniou:20}, see also \cite[Remark 2.2]{Cockburn.Di-Pietro.ea:16}, 
the HHO discretization of the Laplace operator introduced in \eqref{lmDiffHHO} is coercive \ie for all $T \in \Th$ and all $\vTF \in \underline{\boldsymbol{V}}_T^{k}$, it holds
$$ \intT \, {\GhT^k} \vTF : {\GhT^k} \vTF   
    + \intpT \frac{1}{h_F} \, \left| \bm{\pi}_{\partial T}^k(\vT - \vpT) \right|^2 \gtrsim \intT \, \nabla \vT : \nabla \vT   
    + \intpT \frac{1}{h_F} \, \left| (\vT - \vpT) \right|^2, $$
where the hidden (positive) constant is independent of $h$ and $T$.
As a consequence, we are able to conclude that
both $\sd^{\lm,\pi^k}_{T} \bigl((\uTF,\rTF,\pTF);\uTF\bigr)$ and $\sd^\mas_{T}\bigl((\uTF,\rTF);\rTF\bigr)$ are non-negative.

Let's now consider the unsteady linear momentum HHO discretization and the unsteady mass HHO discretization 
defined in \eqref{eq:uLmSD} and \eqref{eq:uMasSD}, respectively.
In order to reformulate the time derivative in the momentum equation, we introduce the following identity 
\begin{equation}
\label{eq:dtLM}
\left( \rT \frac{\partial \uT}{\partial t} + \frac{\uT}{2} \frac{\partial \rT }{\partial t} \right) \cdot \uT = \frac{1}{2}\frac{\partial (\rho \uT \cdot \uT)}{\partial t}= \frac{1}{2}\frac{\partial \left|\sqrt{\rho} \uT \right|^2 }{\partial t}. 
\end{equation}
Setting $\vTF = \uTF$ in $\sd^{\ulm,\pi^k}_{T}$ and $\zTF = \rTF$ in $\sd^\umas_{T}$, using \eqref{eq:dtLM}
and summing over all mesh elements, we arrive at the following stability argument
\begin{alignat}{1}
\frac{d}{dt} \sum_{T \in \Th}\intT \bigl| \sqrt{\rho_T} \uT \bigr|^2 & = -2 \sum_{T \in \Th}\sd^{\lm,\pi^k}_{T} \bigl((\uTF,\rTF,\pTF);\uTF\bigr) \leq 0, \nonumber \\
\frac{d}{dt} \sum_{T \in \Th}\left( \intT \bigl( \rho_T \bigr)^2 + \frac{h_T}{(k+1)^2} \intpT \bigl( \rT -\rpT \bigr)^2 \right)& = -2\sum_{T \in \Th}\sd^\mas_{T}\bigl((\uTF,\rTF);\rTF\bigr)  \leq 0.   
\end{alignat}

\subsection{ESDIRK temporal discretization} \label{sec:TempDiscr}
The temporal discretization must take into account that the HHO space discretization of the variable density problem 
is formulated in terms of primitive variables $\uVec,\rho$ rather then conservative variables $\rho \uVec, \rho$.
This latter choice allows to obtain a $\boldsymbol{H}(\operatorname{div})$-conforming velocity field and achieve pressure-robustness. 
In order to rewrite the variable density problem as a system of Ordinary Differential Equations (ODEs), 
we first introduce the algebraic problem corresponding to the semi-discrete HHO formulation. 
For the sake of notation, in this section, we disregard the time derivative stabilization in \eqref{eq:uMasSD}. 

The unknowns for a mesh cell $\elem\in\Th$ correspond to the coefficients of the expansions of the velocity, density and pressure in the selected local bases.
Assuming that the unknowns are ordered so that cell velocities come first and boundary velocities next, these coefficients are collected in the vectors
\begin{equation*}
  \text{
    $\UVEC{U}_T=\begin{bmatrix} \VEC{U}_T \\ \VEC{U}_{\partial T} \end{bmatrix}$,
    $\UVEC{R}_T=\begin{bmatrix} \VEC{R}_T \\ \VEC{R}_{\partial T} \end{bmatrix}$,
    and $\UVEC{P}_T=\begin{bmatrix} \VEC{P}_T \\ \VEC{P}_{\partial T} \end{bmatrix}$,
  }
\end{equation*}
where the block partition is the one naturally induced by the selected ordering of unknowns.
For convenience we collect element and face unknowns in the vectors
\begin{equation*}
  \text{
    ${\VEC{W}}_T=\begin{bmatrix} \VEC{U}_T \\ \VEC{R}_{T} \\ \VEC{P}_T \end{bmatrix}$, 
    $\;{\VEC{W}}_T^{!p}=\begin{bmatrix} \VEC{U}_T \\ \VEC{R}_{T} \\ \VEC{0} \end{bmatrix}$, 
    $\;{\VEC{W}}_{\partial T}=\begin{bmatrix} \VEC{U}_{\partial T} \\ \VEC{R}_{\partial T} \\ \VEC{P}_{\partial T} \end{bmatrix}$,$\;$
     and we define $\UVEC{W}_T \coloneqq \begin{bmatrix} \VEC{W}_T \\ \VEC{W}_{\partial T} \end{bmatrix}$.
  }
\end{equation*}
Notice that the vector of element unknowns omitting the pressure in ${\VEC{W}}_T^{!p}$ have been introduced to conveniently express time derivatives
in what follows. 
We also introduce the notation $\UVEC{W}_{\Th}$ for the global vector collecting all element and face unknowns.

Let's introduce the vector representations
$$\UVEC{D}_{T}^\lm=\begin{bmatrix}\VEC{D}_{T}^\lm\\ \VEC{D}_{\partial T}^\lm\end{bmatrix}, \;
\UVEC{D}_{T}^\mas=\begin{bmatrix}\VEC{D}_{T}^\mas\\ \VEC{D}_{\partial T}^\mas\end{bmatrix}, \;
\UVEC{D}_{T}^\vol=\begin{bmatrix}\VEC{D}_{T}^\vol\\ \VEC{D}_{\partial T}^\vol\end{bmatrix},
$$
of the local spatial discretizations
$\sd^\lm_{T}, \, \sd^\mas_{T}, \, \sd^\vol_{T}$
defined in \eqref{HHObc}.
The block partition is the one induced by mimicking the selected ordering of unknowns for the expansion bases.
For convenience we collect element and face components in the vectors
$$
\text{
$\VEC{D}_{T}^s=\begin{bmatrix} \VEC{D}_{T}^\lm \\ \VEC{D}_{T}^\mas \\ \VEC{D}_{T}^\vol \end{bmatrix}$, \;
$\VEC{D}_{\partial T}^s=\begin{bmatrix} \VEC{D}_{\partial T}^\lm \\ \VEC{D}_{\partial T}^\mas \\ \VEC{D}_{\partial T}^\vol \end{bmatrix}$,
    \quad and we define $\UVEC{D}_T^s = \begin{bmatrix} \VEC{D}_T^s \\ \VEC{D}_{\partial T}^s \end{bmatrix}$.}
$$

The algebraic counterpart of problem \eqref{eq:hhoGlobalProblem} reads: find $\UVEC{W}_{\Th}$ such that
\begin{equation}
\label{HHOsemiAlg}
\sum_{T\in\Th}\MAT{M}_T \frac{d \VEC{W}_T^{!p}}{d t} + \sum_{T\in\Th}\VEC{D}_T^s(\UVEC{W}_T) = \VEC0,  \quad \text{and}  \quad  \sum_{T\in\Th}\VEC{D}_{\partial T}^s(\UVEC{W}_T) = \VEC0,
\end{equation} 
where
$$
\MAT{M}_T(\VEC{W}_T) = 
\begin{bmatrix} 
\int_T \rT \; \widetilde{\sf{v}}_T \otimes \widetilde{\sf{v}}_T & \int_T \frac{\uT}{2} \; \widetilde{\sf{v}}_T \otimes \widetilde{\sf{z}}_T &  \MAT0 \\
\MAT0 & \int_T  \widetilde{\sf{z}}_T \otimes \widetilde{\sf{z}}_T &  \MAT0 \\
\MAT0 & \MAT0 &  \MAT{I} \\
\end{bmatrix},
$$
and 
$\widetilde{\sf{v}}_T$,
$\widetilde{\sf{z}}_T$,
$\widetilde{\sf{q}}_T$
are the vectors collecting local cell bases, \ie
\begin{equation*}
\begin{array}{c}
\SPAN\left\{\widetilde{\sf{v}}^{T}\right\} = \Poly{k+1}(T)^d\\
\SPAN\left\{\widetilde{\sf{z}}_{T}\right\} = \Poly{k}(T)\\
\SPAN\left\{\widetilde{\sf{q}}_{T}\right\} = \Poly{k}(T)\\
\end{array} \quad \text{and, for any $\xVec \in T$} \quad
\begin{array}{c} 
\uT |_{\xVec} = \VEC{U}_T \cdot \widetilde{\sf{v}}_T |_{\xVec} \\ 
\rT |_{\xVec} = \VEC{R}_T \cdot \widetilde{\sf{z}}_T |_{\xVec} \\ 
\pT |_{\xVec} = \VEC{P}_T \cdot \widetilde{\sf{q}}_T |_{\xVec} \\ 
\end{array}.
\end{equation*}

Problem \eqref{HHOsemiAlg} can be finally rewritten as a system of ODEs: find $\UVEC{W}_{\Th}$ such that 
\begin{equation}
\label{HHOsemiAlgODE}
\sum_{T\in\Th}\frac{d \VEC{W}_T^{!p}}{d t} +\sum_{T\in\Th} \MAT{M}_T^{-1} \VEC{D}^s_T(\UVEC{W}_T) = \VEC0,  \quad \text{and}  \quad  \sum_{T\in\Th}\VEC{D}_{\partial T}^s(\UVEC{W}_T) = \VEC0,
\end{equation} 
where 
$$
\MAT{M}_T^{-1}(\VEC{W}_T) =   \begingroup 
\setlength\arraycolsep{1pt}
\begin{bmatrix} 
\left(\int_T \rT \widetilde{\sf{v}}_T \otimes \widetilde{\sf{v}}_T\right)^{-1} & 
-\left(\int_T \rT \widetilde{\sf{v}}_T \otimes \widetilde{\sf{v}}_T\right)^{-1} \left(\int_T \frac{\uT}{2} \widetilde{\sf{v}}_T \otimes \widetilde{\sf{z}}_T\right)
\left(\int_T  \widetilde{\sf{z}}_T \otimes \widetilde{\sf{z}}_T\right)^{-1} 
 &  \MAT{0} \\
\MAT0 & (\int_T  \widetilde{\sf{z}}_T \otimes \widetilde{\sf{z}}_T)^{-1} &  \MAT0 \\
\MAT0 & \MAT0 &  \MAT{I} \\
\end{bmatrix}.
   \endgroup 
$$

Based on multistage ESDIRK time marching strategies, the solutions $\UVEC{W}_{\Th} |_{t+\delta t}$
are computed considering an approximated time integration 
formula for the global residuals in \eqref{HHOsemiAlgODE} over the time interval $\delta t$:
\begin{equation}
\label{eq:HHOdiscrTemp}
\begin{aligned}
\sum_{T\in\Th}\frac{\delta \VEC{W}_T^{!p}}{\delta t} +\sum_{T\in\Th} \sum_{i = 1}^{s} b_i \,\MAT{M}_T^{-1}(\VEC{W}_T|_{t+\delta t_i}) \VEC{D}_T^s(\UVEC{W}_T|_{t+\delta t_i}) &= \VEC0, \\
\sum_{T\in\Th} \sum_{i = 1}^{s} b_i \, \VEC{D}_{\partial T}^s(\UVEC{W}_T|_{t+\delta t_i}) &= \VEC0, 
\end{aligned}
\end{equation}
where {$\delta \VEC{W}_X \coloneqq \UVEC{W}_X |_{t+\delta t} - \UVEC{W}_X |_t$}, and, for all $T \in \Th$,  $\UVEC{W}_T|_{t+\delta t_i}$ are the solutions of the ESDIRK $i$th-stage,
with $\delta t_i = \delta t \; \sum_{j=1}^i a_{ij}$.
The number of stages $s$ and the real coefficients $a_{ij}$, with $i = 1,\dots,s$ and $j = 1,\dots,i$, uniquely define an ESDIRK formulation.
In particular, for all schemes belonging to the ESDIRK family, it holds that $a_{11}=0$ and $b_i = a_{si}$, for $i = 1,\dots,s$.
As a consequence, since $\delta t_1 = 0 $ and $\delta t_s = \delta t$,
the approximated time integration formula in \eqref{eq:HHOdiscrTemp} is equivalent to the last ESDIRK stage
and $\UVEC{W}_{\Th}|_{t+\delta t}$ is the last stage solution.
The stages solutions are built incrementally, starting from the first stage, moving to second, and so on up to the last stage.
To complete the presentation of the time marching strategy we need to specify how the ESDIRK $i$th-stage solutions are computed. 

Given $\UVEC{W}_{\Th}|_{t+\delta t_j}$, with $j=1,\dots,i$, the space-time discretization of the ESDIRK $i$th-stage reads
\begin{subequations}
\label{eq:HHOESDIRKi}
\begin{alignat}{1}
\VEC{D}_{i,T}^{st}(\UVEC{W}_T(t))            & \coloneqq \frac{\delta_i \VEC{W}_T^{!p}}{\delta t} + \sum_{j = 1}^{i} a_{ij} \,\MAT{M}_T^{-1}(\VEC{W}_T|_{t+\delta t_j}) \VEC{D}_T^s(\UVEC{W}_T|_{t+\delta t_j}), \\
\VEC{D}_{i,\partial T}^{st}(\UVEC{W}_T(t)) & \coloneqq                                       \sum_{j = 1}^{i} a_{ij} \, \VEC{D}_{\partial T}^s(\UVEC{W}_T|_{t+\delta t_j}),  
\end{alignat}
\end{subequations}
where $\delta_i \VEC{W}_X \coloneqq \VEC{W}_X |_{t+\delta t_i} - \VEC{W}_X |_t$.
We remark that, the spatial discretizations $\UVEC{D}_T^s(\UVEC{W}_T(t))$ 
in \eqref{eq:HHOdiscrTemp}-\eqref{eq:HHOESDIRKi} are written omitting the dependence from the boundary conditions and forcing terms. 
In doing so, we implicitly assume that $\boldsymbol{f}$, $\boldsymbol{g}_\neu, \boldsymbol{g}_\dir $ and $g_\In$
are evaluated at the same time points as the variables $\uTF, \rTF, \pTF$.

The ESDIRK $i$th-stage problem, that is the time discrete counterpart of problem \eqref{eq:hhoGlobalProblem},
reads:
Given $\UVEC{W}_{\Th}|_{t+\delta t_j}$, with $j=1,\dots,i-1$, 
find $\UVEC{W}_{\Th}|_{t+\delta t_i}$ such that
\begin{equation}
\label{eq:HHOESDIRKresEq}
\sum_{T\in\Th} \VEC{D}_{i,T}^{st} (\UVEC{W}_T(t)) = \VEC0 \quad \text{and} \quad \sum_{T\in\Th} \VEC{D}_{i,\partial T}^{st}(\UVEC{W}_T(t))= \VEC0. \\
\end{equation}
where the left hand sides are the global element and face residuals, respectively, of the ESDIRK $i$th-stage.
The solution of the above problem can be sought by means of Newton's method, as detailed in the next section.

\subsection{Algebraic expression for Newton's method and static condensation} \label{sec:StaticCond}
In order to streamline the computation of the Jacobian matrix we scale the ESDIRK residual equations in \eqref{eq:HHOESDIRKresEq} 
by $\frac{1}{a_{ii}}$, to obtain the following equivalent reformulation of the $i$-th stage space-time discretization
\begin{subequations}
\begin{alignat}{1}
\VEC{\widehat{D}}_{i,T}^{st}(\UVEC{W}_T(t))   & \coloneqq \frac{\delta_i \VEC{W}_T^{!p}}{a_{ii} \, \delta t} + \MAT{M}_T^{-1}(\VEC{W}_T|_{t+\delta t_i}) \VEC{D}_T^s(\UVEC{W}_T|_{t+\delta t_i}) + \sum_{j = 1}^{i-1} \frac{a_{ij}}{a_{ii}} \,\MAT{M}_T^{-1}(\VEC{W}_T|_{t+\delta t_j}) \VEC{D}_T^s(\UVEC{W}_T|_{t+\delta t_j})\label{eq:HHOdiscrSpaceTimeT} \\ 
\VEC{\widehat{D}}_{i,\partial T}^{st}(\UVEC{W}_T(t)) & \coloneqq  \VEC{D}_{\partial T}^s(\UVEC{W}_T|_{t+\delta t_i}) + \sum_{j = 1}^{i-1} \frac{a_{ij}}{a_{ii}} \, \VEC{D}_{\partial T}^s(\UVEC{W}_T|_{t+\delta t_j}) \label{eq:HHOdiscrSpaceTimepT} 
\end{alignat}
\end{subequations}

Newton's method for solving problem \eqref{eq:HHOESDIRKresEq} reads: Find $\delta_i \UVEC{W}_{\Th}$ such that
\begin{equation}
\label{eq:Newton}
-\sum_{T \in \Th} 
\left(
\begin{bmatrix}
\MAT{M}_{i,T} \VEC{\widehat{D}}_{i,T}^{st} \\
\VEC{\widehat{D}}_{i,\partial T}^{st}
\end{bmatrix}
\right)
=
\sum_{T \in \Th} 
\left(
\begin{bmatrix}
\MAT{M}_{i,T} \MAT{\widehat{J}}_{TT} & \MAT{M}_{i,T} \MAT{\widehat{J}}_{T \partial T}\\
\MAT{\widehat{J}}_{\partial T T} &  \MAT{\widehat{J}}_{\partial T \partial T}
\end{bmatrix}
\begin{bmatrix}
\delta_i\VEC{W}_T \\
\delta_i\VEC{W}_{\partial T}
\end{bmatrix}
\right),
\end{equation}
where $\MAT{\widehat{J}}_{X Y} \coloneqq \frac{\partial \VEC{\widehat{D}}^{st}_{i,X}(\UVEC{W}_T(t))}{\partial \VEC{W}_Y|_{t+\delta t_i}}$ 
are the blocks of the local Jacobian matrix
and $\MAT{M}_{i,T} \coloneqq \MAT{M}_T(\VEC{W}_T|_{t+\delta t_i})$, replace
$\UVEC{W}_{\Th}|_{t+\delta t_i} \gets \UVEC{W}_{\Th}|_{t+\delta t_i} + \delta_i \UVEC{W}_{\Th}$
until $\delta_i \UVEC{W}_{\Th}$ is small enough.
Note that the each of the element residual equations in \eqref{eq:Newton} has been multiplied by $\MAT{M}_{i,T}$ 
in order to further simplify the code implementation of the Jacobian matrix.
In particular, since 
$$
\frac{\partial \MAT{M}_{i,T}^{-1}}{\partial \VEC{W}_T |_{t+\delta t_i}} = -\MAT{M}_{i,T}^{-1} \frac{\partial \MAT{M}_{i,T}}{\partial \VEC{W}_T |_{t+\delta t_i}} \MAT{M}_{i,T}^{-1},
\quad \text{and} \quad \frac{\partial \MAT{M}_{i,T}^{-1}}{\partial \VEC{W}_{\partial T} |_{t+\delta t_i}} = \MAT{0},
$$
the equations systems in \eqref{eq:Newton} can be rewritten and follows
\begin{equation}
\label{eq:NewtonBis}
-\sum_{T \in \Th} 
\left(
\begin{bmatrix}
\MAT{M}_{i,T} \VEC{\widehat{D}}_{i,T}^{st} \\
\VEC{\widehat{D}}_{i,\partial T}^{st}
\end{bmatrix}
\right)
=
\sum_{T \in \Th} 
\left(
\begin{bmatrix}
\MAT{M}_{i,T} \MAT{\widehat{J}}_{TT} & \MAT{J}_{T \partial T}\\
\MAT{J}_{\partial T T} &  \MAT{J}_{\partial T \partial T}
\end{bmatrix}
\begin{bmatrix}
\delta_i\VEC{W}_T \\
\delta_i\VEC{W}_{\partial T}
\end{bmatrix}
\right),
\end{equation}
where
$$
\MAT{M}_{i,T} \MAT{\widehat{J}}_{TT} = \frac{\MAT{M}_{i,T}}{a_{ii} \, \delta t_i} \frac{\partial \VEC{W}_T^{!p} |_{t+\delta t_i}}{\partial \VEC{W}_T |_{t+\delta t_i}} 
     - \frac{\partial \MAT{M}_{i,T}}{\partial \VEC{W}_T |_{t+\delta t_i}} \MAT{M}_{i,T}^{-1}\VEC{D}^{s}_{T}(\UVEC{W}_T|_{t+\delta t_i})
     + \MAT{J}_{TT}
$$
and $\MAT{J}_{X Y} = \frac{\partial \VEC{D}^{s}_{X}(\UVEC{W}_T|_{t+\delta t_i})}{\partial \VEC{W}_Y |_{t+\delta t_i}}$.
Notice that the spatial discretizations at the $j$th-stages, $j=1,...,i{-}1$, that is the 
last terms in \eqref{eq:HHOdiscrSpaceTimeT} and \eqref{eq:HHOdiscrSpaceTimepT}, 
do not appear in the Jacobian matrix related to the $i$th-stage.

Static condensation is applied to the equations systems in \eqref{eq:NewtonBis} in a cell-by-cell fashion.
The only unknowns are globally coupled are those collected in the subvector $\delta_i \VEC{W}_{\partial T}$
while the remaining unknowns collected in $\delta_i \VEC{W}_T$
can be eliminated by expressing them in terms of the former.
After performing this local elimination, each Newton method iteration amounts at solving
\begin{equation*}
  {-} \sum_{\elem \in \Th}
\left(
    \VEC{\widehat{D}}^{st}_{i,\partial T}
  {-} 
 \MAT{J}_{\partial T  T} 
 \bigl( \MAT{M}_{i,T} \MAT{\widehat{J}}_{T  T} \bigr)^{-1}
 \bigl( \MAT{M}_{i,T} \VEC{\widehat{D}}^{st}_{i,T} \bigr)
\right)
 {=}
  \sum_{\elem \in \Th}\bigl(\MAT{S}_T^{} \delta_i \VEC{W}_{\partial T}  \bigr),
\end{equation*}
where $\MAT{S}_T^{}$ denotes the Schur complement of the top left block of the left-hand side matrix in \eqref{eq:NewtonBis}, that is,
\begin{equation*}
  \MAT{S}_T^{}
  \coloneqq 
  \MAT{J}_{\partial T  \partial T} 
    -
 \MAT{J}_{\partial T  T} 
\bigl(\MAT{M}_{i,T}\MAT{\widehat{J}}_{T T}\bigr)^{-1}          
\MAT{J}_{T \partial T}   
\end{equation*}
When increasing the polynomial degree $k$, the dimension of polynomial spaces over mesh cells grows much faster that the dimension of polynomial spaces over mesh faces.
Accordingly, the possibility to statically condense the cells unknowns
is the key for achieving efficiency when high-order accurate formulation are employed.

\section{Numerical results}
Let us briefly describe the numerical setup common to the test cases presented in what follows.
Numerical integration over mesh cells and mesh faces is performed over standardized reference cells based on Gaussian quadrature rules.
Reference-to-physical-frame mappings from reference entities to mesh entities are defined by means of Lagrange polynomials.
Discrete polynomial spaces over mesh cells and mesh faces are spanned by orthonormal modal bases.
Specifically, we rely on physical frame polynomial spaces over mesh cells
and reference frame polynomial spaces over mesh faces, as proposed in \cite{Botti.Di-Pietro:18}.
Orthogonalization in the physical frame is performed cell-by-cell starting from a
monomial basis defined in a local reference frame aligned with the principal axes of inertia of each mesh cell, as described in \cite{Bassi.Botti.ea:12}.
Note that, on simplicial meshes, since planar faces and affine mappings go hand-in-hand, 
reference and physical frame polynomials provide the same accuracy, see \cite{Botti:12} for details.

We remark that Newton's method is solved up to machine precision in order to guarantee exact
conservation of volume at the discrete level, accordingly the linear solver stopping criterion
does not influence the numerical results.
In order to further improve the performance of the iterative linear solver required for the solution
of each Newton step we rely on the $p$-multilevel preconditioner strategy proposed in \cite{BottiDiPietroHHOpMG2021}.
In particular, we apply one sweep of $p$-multigrid V-cycle as a preconditioner for the FGMRES (flexible GMRES) iteration.
As a smoothing strategy within the multigrid V-cycle preconditioner we rely on a few iterations of GMRES.
All applications of prolongation and restriction operators involved in the V-cycle
iteration are performed matrix-free, that is, without assembling the global sparse matrices
associated to the operators, see \cite{BottiDiPietroHHOpMG2021} for details.
Static condensation is performed cell-by-cell at matrix assembly, 
meaning that only statically condensed global matrices are stored in memory.
It is worth to mention that the code can be executed in parallel based on a distributed memory paradigm                                                                                
and relies on the iterative solvers and preconditioners provided by the PETSc library \cite{petscTR}.
From the computational cost viewpoint, as is often the case when considering fully implicit 
time-marching strategies for incompressible fluid flow computations, the                              
computational expense associated with linear systems solutions dominates over matrix assembly CPU-times.
In practice, the ESDIRK-HHO formulation coupled with the $p$-multigrid solution strategy is best efficient when 
considering higher-order polynomials and coarse meshes.

In what follows we utilize the notation HHO-$\pi^k$ and HHO-$\pi^{k+1}$ to identify the two 
variants of the variable density HHO formulation introduced in see Sec.~\ref{sec:SpaceDiscr}. 
In particular, HHO-$\pi^k$ and HHO-$\pi^{k+1}$ are based on the momentum equation spatial discretization $\sd^{\lm,\pi^p}$ with $p{=}k$ 
and $p{=}k{+}1$, respectively. Moreover, for the sake of reporting convergence rates, 
we indicate with $\mathfrak{E}$ the error in $L^2$-norm computed against the analytical solution.
Let $\uVec_h \in \Poly{k+1}(\Th) \subset L^2(\Omega)$ be the discrete velocity obtained gluing together the cell components, 
the velocity error $\mathfrak{E} \; \uVec_h$ reads
$$
\mathfrak{E} \; \uVec_h \coloneqq \| \uVec - \uVec_h \|_{L^2 (\Omega)^d} = \left( \sum_{T \in \Th} \int_T \left( \uVec - \uVec_T \right) \cdot \left( \uVec - \uVec_T \right)\right)^{\frac{1}{2}}, 
$$
and the velocity gradient error $\mathfrak{E} \; \nabla \uVec_h$ is computed as follows 
$$
\mathfrak{E} \; \nabla \uVec_h \coloneqq \| \nabla \uVec - \nabla \uVec_h \|_{L^2 (\Omega)^{d \times d}} 
 = \left( \sum_{T \in \Th} \int_T \left( \nabla \uVec - \nabla \uVec_T \right) : \left(  \nabla \uVec - \nabla \uVec_T \right)\right)^{\frac{1}{2}}.
$$
The pressure error $\mathfrak{E} \; p_h $, the density error $\mathfrak{E} \; \rho_h $ and the divergence error $\mathfrak{E} \; \nabla \cdot \uVec_h $ are computed in a similar fashion.

\subsection{Kovasznay test case} \label{test:kova}
\begin{table}[!htb]
\small
\centering
\begin{tabular}{cccccccccc}
$\cardc{\Th}$ & $\mathfrak{E}\,{\uVec_h} $ & rate  & $\mathfrak{E}\,{\nabla \uVec_h}$    & rate & $\mathfrak{E}\,{p_h} $ & rate  &  $\mathfrak{E}\,{\rho_h}$ & $\mathfrak{E}\,{\nabla {\cdot} \uVec_h}$  \\
\hline
\multicolumn{9}{c}{$k=0$} \\
\hline
50         & 0.677   & --           & 5.818    & --         & 0.666103   & --           & 9.8e-16  & 1.8-16 \\ 
192        & 0.372   & 0.89         & 4.158    & 0.50       & 0.330301   & 1.04         & 8.5e-16  & 2.5-16 \\ 
810        & 0.188   & 0.95         & 2.267    & 0.84       & 0.153069   & 1.07         & 1.4e-15  & 2.7-16 \\ 
3126       & 0.100   & 0.93         & 1.244    & 0.89       & 0.0757808  & 1.04         & 1.3e-15  & 2.6-16 \\ 
12706      & 0.0504  & 0.98         & 0.629    & 0.97       & 0.0354129  & 1.09         & 1.6e-15  & 2.6-16 \\ 
\hline                                                                                                        
\multicolumn{9}{c}{$k=1$} \\                                                                                  
\hline                                                                                                        
50         & 0.198        & --       & 2.986   & --         & 0.111    & --        & 8.2e-14  & 3.4e-16 \\ 
192        & 0.0483       & 2.10     & 1.160   & 1.41       & 0.0238   & 2.29      & 1.5e-14  & 3.6e-16 \\ 
810        & 0.00793      & 2.51     & 0.380   & 1.55       & 0.00440  & 2.35      & 4.5e-14  & 3.2e-16   \\ 
3126       & 0.00106      & 2.98     & 0.112   & 1.80       & 0.000881 & 2.38      & 7.0e-14  & 3.3e-16 \\ 
12706      & 0.000142     & 2.86     & 0.0307  & 1.85       & 0.000202 & 2.10      & 9.1e-14  & 3.2e-16 \\ 
\hline                                                                                                        
\multicolumn{9}{c}{$k=2$} \\                                                                                  
\hline                                                                                                        
50         & 0.0418   & --       & 1.041    & --           & 0.0286   & --         & 2.3-14  & 4.3e-16 \\ 
192        & 0.00417  & 3.43     & 0.185    & 2.56         & 0.00227  & 3.77       & 5.3-14  & 3.8e-16 \\ 
810        & 0.000303 & 3.64     & 0.0276   & 2.65         & 0.000202 & 3.36       & 4.3-14  & 3.8e-16 \\ 
3126       & 2.53e-05 & 3.67     & 0.00460  & 2.65         & 2.53e-05 & 3.07       & 3.6-14  & 3.8e-16 \\ 
12706      & 1.64e-06 & 3.90     & 0.000589 & 2.93         & 3.23e-06 & 2.94       & 1.3-13  & 3.8e-16 \\ 
\hline                                                                                                        
\multicolumn{9}{c}{$k=3$} \\                                                                                  
\hline                                                                                                        
50         & 0.00827  & --       & 0.266    & --           & 0.00198  & --      & 5.4e-14 & 4.0e-16    \\ 
192        & 0.000383 & 4.57     & 0.0261   & 3.45         & 0.000150 & 3.83    & 2.9e-14 & 4.2e-16    \\ 
810        & 2.02e-05 & 4.09     & 0.00243  & 3.30         & 1.23e-05 & 3.47    & 2.3e-14 & 3.6e-16    \\ 
3126       & 6.35e-07 & 5.13     & 0.000163 & 4.00         & 8.05e-07 & 4.05    & 5.9e-14 & 3.9e-16    \\ 
12706      & 2.29e-08 & 4.74     & 1.17e-05 & 3.76         & 5.13e-08 & 3.93    & 7.9e-14 & 3.8e-16    \\ 
\hline                                                                                                        
\multicolumn{9}{c}{$k=4$} \\                                                                                  
\hline                                                                                                        
50         & 0.000933 & --    & 0.0453   & --           & 0.000487 & --     & 2.4e-14 & 4.8e-16  \\ 
192        & 2.31e-05 & 5.49  & 0.00227  & 4.45         & 1.19e-05 & 5.52   & 4.2e-14 & 4.8e-16  \\ 
810        & 5.19e-07 & 5.28  & 9.29e-05 & 4.45         & 4.28e-07 & 4.62   & 5.3e-14 & 4.3e-16  \\ 
3126       & 1.17e-08 & 5.62  & 4.15e-06 & 4.60         & 1.47e-08 & 4.99   & 1.0e-13 & 4.5e-16  \\ 
12706      & 1.98e-10 & 5.81  & 1.37e-07 & 4.86         & 5.14e-10 & 4.79   & 5.3e-13 & 4.4e-16  \\ 
\end{tabular}
\caption{Kovasnzay test case: spatial convergence rates for $k=0,1,2,3,4$ HHO-$\pi^k$ formulations over a Delaunay type $h$-refined triangular elements mesh sequence. 
\label{tab:KovaNumResPk}}
\end{table}

\begin{table}[!htb]
\small
\centering
\begin{tabular}{cccccccccc}
$\cardc{\Th}$ & $\mathfrak{E}\,{\uVec_h} $ & rate  & $\mathfrak{E}\,{\nabla \uVec_h}$    & rate & $\mathfrak{E}\,{p_h} $ & rate  &  $\mathfrak{E}\,{\rho_h}$ & $\mathfrak{E}\,{\nabla {\cdot} \uVec_h}$  \\
\hline
\multicolumn{9}{c}{$k=0$} \\
\hline
50         & 0.456       & --       & 4.828       & --      & 0.217       & --          & 6.6e-16  & 1.6e-16    \\ 
192        & 0.123       & 1.95     & 2.853       & 0.78    & 0.0999      & 1.16        & 7.7e-16  & 3.3e-16    \\ 
810        & 0.0291      & 2.00     & 1.390       & 1.00    & 0.0439      & 1.14        & 1.0e-15  & 2.4e-16    \\ 
3126       & 0.00828     & 1.86     & 0.690       & 1.04    & 0.0218      & 1.03        & 1.1e-15  & 2.7e-16    \\ 
12706      & 0.00207     & 1.97     & 0.340       & 1.01    & 0.0109      & 0.99        & 1.5e-15  & 2.7e-16    \\ 
\hline                                                                                                        
\multicolumn{9}{c}{$k=1$} \\                                                                                  
\hline                                                                                                        
50         & 0.105       & --      &  2.280       & --      & 0.0536      & --          & 3.1e-14  & 3.0e-16   \\ 
192        & 0.0116      & 3.28    &  0.550       & 2.11    & 0.00629     & 3.19        & 8.5e-15  & 3.8e-16   \\ 
810        & 0.00139     & 2.94    &  0.135       & 1.95    & 0.00114     & 2.36        & 3.2e-14  & 3.3e-16   \\ 
3126       & 0.000197    & 2.90    &  0.0370      & 1.92    & 0.000286    & 2.06        & 3.3e-14  & 3.2e-16   \\ 
12706      & 2.46e-05    & 2.97    &  0.00904     & 2.01    & 7.27e-05    & 1.95        & 2.2e-14  & 3.2e-16   \\ 
\hline                                                                                                        
\multicolumn{9}{c}{$k=2$} \\                                                                                  
\hline                                                                                                        
50         & 0.0319      & --      & 0.841        & --      & 0.0186      & --          & 5.0e-14  & 4.4e-16      \\
192        & 0.00168     & 4.37    & 0.107        & 3.05    & 0.000739    & 4.80        & 7.2e-14  & 4.5e-16      \\
810        & 8.75e-05    & 4.11    & 0.0124       & 3.00    & 5.51e-05    & 3.61        & 2.9e-14  & 3.8e-16      \\       
3126       & 5.65e-06    & 4.06    & 0.00161      & 3.02    & 6.36e-06    & 3.20        & 3.6e-14  & 4.0e-16      \\ 
12706      & 3.56e-07    & 3.94    & 0.000204     & 2.95    & 7.84e-07    & 2.99        & 5.2e-14  & 3.9e-16      \\ 
\hline                                                                                                        
\multicolumn{9}{c}{$k=3$} \\                                                                                  
\hline                                                                                                        
50         & 0.00285     & --      & 0.136        & --      & 0.00130     & --          & 6.1e-14  & 4.7e-16  \\
192        & 9.52e-05    & 5.05    & 0.00991      & 3.90    & 4.96e-05    & 4.86        & 1.0e-13  & 4.7e-16   \\
810        & 3.14e-06    & 4.74    & 0.000663     & 3.76    & 2.10e-06    & 4.39        & 4.4e-14  & 3.6e-16   \\       
3126       & 1.23e-07    & 4.80    & 5.04e-05     & 3.81    & 1.34e-07    & 4.07        & 6.6e-14  & 4.0e-16    \\ 
12706      & 3.90e-09    & 4.92    & 3.15e-06     & 3.95    & 8.22e-09    & 3.99        & 2.8e-13  & 3.8e-16   \\ 
\hline                                                                                                        
\multicolumn{9}{c}{$k=4$} \\                                                                                  
\hline                                                                                                        
50         & 0.000739     & --     & 0.0429       & --      & 0.000349    & --           & 1.8e-14 & 4.9e-16     \\
192        & 9.45e-06    & 6.48    & 0.00123      & 5.27    & 4.36e-06    & 6.52         & 2.4e-14 & 5.0e-16     \\
810        & 1.49e-07    & 5.77    & 3.87e-05     & 4.81    & 1.04e-07    & 5.18         & 5.3e-14 & 4.1e-16     \\
3126       & 2.49e-09    & 6.06    & 1.32e-06     & 5.00    & 3.36e-09    & 5.09         & 1.3e-13 & 4.5e-16     \\
12706      & 3.95e-11    & 5.91    & 4.18e-08     & 4.93    & 1.04e-10    & 4.95         & 2.0e-13 & 4.4e-16     \\
\end{tabular}
\caption{Kovasnzay test case: spatial convergence rates for $k=0,1,2,3,4$ HHO-$\pi^{k+1}$ formulations over a Delaunay type $h$-refined triangular elements mesh sequence. 
\label{tab:KovaNumRes}}
\end{table}

As a first validation step we seek to numerically demonstrate pure advection of the density 
by tackling the well-known Kovasznay flow problem \cite{Kovasznay:48}.
The analytical solution of the incompressible Navier--Stoke equations,
which describes steady flow behind a grid made of equally spaces parallel rods for a fluid with constant density, reads as follows 
\begin{equation}\label{eq:kovasznay_solution}
       \uVec 
            = \left[1-\eulern^{\kappa x}\cos{\rbrackets{2\pi y}}\right] \iVec +
             \dfrac{\kappa }{2\pi}\eulern^{\kappa x}\sin{\rbrackets{2\pi y}} \jVec, \qquad 
        p =  p_{0}-\dfrac{1}{2}\eulern^{2\kappa x}, \qquad
     \rho = 1,
\end{equation}
where $p_{0}\in\Real$ is an arbitrary constant and the parameter $\kappa$ depends on the Reynolds number  
\begin{equation*}
    \kappa  =   \dfrac{\Reynolds}{2}    -\sqrt{\dfrac{\Reynolds^{2}}{4} +4\pi^{2}}.
\end{equation*}
We define the Kovasznay flow problem on the 
bi-unit square computational domain $\Omega = \rbrackets{-0.5,1.5}{\times}\rbrackets{0,2}$
and derive boundary conditions from the analytical solution \eqref{eq:kovasznay_solution} at $\Reynolds {=} \frac{1}{\mu} {=} 40$.
In particular, a Neumann boundary condition is imposed at the outflow boundary (right side of the square domain) 
while Dirichlet boundary conditions are imposed on the remaining boundaries. 
We apply the variable density HHO-$\pi^k$ and HHO-$\pi^{k+1}$ formulations to a $h$-refined mesh sequence composed 
of regular triangular elements and compute numerical solutions
for several polynomial degrees $k=0,1,2,3,4$.
Triangular elements meshes are generated by means of the Delaunay algorithm and, at each step of the mesh sequence, 
a four fold increase of the mesh cardinality is approximately targeted.

Tables \ref{tab:KovaNumResPk} and \ref{tab:KovaNumRes} report the results obtained by means of HHO-$\pi^{k}$ and HHO-$\pi^{k+1}$ formulations, respectively.
Errors in $L^2$-norm and $h$-convergence rates are computed testing against the analytical solution and 
tabulated for the approximated velocity, velocity gradients and pressure fields. 
A $k{+}1$ rate of error reduction is observed for the pressure error in $L^2$-norm.
Since the flow field is diffusion dominated at this Reynolds number,
the velocity and velocity gradient errors in the $L^2$-norm are expected to exhibit convergence rates of $k{+}2$ and $k{+}1$, respectively.
The major difference between HHO-$\pi^{k+1}$ and HHO-$\pi^{k}$ is observed at $k=0$,
note that both the velocity and the velocity gradient are first order accurate in case of HHO-$\pi^{k}$ while 
HHO-$\pi^{k+1}$ is able to reach second order accuracy for the velocity unknown.
When considering $k>0$, both HHO-$\pi^{k+1}$ and HHO-$\pi^{k}$ are able to reach the theoretical convergence rates.
Nonetheless, HHO-$\pi^{k+1}$ is the best accurate scheme and the asymptotic convergence rates are usually attained on coarser meshes.

The errors in $L^2$-norm for the velocity divergence and the density, reported in Tables \ref{tab:KovaNumResPk}-\ref{tab:KovaNumRes}, show that,
since the incompressibility constraint is enforced up to machine precision, the velocity field is pointwise divergence-free and,
since the density is purely advected, a unit density field is obtained irrespectively from the discretization parameters.
As expected, the density solution at the steady state matches, up to numerical precision, the unit density field employed as initial guess.

\subsection{Guermond and Quartapelle manufactured solution} \label{test:GQ}
In order to validate spatial and temporal convergence rates for the HHO-ESDIRK formulation 
we consider the variable density manufactured solution proposed by \C{Guermond and Quartapelle~}\cite{Guermond2000}.
The analytical velocity, pressure and density fields read
\begin{equation*}
\begin{aligned}
    \uVec &= (-y \iVec + x \jVec) \cos{\rbrackets{t}},  \\
    p &= \sin{\rbrackets{x}}\sin{\rbrackets{y}}\sin{\rbrackets{t}}, \\
    \rho &= 2 + x \cos{\left({\sin{\rbrackets{t}}} \right)}+ y \sin{\left({\sin{\rbrackets{t}}} \right)}.
\end{aligned}
\end{equation*}
and the forcing term $\boldsymbol{f}$ within the momentum equation is set according to the analytical solution.
Dirichlet and Neumann boundary conditions are imposed on the boundary of the unit square domain $\Omega = \left(0,1\right)^2$, 
in particular two of the four edges are of Dirichlet type and the other two are of Neumann type. 
Time integration is carried out in the interval $(0,\frac{7}{4}\pi]$.
Accordingly, the errors in $L^2$ norm for the velocity, the velocity gradients, the velocity divergence, the pressure 
and the density are computed based on the exact solution at the final time $\tF = \frac{7}{4}\pi$. 
The velocity and density fields are initialized interpolating the exact solution.
We rely on triangular cells meshes obtained subdividing in two right triangles the square cells of $N^2$ Cartesian grids, with $N$ cells per Cartesian direction.
Accordingly, the mesh cardinality is $\card{\Th}=2\,N^2$ when $h = \frac{\sqrt{2}}{N}$.

\begin{table}[!htbp]
\small
\centering
\begin{tabular}{cccccccccc}
$\cardc{\Th}$ &$\mathfrak{E}\,{\uVec_h} $ & rate  & $\mathfrak{E}\,{\nabla \uVec_h}$    & rate & $\mathfrak{E}\,{\rho_h}$  & rate & $\mathfrak{E}\,{p_h} $ & rate & $\mathfrak{E}\,{\nabla {\cdot} \uVec_h}$  \\
\hline
\multicolumn{10}{c}{$k=0$} \\
\hline
32      & 0.00773 & --      & 0.0297   & --         & 0.0628  & --         & 0.0327  & --         & 5.4e-16    \\ 
128     & 0.00397 & 0.96    & 0.0151   & 0.97       & 0.0328  & 0.93       & 0.0168  & 0.96       & 4.9e-16    \\ 
512     & 0.00201 & 0.98    & 0.00760  & 1.00       & 0.0172  & 0.93       & 0.00870 & 0.95       & 4.6e-16    \\ 
2048    & 0.001017 & 0.99   & 0.00379  & 1.00       & 0.00897 & 0.94       & 0.00444 & 0.97       & 4.9e-16    \\ 
8192    & 0.000511 & 0.99   & 0.00189  & 1.00       & 0.00462 & 0.96       & 0.00225 & 0.98       & 4.9e-16    \\ 
32768   & 0.000256 & 1.00   & 0.000943 & 1.00       & 0.00236 & 0.97       & 0.00113 & 0.99       & 4.8e-16    \\ 
131072  & 0.000128 & 1.00   & 0.000471 & 1.00       & 0.00120 & 0.98       & 0.000567 & 1.00      & 4.8e-16    \\ 
\hline                                                                                                                    
\multicolumn{10}{c}{$k=1$} \\                                                                                              
\hline                                                                                                                    
32           & 3.79e-06 & --    & 0.000124 & --            & 8.69e-07 & --         & 0.00122  & --         & 7.2e-16 \\ 
128          & 2.49e-07 & 3.93  & 1.62e-05 & 2.94          & 6.37e-08 & 3.77       & 0.000307 & 1.99       & 7.3e-16 \\ 
512          & 1.59e-08 & 3.97  & 2.07e-06 & 2.97          & 4.38e-09 & 3.86       & 7.69e-05 & 2.00       & 6.9e-16 \\ 
2048         & 1.01e-09 & 3.97  & 2.61e-07 & 2.99          & 4.12e-10 & 3.41       & 1.92e-05 & 2.00       & 6.8e-16 \\ 
8192         & 1.94e-10 & 2.39  & 3.29e-08 & 2.99          & 2.55e-10 & 0.69       & 4.81e-06 & 2.00       & 6.9e-16 \\ 
32768        & 1.87e-10 & 0.05  & 5.15e-09 & 2.68          & 2.71e-10 & -0.09      & 1.20e-06 & 2.00       & 6.8e-16 \\ 
\hline                                                                                                                    
\multicolumn{10}{c}{$k=2$} \\                                                                                              
\hline                                                                                                                    
32           & 2.06e-10 & --   & 2.78e-09 & --       & 1.77e-10 & --      & 5.65e-05 & --         & 6.8e-16   \\ 
128          & 1.90e-10 & --   & 3.66e-09 & --       & 2.05e-10 & --      & 7.07e-06 & 3.00       & 6.8e-16   \\ 
512          & 1.87e-10 & --   & 3.19e-09 & --       & 2.32e-10 & --      & 8.83e-07 & 3.00       & 6.8e-16   \\ 
2048         & 1.86e-10 & --   & 3.10e-09 & --       & 2.53e-10 & --      & 1.10e-07 & 3.00       & 6.8e-16   \\ 
8192         & 1.86e-10 & --   & 3.10e-09 & --       & 2.70e-10 & --      & 1.44e-08 & 2.94       & 6.9e-16   \\ 
32768        & 1.86e-10 & --   & 3.10e-09 & --       & 3.07e-10 & --      & 4.42e-09 & 1.70       & 6.8e-16   \\ 
\end{tabular}
\caption{Guermond and Quartapelle unsteady test case: spatial convergence rates for $k=0,1,2$ HHO-$\pi^{k}$ formulations over a $h$-refined triangular elements mesh sequence.
         Time integration is conducted with a high-order accurate ESDIRK scheme, the time step is selected such that spatial errors dominate. 
         \label{tab:sconvGQPk}}
\end{table}

\begin{table}[!htbp]
\small
\centering
\begin{tabular}{cccccccccc}
$\cardc{\Th}$ & $\mathfrak{E}\,{\uVec_h} $ & rate  & $\mathfrak{E}\,{\nabla \uVec_h}$    & rate & $\mathfrak{E}\,{\rho_h}$  & rate & $\mathfrak{E}\,{p_h} $ & rate & $\mathfrak{E}\,{\nabla {\cdot} \uVec_h}$  \\
\hline
\multicolumn{10}{c}{$k=0$} \\
\hline
32      & 0.00249  & --         & 0.00721  & --           & 0.0609  & --         & 0.0313  & --         & 5.1e-16 \\ 
128     & 0.00112  & 1.15       & 0.00318  & 1.18         & 0.0317  & 0.94       & 0.0158  & 0.98       & 4.8e-16 \\ 
512     & 0.000533 & 1.07       & 0.00152  & 1.06         & 0.0165  & 0.94       & 0.00801 & 0.98       & 5.1e-16 \\ 
2048    & 0.000259 & 1.04       & 0.000752 & 1.02         & 0.00857 & 0.95       & 0.00404 & 0.99       & 4.8e-16 \\ 
8192    & 0.000128 & 1.02       & 0.000376 & 1.00         & 0.00440 & 0.96       & 0.00203 & 0.99       & 4.9e-16 \\ 
32768   & 6.38e-05 & 1.01       & 0.000189 & 0.99         & 0.00224 & 0.97       & 0.00102 & 1.00       & 4.8e-16 \\ 
131072  & 3.19e-05 & 1.00       & 9.51e-05 & 0.99         & 0.00114 & 0.98       & 0.000509 & 1.00      & 4.9e-16 \\ 
\hline                                                                                                                    
\multicolumn{10}{c}{$k=1$} \\                                                                                              
\hline                                                                                                                    
32          & 3.88e-06 & --     & 0.000126 & --           & 8.70e-07 & --         & 0.00122 & --         & 6.6e-16   \\ 
128         & 2.52e-07 & 3.94   & 1.64e-05 & 2.95         & 6.37e-08 & 3.77       & 0.000307 & 1.99      & 6.6e-16   \\ 
512         & 1.60e-08 & 3.98   & 2.08e-06 & 2.98         & 4.37e-09 & 3.86       & 7.69e-05 & 2.00      & 7.0e-16   \\ 
2048        & 1.02e-09 & 3.98   & 2.62e-07 & 2.99         & 4.12e-10 & 3.41       & 1.92e-05 & 2.00      & 6.8e-16   \\ 
8192        & 1.94e-10 & 2.39   & 3.30e-08 & 2.99         & 2.56e-10 & 0.69       & 4.81e-06 & 2.00      & 6.8e-16   \\ 
32768       & 1.87e-10 & 0.05   & 5.15e-09 & 2.68         & 2.71e-10 & -0.09      & 1.20e-06 & 2.00      & 6.8e-16   \\ 
\hline                                                                                                                    
\multicolumn{10}{c}{$k=2$} \\                                                                                              
\hline                                                                                                                    
32          & 2.06e-10 & --   & 2.78e-09 & --           & 1.77e-10 & --          & 5.65e-05 & --         & 7.0e-16  \\ 
128         & 1.90e-10 & --   & 3.67e-09 & --           & 2.05e-10 & --          & 7.07e-06 & 3.00       & 6.9e-16  \\ 
512         & 1.87e-10 & --   & 3.19e-09 & --           & 2.32e-10 & --          & 8.84e-07 & 3.00       & 7.1e-16  \\ 
2048        & 1.86e-10 & --   & 3.10e-09 & --           & 2.53e-10 & --          & 1.10e-07 & 3.00       & 6.9e-16  \\ 
8192        & 1.86e-10 & --   & 3.10e-09 & --           & 2.70e-10 & --          & 1.44e-08 & 2.94       & 6.8e-16  \\ 
32768       & 1.86e-10 & --   & 3.10e-09 & --           & 3.07e-10 & --          & 4.42e-09 & 1.70       & 6.9e-16  \\ 
\end{tabular}
\caption{Guermond and Quartapelle unsteady test case: spatial convergence rates for $k=0,1,2$ HHO-$\pi^{k+1}$ formulations over a $h$-refined triangular elements mesh sequence.
         Time integration is conducted with a high-order accurate ESDIRK scheme, the time step is selected such that spatial errors dominate. 
         \label{tab:sconvGQ}}
\end{table}
The spatial convergence results are reported in Table \ref{tab:sconvGQPk} and Table \ref{tab:sconvGQ}
for $k=0,1,2$ HHO-$\pi^{k}$ and HHO-$\pi^{k+1}$ formulations, respectively, considering five grids of a triangular mesh sequence with $N=4 \times {2^i}$, $i = 0,\dots,5$.
Time integration is performed relying on a third order accurate six stages ESDIRK scheme and using 320 uniform time steps.
Note that, since the analytical pressure has a trigonometric spatial behavior, 
$k+1$ convergence rates for the pressure error in $L^2$-norm are attained at all polynomial degrees. 
Interestingly, the velocity and density errors in $L^2$-norm tabulated for $k=2$ are grid independent and allow to evaluate the temporal accuracy of the scheme. 
Indeed, since the by product between the density and the velocity is a second degree polynomial in $\xVec$,
$k=2$ computations provide machine precision spatial accuracy for both the density and the velocity variables.
This behavior confirms that the HHO formulations are pressure-robust, note indeed that the velocity error is clearly independent from the pressure error.

Overall HHO-$\pi^{k+1}$ and HHO-$\pi^{k}$ formulations provide almost identical accuracy and convergence rates for $k=1,2$.
The numerical results for $k=1$ show fourth order rate of error reduction for both the velocity and density unknowns, 
until spatial accuracy is on pair with temporal accuracy, resulting in error stagnation. 
We remark that, spatial convergence rates for $k=1$ must be 
carefully evaluated considering that, according to the manufactured solution, 
both the velocity and the density unknowns are linear functions of $\xVec$.
At $k=0$, both the velocity and the velocity gradient exhibiting first order of convergence in $L^2$-norm, 
and, similarly to the Kovasznay test case, HHO-$\pi^{k+1}$ shows improved accuracy with respect to HHO-$\pi^{k}$.

\begin{table}[!htbp]
\centering
\begin{tabular}{ccccccccccc}
\hline                              
\multicolumn{11}{c}{3 stages ESDIRK} \\
\hline                              
k & $ \frac{7/4\pi}{\delta t}$ & $\mathfrak{E}\,{\uVec_h} $ & rate  & $\mathfrak{E}\,{\nabla \uVec_h}$    & rate & $\mathfrak{E}\,{\rho_h} $ & rate &  $\mathfrak{E}\,{p_h}$  & rate & $\mathfrak{E}\,{\nabla {\cdot} \uVec_h}$  \\
\hline                              
\multirow{5}{*}{6} &  20    & 0.000614 & --     & 0.00161 & --           & 0.00202  & --         & 0.00168  & --   &  8.7e-17 \\ %
                   &  40    & 0.000147 & 2.06   & 0.000389 & 2.06        & 0.000509 & 1.99       & 0.000404 & 2.07 &  8.2e-17  \\ %
                   &  80    & 3.62e-05 & 2.03   & 9.53e-05 & 2.03        & 0.000127 & 2.00       & 9.74e-05 & 2.04 &  8.6e-17 \\ %
                   &  160   & 8.98e-06 & 2.01   & 2.36e-05 & 2.01        & 3.19e-05 & 1.99       & 2.39e-05 & 2.02 &  8.9e-17 \\ %
                   &  320   & 2.23e-06 & 2.01   & 5.87e-06 & 2.01        & 7.98e-06 & 2.00       & 5.93e-06 & 2.01 &  8.8e-17 \\ %
\hline                                                                                                                            
\multirow{5}{*}{2} &  20    & 0.000614 & --     & 0.00161  & --          & 0.00201  & --         & 0.00168 & --    &  6.6e-17 \\ %
                   &  40    & 0.000147 & 2.06   & 0.000388 & 2.06        & 0.000501 & 2.01       & 0.000401 & 2.07 &  6.4e-17  \\ %
                   &  80    & 3.62e-05 & 2.03   & 9.53e-05 & 2.03        & 0.000124 & 2.01       & 9.74e-05 & 2.04 &  7.2e-17 \\ %
                   &  160   & 8.98e-06 & 2.01   & 2.36e-05 & 2.01        & 3.10e-05 & 2.00       & 2.42e-05 & 2.01 &  7.3e-17 \\ %
                   &  320   & 2.23e-06 & 2.01   & 5.87e-06 & 2.01        & 7.73e-06 & 2.00       & 6.92e-06 & 1.81 &  6.3e-17 \\ %
\hline                                             
\multicolumn{11}{c}{6 stages ESDIRK} \\
\hline                              
k & $ \frac{7/4\pi}{\delta t}$ & $\mathfrak{E}\,{\uVec_h} $ & rate  & $\mathfrak{E}\,{\nabla \uVec_h}$    & rate & $\mathfrak{E}\,{p_h} $ & rate &  $\mathfrak{E}\,{\rho_h}$  & rate & $\mathfrak{E}\,{\nabla {\cdot} \uVec_h}$ \\
\hline                              
\multirow{5}{*}{6} &  20    & 3.50e-06 & --     & 1.26e-05 & --          & 1.14e-05 & --         & 1.03e-05 & --    & 8.1e-16  \\ %
                   &  40    & 2.83e-07 & 3.63   & 1.43e-06 & 3.14        & 7.87e-07 & 3.87       & 1.22e-06 & 3.08  & 8.4e-16  \\ %
                   &  80    & 2.34e-08 & 3.60   & 1.74e-07 & 3.04        & 5.15e-08 & 3.93       & 1.66e-07 & 2.88  & 8.4e-16  \\ %
                   &  160   & 2.03e-09 & 3.53   & 2.26e-08 & 2.94        & 3.51e-09 & 3.88       & 2.50e-08 & 2.74  & 8.2e-16  \\ %
                   &  320   & 1.86e-10 & 3.45   & 3.11e-09 & 2.86        & 2.57e-10 & 3.77       & 4.06e-09 & 2.62  & 7.4e-16   \\ %
\hline                                                                                                                             
\multirow{5}{*}{2} &  20    & 3.51e-06 & --     & 1.26e-05 & --          & 1.01e-05 & --         & 1.10e-05 & --    & 7.1e-16  \\ %
                   &  40    & 2.83e-07 & 3.63   & 1.43e-06 & 3.14        & 6.93e-07 & 3.87       & 3.78e-06 & 1.54  & 7.0e-16  \\ %
                   &  80    & 2.34e-08 & 3.60   & 1.77e-07 & 3.02        & 4.70e-08 & 3.88       & 3.58e-06 & 0.08  & 7.3e-16   \\ %
                   &  160   & 2.04e-09 & 3.52   & 2.40e-08 & 2.88        & 3.14e-09 & 3.90       & 3.57e-06 & 0.00  & 7.3e-16  \\ %
                   &  320   & 1.89e-10 & 3.43   & 3.40e-09 & 2.82        & 2.05e-10 & 3.93       & 3.57e-06 & 0.00  & 7.2e-16  \\ %
\end{tabular}
\caption{Guermond and Quartapelle unsteady test case: temporal convergence rates for $k=2,6$ HHO-$\pi^{k+1}$ formulations over a triangular elements mesh with $\card{\Th}=128$.
         Time integration is conducted with high-order accurate three and six stages ESDIRK schemes with increasingly small time steps, 
         in particular $\delta t=\frac{7/4\pi}{N}$ with $N=20,40,80,160,320$.
         \label{tab:tconvGQ}}
\end{table}

Temporal convergence results are reported in Table \ref{tab:tconvGQ} considering HHO-$\pi^{k+1}$
(HHO-$\pi^{k}$ is omitted as it provides similar results) and two ESDIRK schemes:
\begin{inparaenum}[i)] 
\item the second order accurate three stages, so-called TR-BDF2, scheme, see \cite{Skvorstov2022} for details,
\item the fourth order accurate six stages scheme proposed in \cite{Skvorstov2010}.
\end{inparaenum}
Both ESDIRK schemes are considered in combination with $k=2$ and $k=6$ formulations on a triangular elements mesh with $N=8$ and $\card{\Th}=128$, 
the former provides machine precision spatial accuracy for the velocity and density variables 
while the latter guarantees negligible spatial errors for all variables, including the pressure.
Note that, when considering the $k=2$ spatial discretization with third order accurate ESDIRK, the pressure error stagnates while, 
thanks to pressure-robustness, the velocity and density errors are almost identical to their $k=6$ counterparts.

Second order accuracy is achieved for all variables when considering the three stages ESDIRK scheme. 
The six stages ESDIRK scheme provides fourth order accuracy for the density, third order accuracy for the velocity and the pressure and
something in between, around three and a half order temporal accuracy, for the velocity gradients.
Overall the numerical results are satisfactory and the convergence rates are comparable with the outcomes of Cai \ea~\cite{Cai2025}, 
a survey where several ESDIRK methods, including those employed in this work, where put to the test.
We remark that, in order to achieve higher than second order temporal convergence rates, 
we rely on Runge-Kutta schemes especially conceived for index two Differential-Algebraic Equation systems (DAEs).
This latter choice is mandatory when dealing with non-passive unsteady boundary conditions.

\subsection{Rayleigh-Taylor instability} \label{sec:RTI}
In order apply the ESDIRK-HHO formulation to realistic variable density incompressible flow problems we consider 
a low and a high Atwood number Rayleigh-Taylor Instability (RTI). 
The test case, originally proposed by \C{Tryggvason~}\cite{Tryggvason1988}, is set up as proposed by \C{Guermond \ea~}\cite{Guermond2000}.

We consider the time evolution of two layers of immiscible fluids initially at rest in the rectangular domain 
$\Omega = (-\frac{d}{2},\frac{d}{2})\times(-2d,2d)$, where $d$ is the reference length.
Evolution of the flow field is driven by the downward gravity field 
as the heavy fluid lays on top of the light fluid.
The density ratio $\frac{\rho_{\rm{t}}}{\rho_{\rm{b}}}$ is set to three and seven,
where $\rho_{\mathrm{t}}$ and $\rho_{\mathrm{b}}{=}1$ are the density of the top and bottom fluids, respectively.
Accordingly, the Atwood number, 
defined as $\Atwood =(\rho_{\mathrm{t}}-\rho_{\mathrm{b}})/(\rho_{\mathrm{t}}+\rho_{\mathrm{b}})$, 
is ${\rm lAt}{=}0.5$ (low Atwood) and ${\rm hAt}{=}0.75$ (high Atwood). 
Following \cite{Guermond2000}, a smooth transition between the two fluids is guaranteed relying on the following expressions 
for the initialization of the density field 
\begin{equation}
\label{eq:rhoRTI}
\begin{aligned}
\frac{\rho_{\rm lAt}(\xVec,t=0)}{\rho_{\mathrm{b}}}&=2+\mathrm{tanh}\left(\frac{\xVec \cdot \jVec -\eta(\xVec)}{0.01\,d}\right),& &\eta(\xVec) = -0.1\,d\,\mathrm{cos}\left(\frac{2\pi}{d} \xVec \cdot \iVec\right)&\\
\frac{\rho_{\rm hAt}(\xVec,t=0)}{\rho_{\mathrm{b}}}&=4+3\,\mathrm{tanh}\left(\frac{\xVec \cdot \jVec -\gamma(\xVec)}{0.01\,d}\right),& &\gamma(\xVec) = -0.01\,d\,\mathrm{cos}\left(\frac{2\pi}{d} \xVec \cdot \iVec\right)&
\end{aligned}
\end{equation}
where $\eta$ and $\gamma$ define the initial position of the perturbed interface in the low and the high Atwood configuration, respectively.
The unit vectors $\iVec$ and $\jVec$ have horizontal orientation pointing right and vertical orientation pointing up, respectively.
The $L^2$ projection of the analytical density distribution in Eq.~\eqref{eq:rhoRTI} provides the initial density field
and the source term in the momentum equation reads $ \boldsymbol{f} = - \rho g \jVec$.

Assuming the symmetry of the initial condition is maintained during the time evolution,
the computational domain has been restricted to $\Omega = (0,\frac{d}{2})\times(-2d,2d)$.
We consider two uniform triangular elements grids obtained splitting in two, along the diagonal,
the quadrilateral elements of a $32{\times}256$ and a $256{\times}2048$ Cartesian grid, 
resulting in $\card{\Th} = 16~384$ and $\card{\Th} = 1~048~576$ for the coarse and the fine grid, respectively.
Symmetry boundary conditions, implemented as homogeneous Dirichlet in the normal direction 
and homogeneous Neumann in the tangential direction, 
are imposed on the computational domain boundary $\partial \Omega$.
Time integration is conducted by means 
of the second order accurate three stages TR-BDF2 ESDIRK scheme, see \cite{Skvorstov2022}, using $3~200$ uniform time steps. 
The time integration intervals are $(0,4]$ and $(0,3.75)$, respectively, in the low and the high Atwood number configuration.

\subsubsection{Low Atwood number computations}
In this section we present and discuss the numerical results in the low Atwood number configuration focusing on the evolution of the density field.
The computations are performed with both the HHO-$\pi^k$ and the HHO-$\pi^{k+1}$ formulations,
the polynomial degree is $k=6$ for the coarse grid and $k=1$ for the fine grid. 
We remark that, even if $k>0$ discretizations are not bounds preserving due to the onset of spurious oscillations at sharp interfaces, the moderate 
density ratio associated with the low Atwood number setup allows to disregard oscillation control strategies.

\begin{figure}[!htb]
\centering
\begin{tabular}{ccccccc}
\includegraphics[trim=40 0 80 0,clip,width=0.108\textwidth]{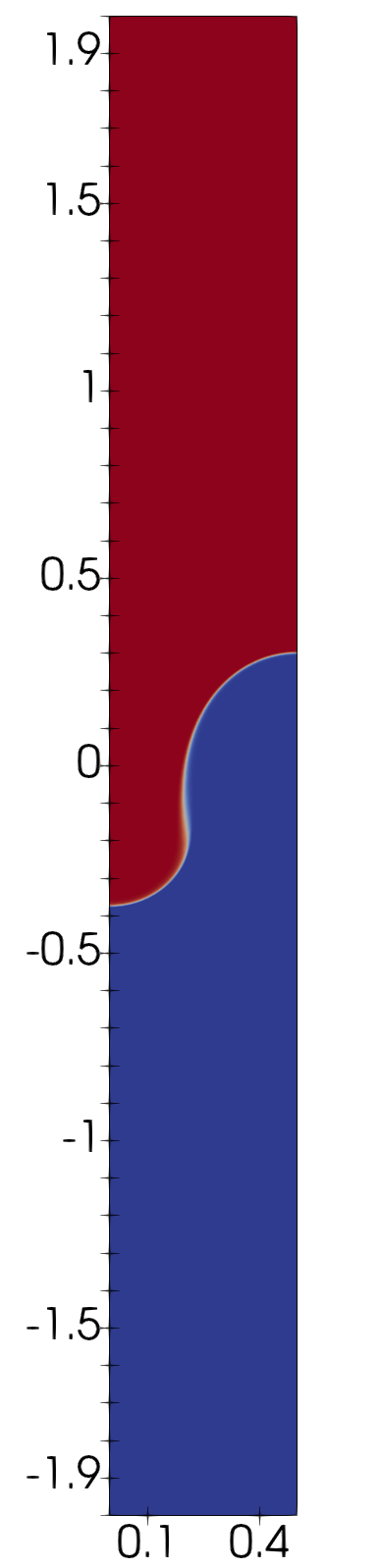} & 
\includegraphics[trim=40 0 80 0,clip,width=0.108\textwidth]{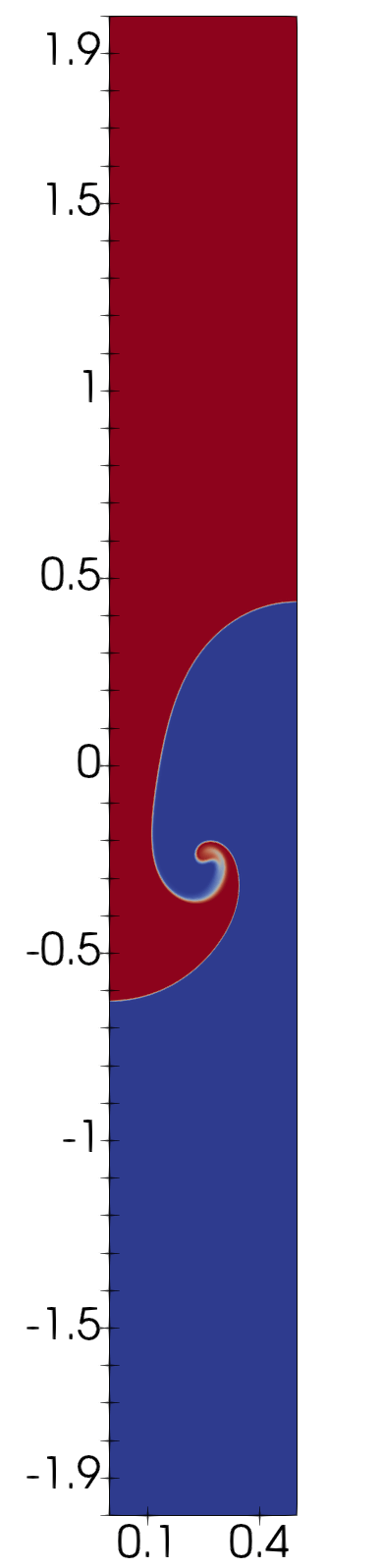} & 
\includegraphics[trim=40 0 80 0,clip,width=0.108\textwidth]{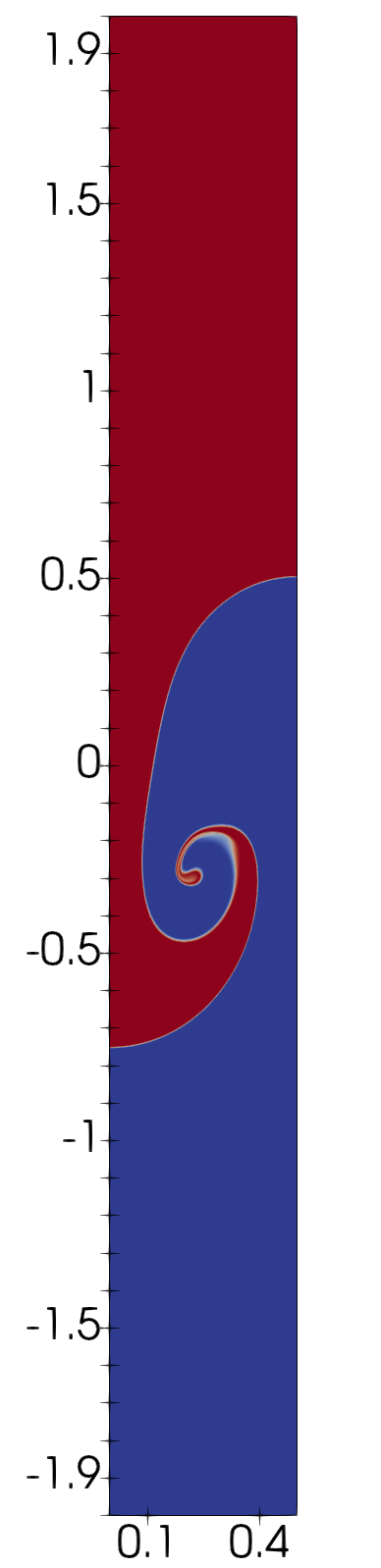} & 
\includegraphics[trim=40 0 80 0,clip,width=0.108\textwidth]{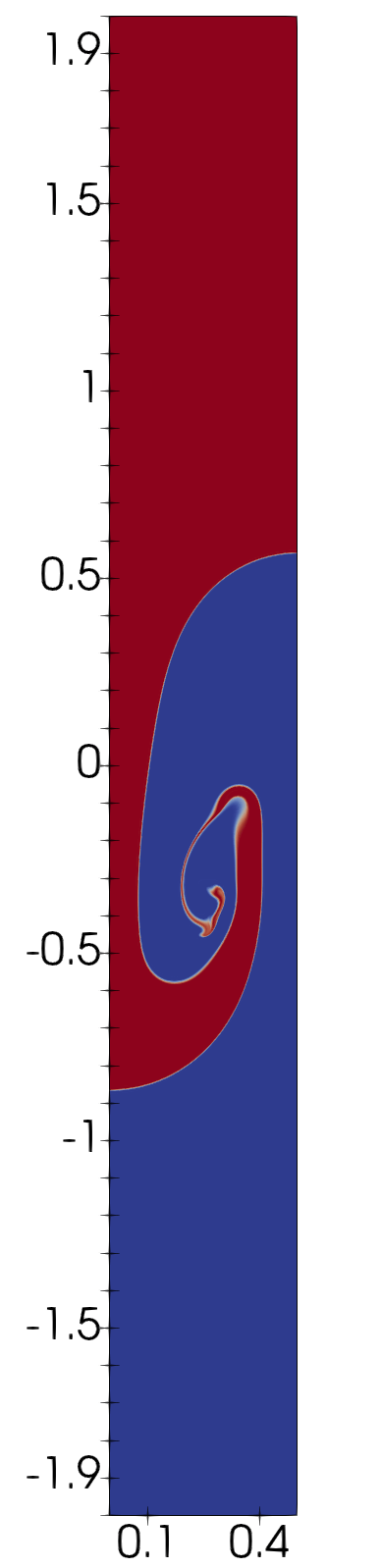} & 
\includegraphics[trim=40 0 80 0,clip,width=0.108\textwidth]{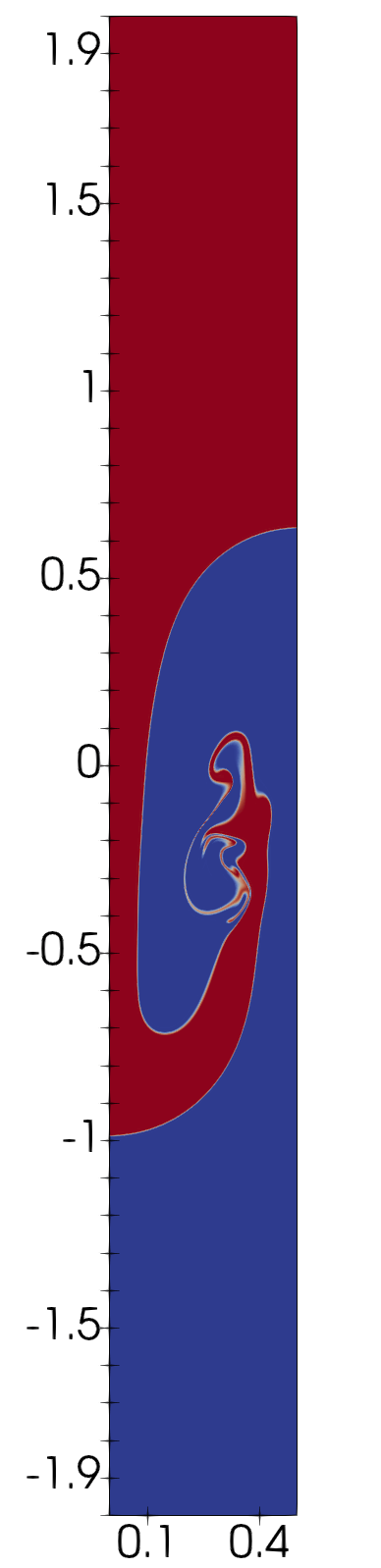} &
\includegraphics[trim=40 0 80 0,clip,width=0.108\textwidth]{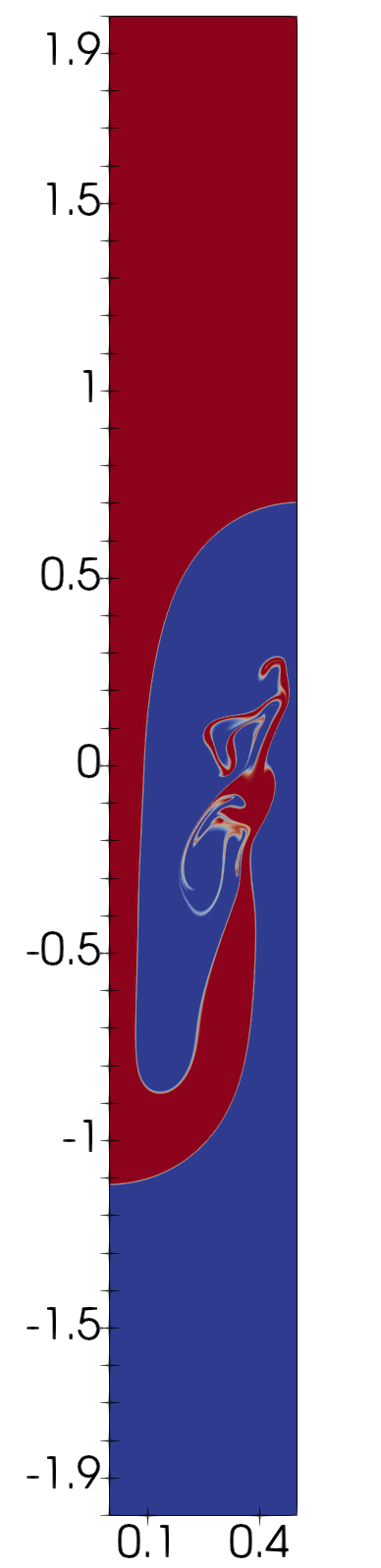} &
\includegraphics[trim=40 0 80 0,clip,width=0.108\textwidth]{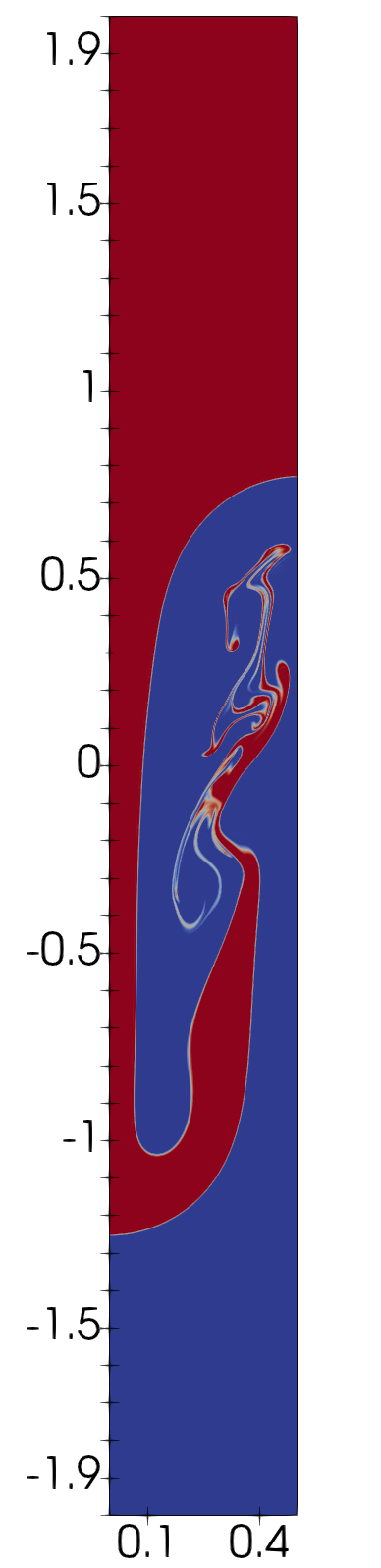} \\
\includegraphics[trim=40 0 80 0,clip,width=0.108\textwidth]{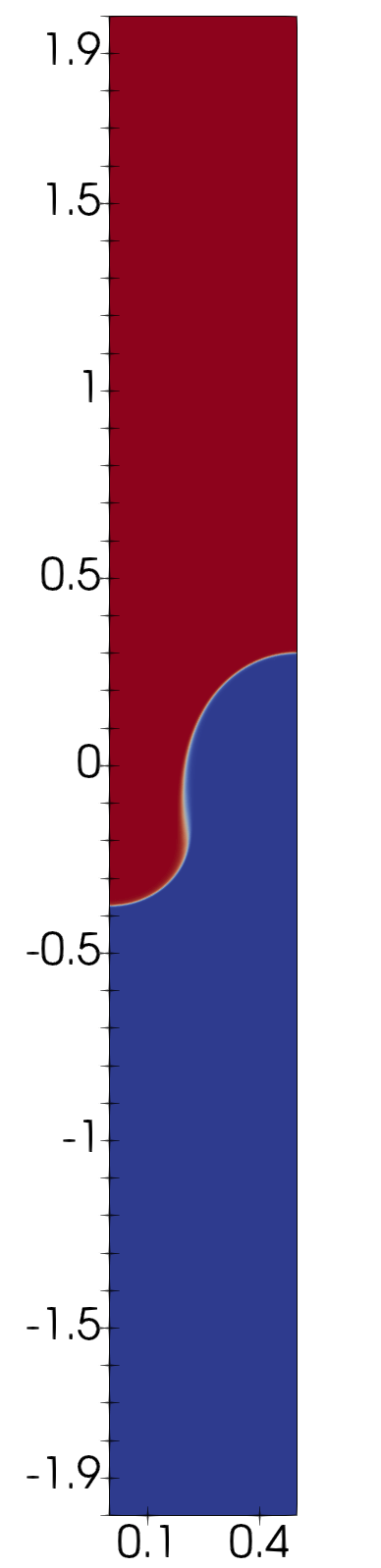} & 
\includegraphics[trim=40 0 80 0,clip,width=0.108\textwidth]{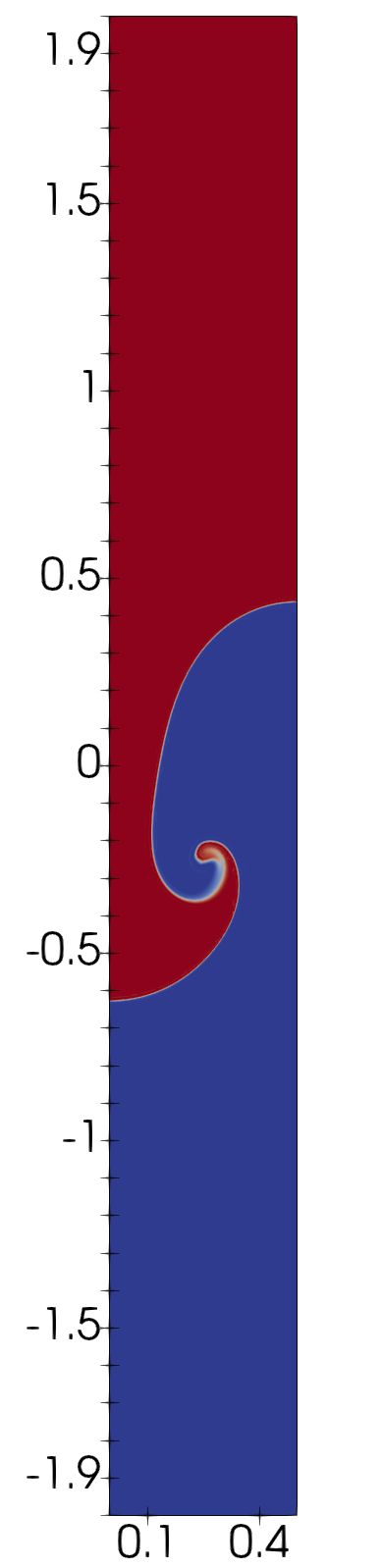} & 
\includegraphics[trim=40 0 80 0,clip,width=0.108\textwidth]{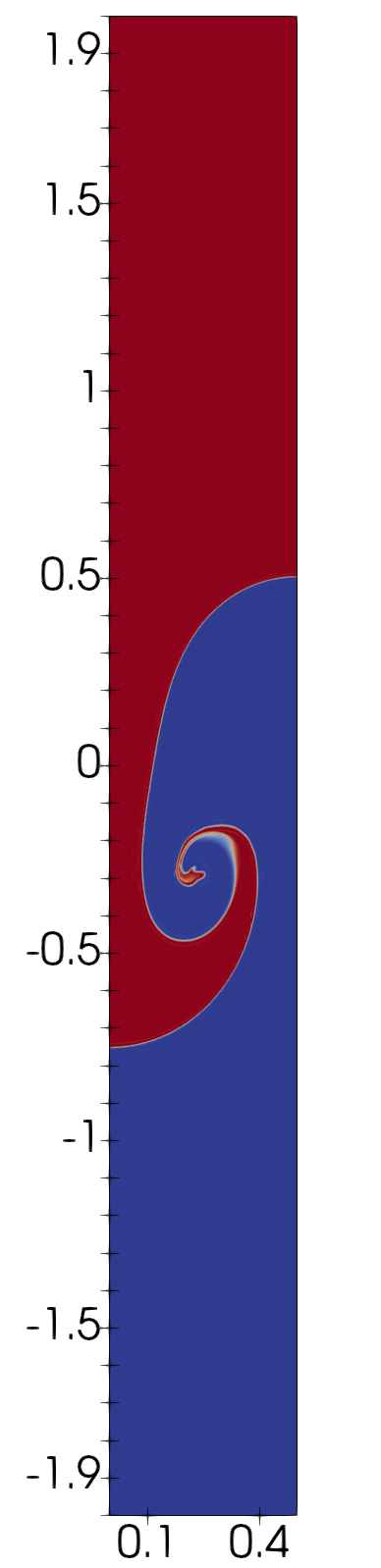} & 
\includegraphics[trim=40 0 80 0,clip,width=0.108\textwidth]{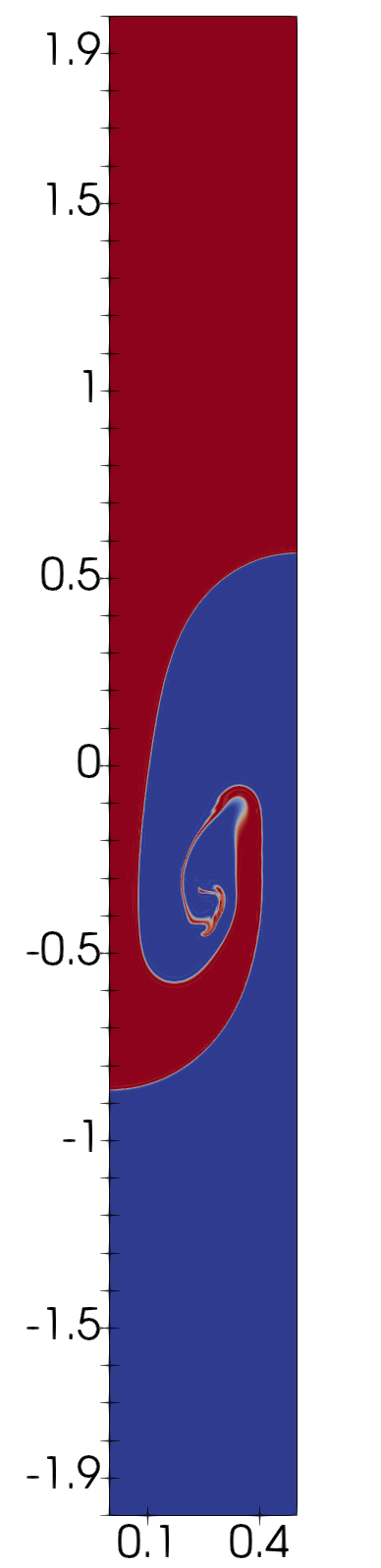} & 
\includegraphics[trim=40 0 80 0,clip,width=0.108\textwidth]{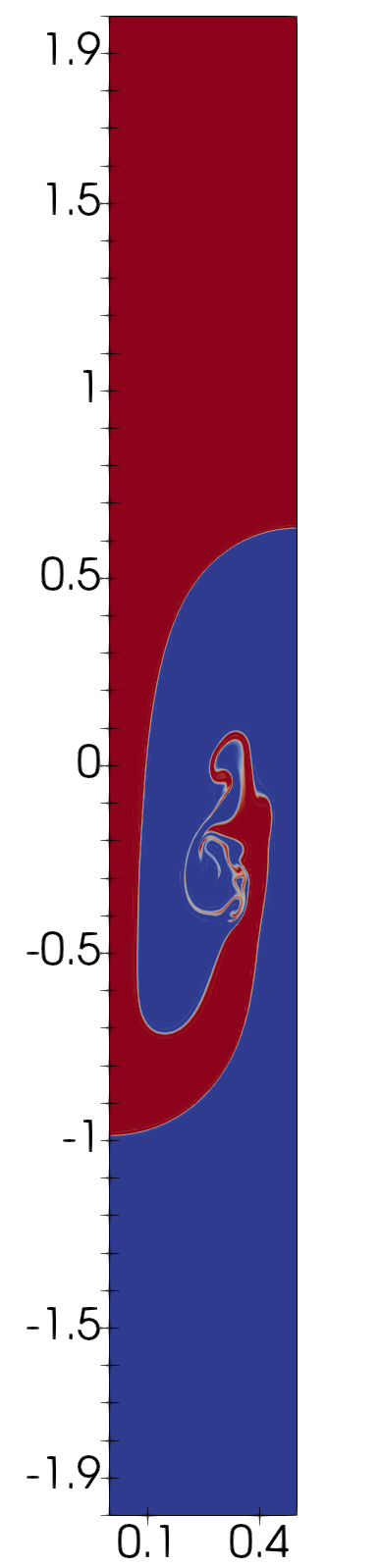} &
\includegraphics[trim=40 0 80 0,clip,width=0.108\textwidth]{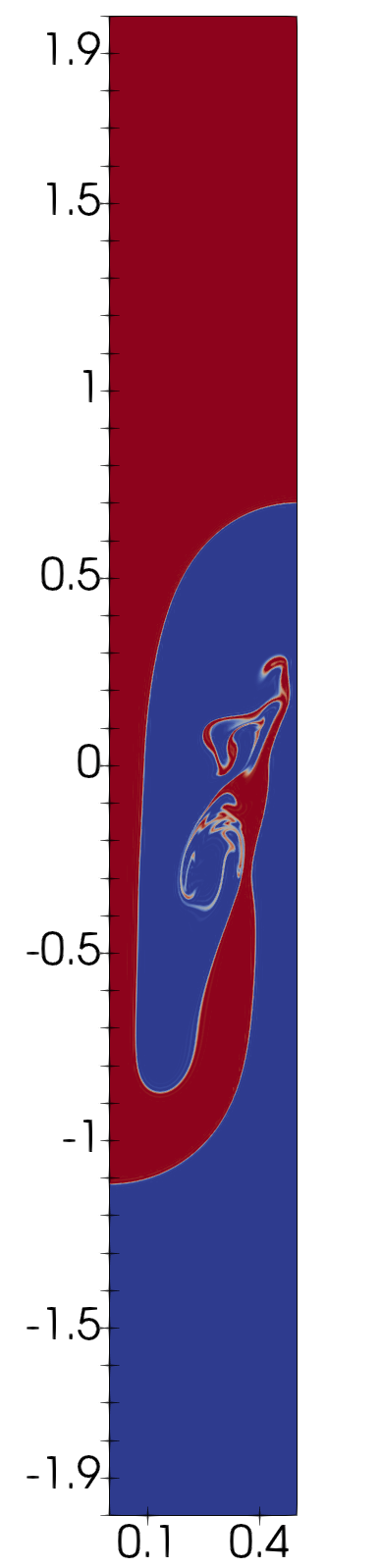} &
\includegraphics[trim=40 0 80 0,clip,width=0.108\textwidth]{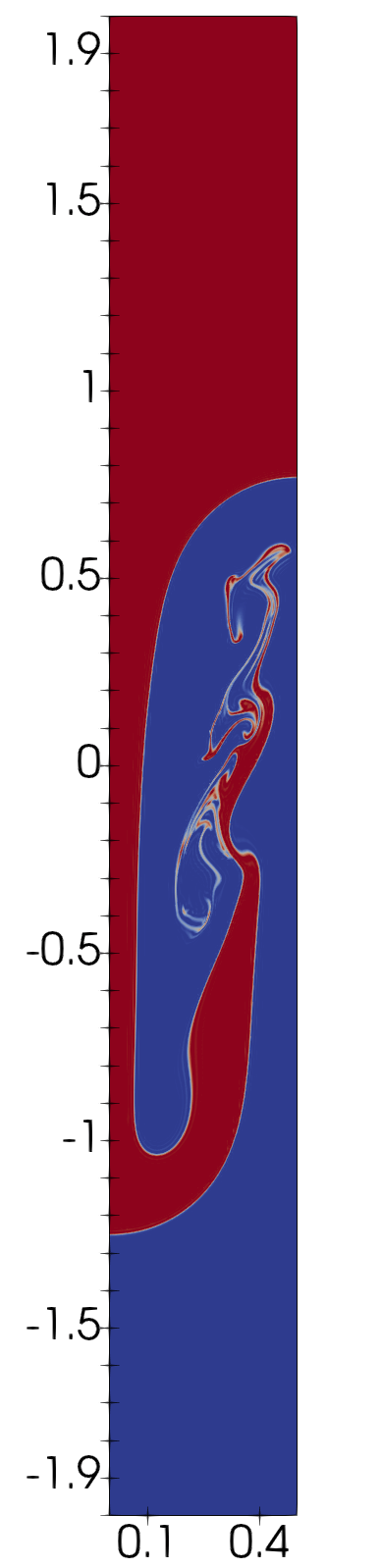} 
\end{tabular}
\caption{RTI at $\Atwood{=}0.5$ and $\Reynolds{=}1~000$ with HHO-$\pi^k$. 
\emph{From left to right}, evolution of the density field at selected time points 
$t_{\mathrm{Trg}} = t \sqrt{\Atwood} \simeq 1, 1.5, 1.75, 2, 2.25, 2.5, 2.75$. 
\emph{Top and bottom row}, results for $k{=}1$ over fine grid and $k{=}6$ over the coarse grid, respectively. 
         \label{fig:RTI1000}}
\end{figure}

\begin{figure}[!htb]
\centering
\begin{tabular}{ccccccc}
\includegraphics[trim=40 0 80 0,clip,width=0.108\textwidth]{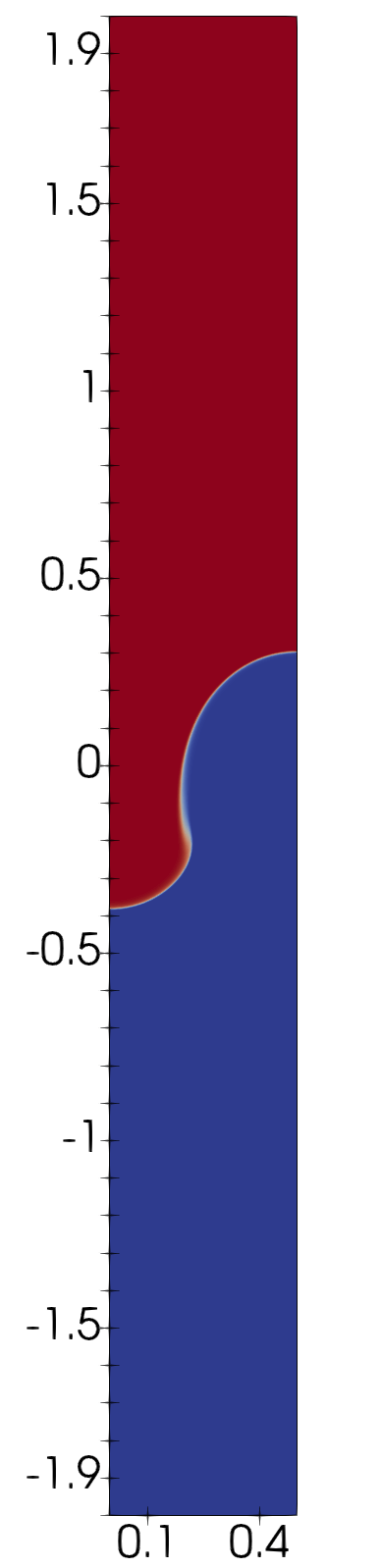} & 
\includegraphics[trim=40 0 80 0,clip,width=0.108\textwidth]{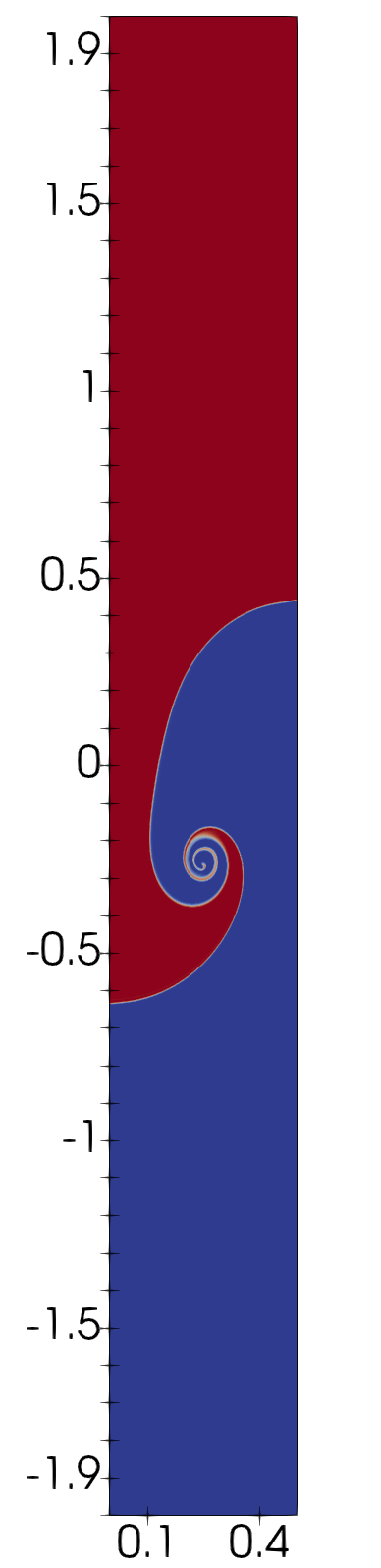} & 
\includegraphics[trim=40 0 80 0,clip,width=0.108\textwidth]{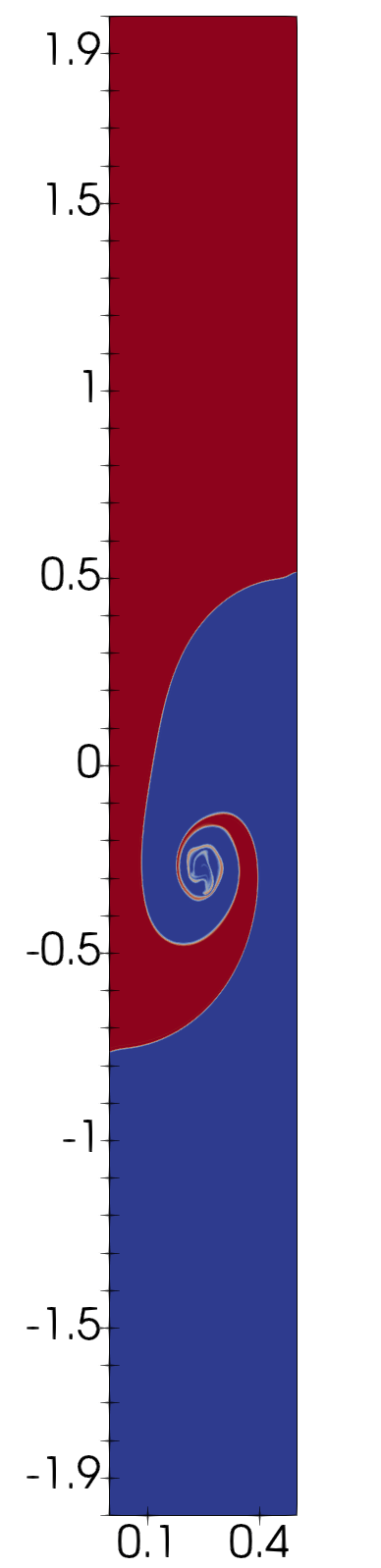} & 
\includegraphics[trim=40 0 80 0,clip,width=0.108\textwidth]{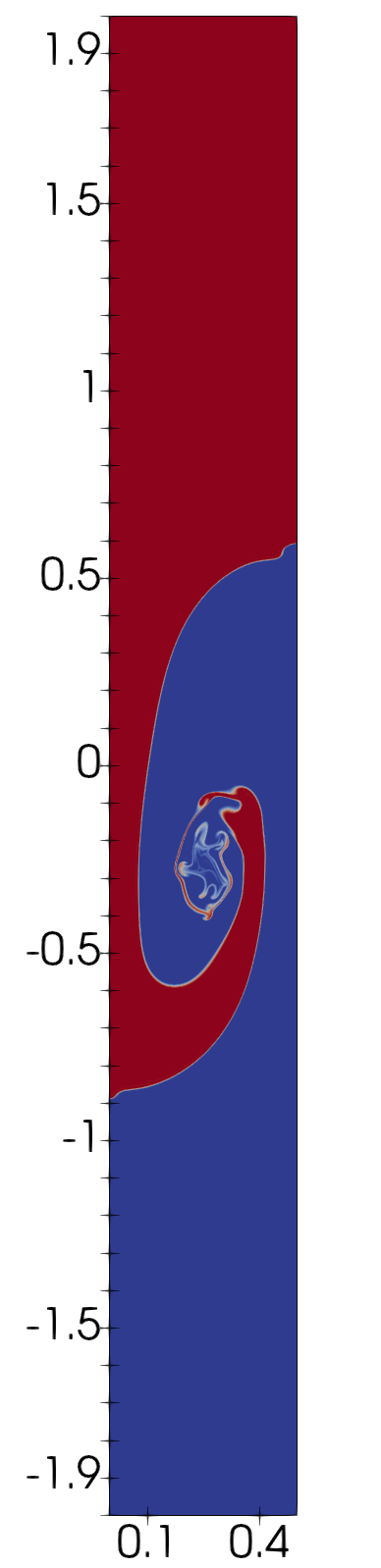} & 
\includegraphics[trim=40 0 80 0,clip,width=0.108\textwidth]{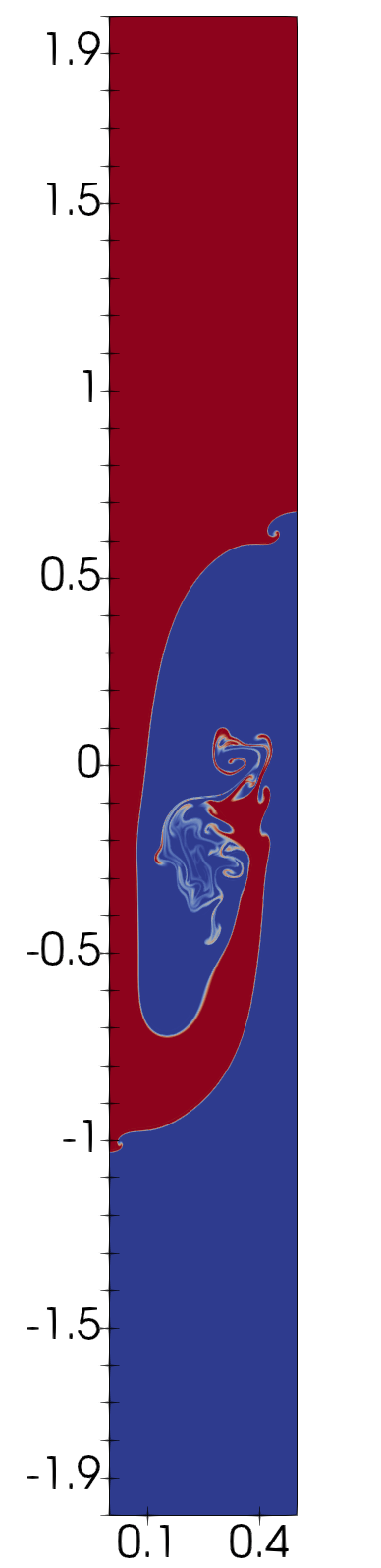} &
\includegraphics[trim=40 0 80 0,clip,width=0.108\textwidth]{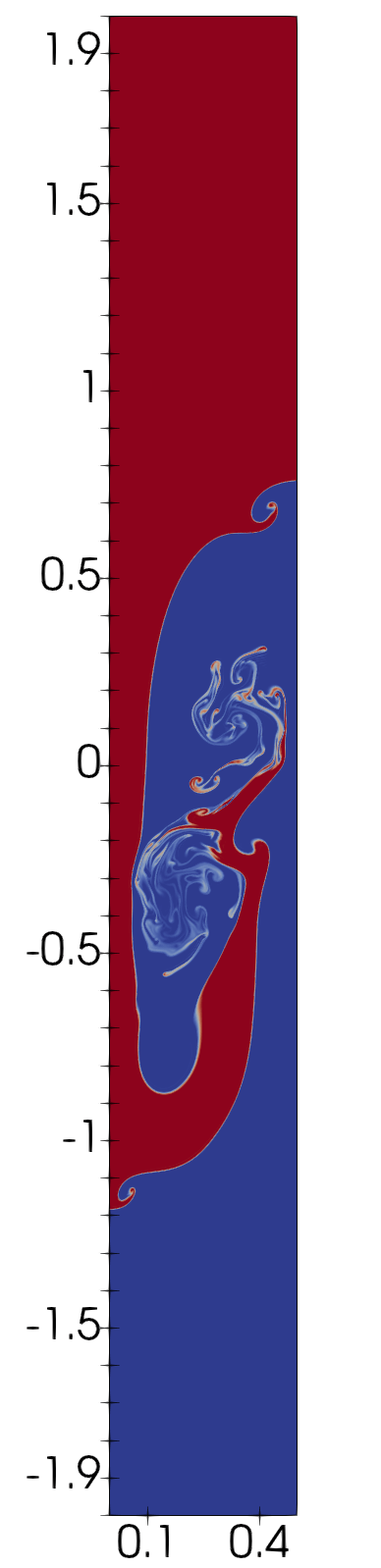} &
\includegraphics[trim=40 0 80 0,clip,width=0.108\textwidth]{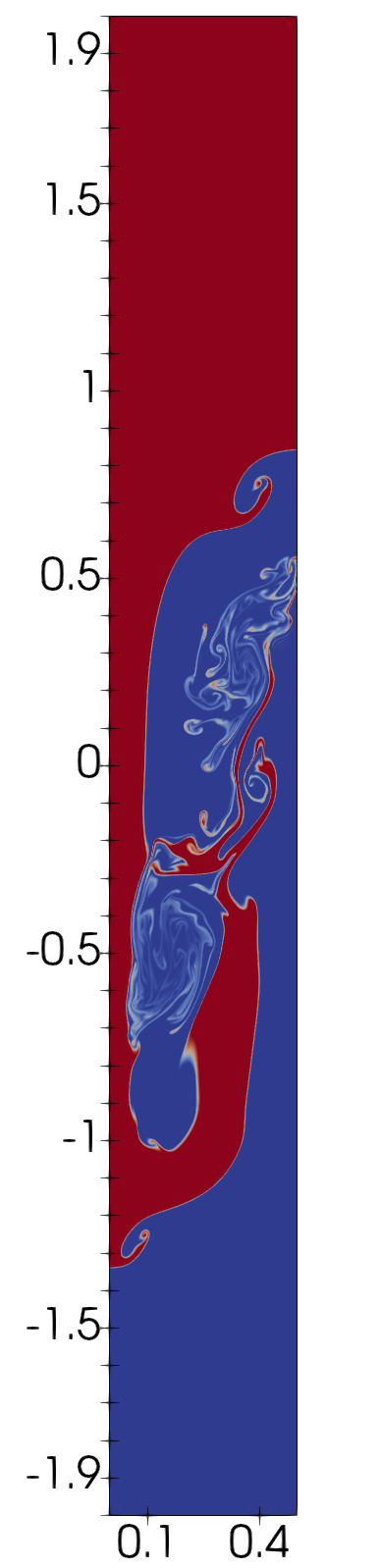} \\
\includegraphics[trim=40 0 80 0,clip,width=0.108\textwidth]{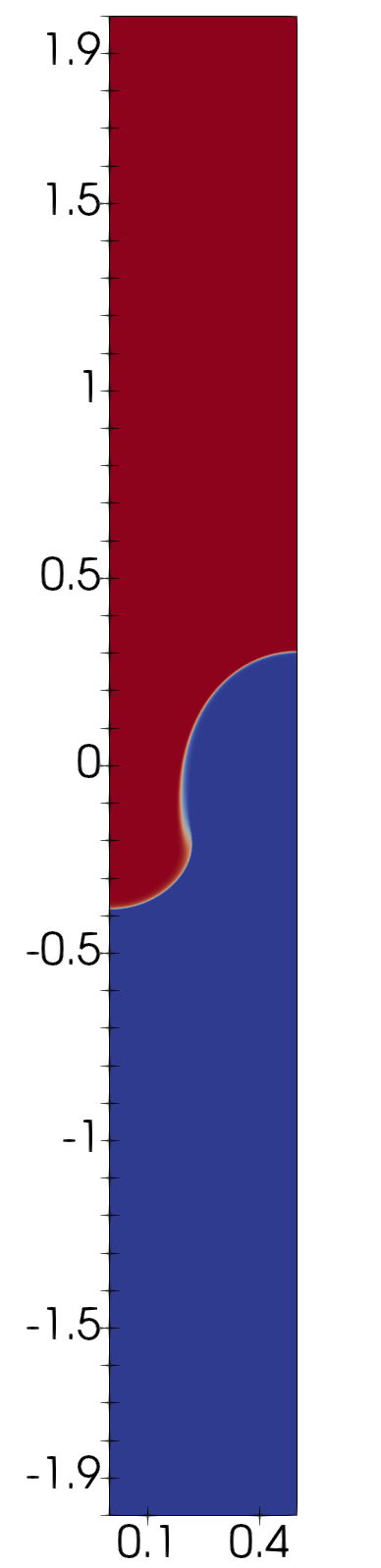} & 
\includegraphics[trim=40 0 80 0,clip,width=0.108\textwidth]{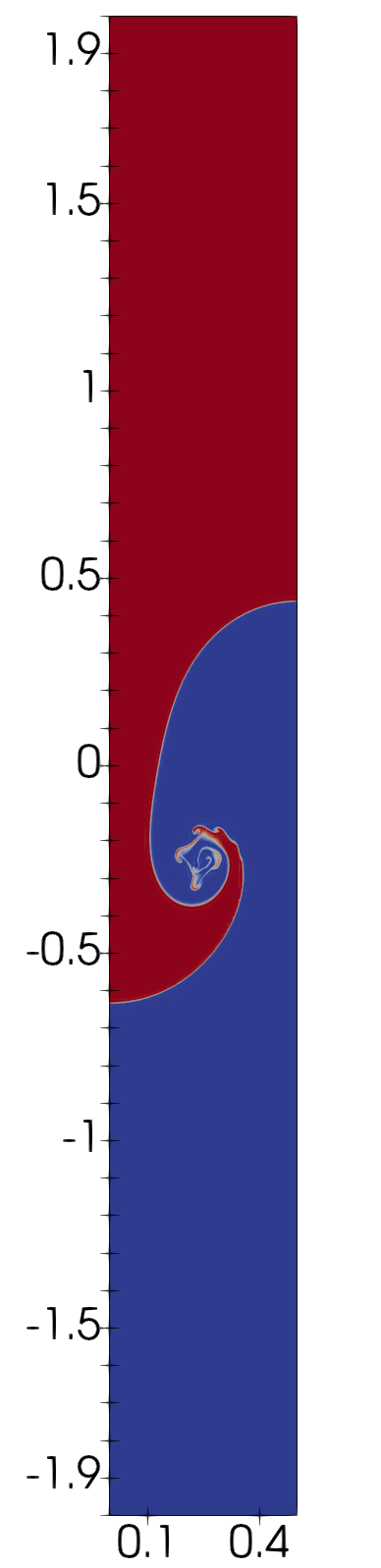} & 
\includegraphics[trim=40 0 80 0,clip,width=0.108\textwidth]{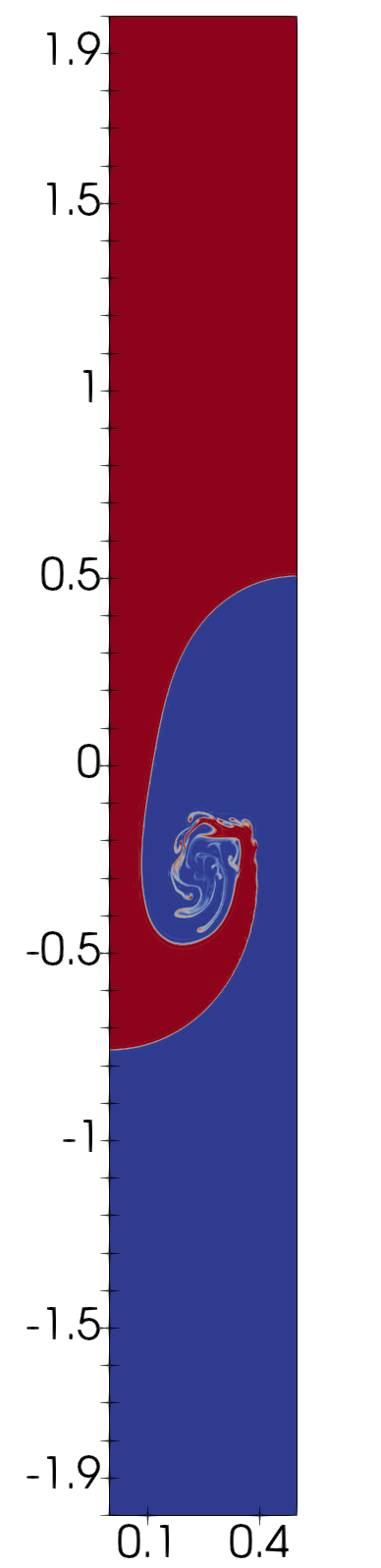} & 
\includegraphics[trim=40 0 80 0,clip,width=0.108\textwidth]{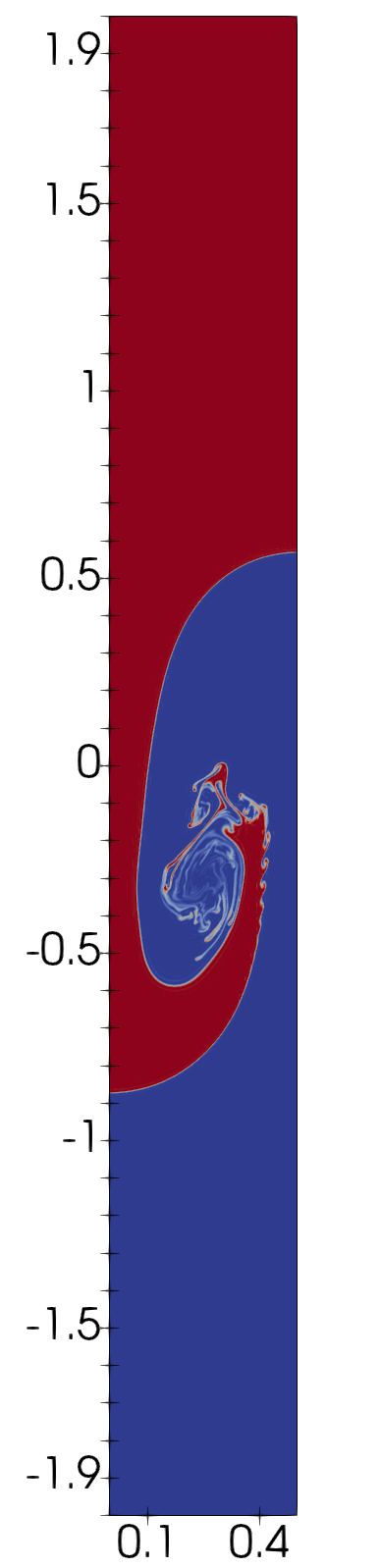} & 
\includegraphics[trim=40 0 80 0,clip,width=0.108\textwidth]{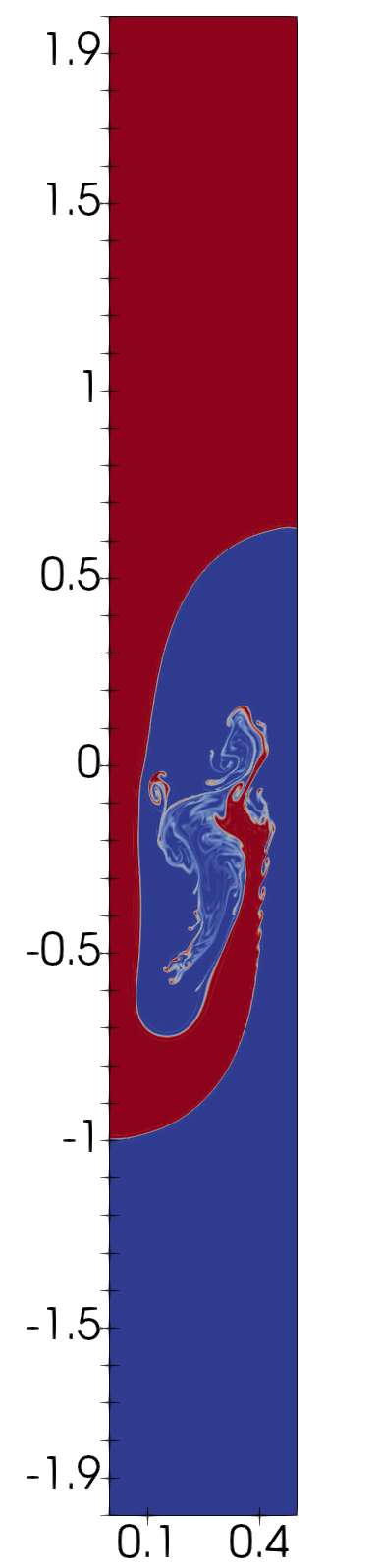} &
\includegraphics[trim=40 0 80 0,clip,width=0.108\textwidth]{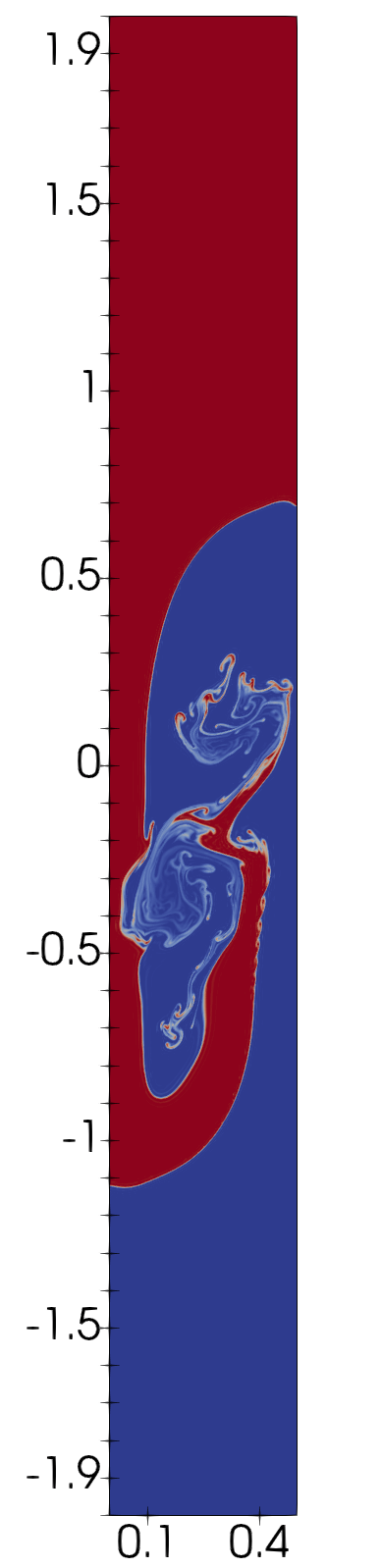} &
\includegraphics[trim=40 0 80 0,clip,width=0.108\textwidth]{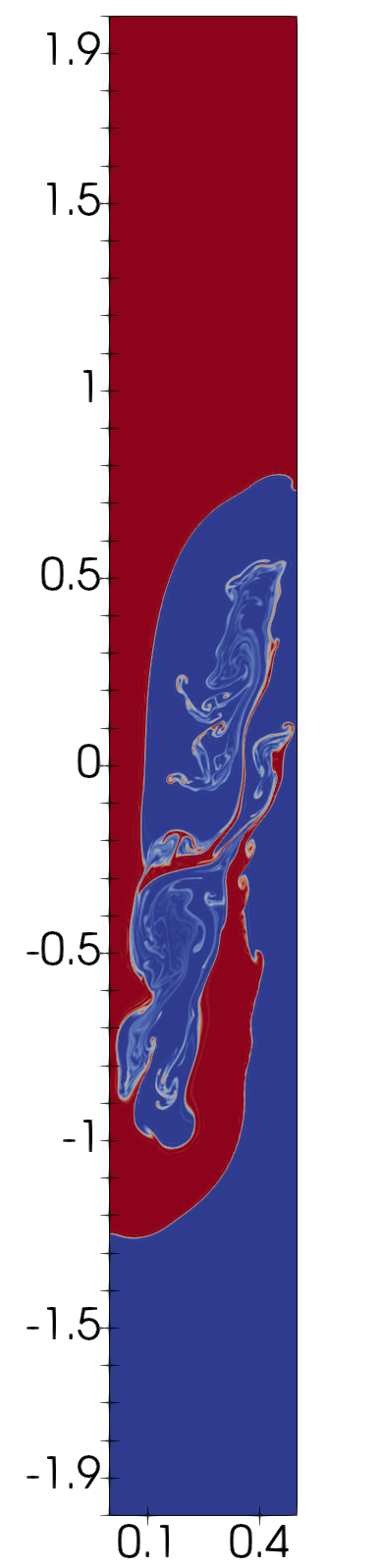} 
\end{tabular}
\caption{RTI at $\Atwood{=}0.5$ and $\Reynolds{=}5~000$ with HHO-$\pi^k$. 
\emph{From left to right}, evolution of the density field at selected time points 
$t_{\mathrm{Trg}} = t \sqrt{\Atwood} \simeq 1, 1.5, 1.75, 2, 2.25, 2.5, 2.75$. 
\emph{Top and bottom row}, results for $k=1$ over the fine grid and $k=6$ over the coarse grid, respectively.
         \label{fig:RTI5000}}
\end{figure}

The Reynolds number is defined as $\Reynolds = \frac{\rho_{\mathrm{b}} d^{\frac{3}{2}} g^{\frac{1}{2}}  }{\mu}$, 
note that $d^{\frac{1}{2}} g^{\frac{1}{2}}$ is the reference velocity.
In practice we consider unit reference length, unit gravity and unit $\rho_{\mathrm{b}}$, 
so that also the reference velocity is equal to one. 
Thus, we set $\mu=\frac{1}{\Reynolds}$ and we run computations considering two Reynolds numbers: $\Reynolds {=} 1~000$ and $\Reynolds {=} 5~000$.
The time evolution of the density field for HHO-$\pi^k$ is shown in Figures \ref{fig:RTI1000} and \ref{fig:RTI5000} 
at times $t_{\mathrm{Trg}} \simeq 1, 1.5, 1.75, 2, 2.25, 2.5, 2.75$ 
in the time scale of Tryggvason, where $t_{\mathrm{Trg}} = t \sqrt{\Atwood}$. 
In practice, for a given a value of $t_{\mathrm{Trg}}$, we select the closest available time point among the $3~200$ available 
by rounding off to the closest integer the quantity $\frac{t_{\mathrm{Trg}}}{\sqrt{\Atwood} \, \delta t}$, where $\delta t = \frac{t_F}{3~200}$.

At $\Reynolds{=}1~000$ the density fields provided by $k{=}1$ HHO-$\pi^k$ over the fine grid and $k{=}6$ HHO-$\pi^k$ over the coarse grid are closely matched 
and the results are in very good agreement with those obtained by \C{Guermond \ea~}\cite{Guermond2009}.
As opposite, at $\Reynolds{=}5~000$, some differences can be appreciated between the density fields provided by $k{=}1$ and $k{=}6$ approximations.
In particular, $k{=}6$ results show early breakup of the spiraling filament while $k{=}1$ results show interface sprouts 
in the proximity of the computational domain boundaries, especially at later stages of the flow field evolution.
Moreover, at $k{=}6$, the peripheral curling filaments interact with the bulk of the descending plume, 
a phenomena which does not take place in our results of reference \cite{Guermond2009}, where an entropy viscosity stabilization was employed.
Since high Reynolds number computations are associated with sharper interfaces featuring strong variations of all relevant physical quantities,
the aforementioned discrepancies can be ascribed to the lack of oscillation control.
In order to highlight the benefits of $p$-multigrid solutions strategies combined with static condensation,
it is worth to mention that $k{=}6$ HHO over the coarse grid provides 
a fourteen fold decrease of the global Jacobian matrix dimension compared to 
$k{=}1$ HHO over the fine grid, leading to significant advantages in terms of computation time. 

\begin{figure}[!htb]
\centering
\begin{tabular}{ccccccc}
\includegraphics[trim=50 0 100 0,clip,width=0.146\textwidth]{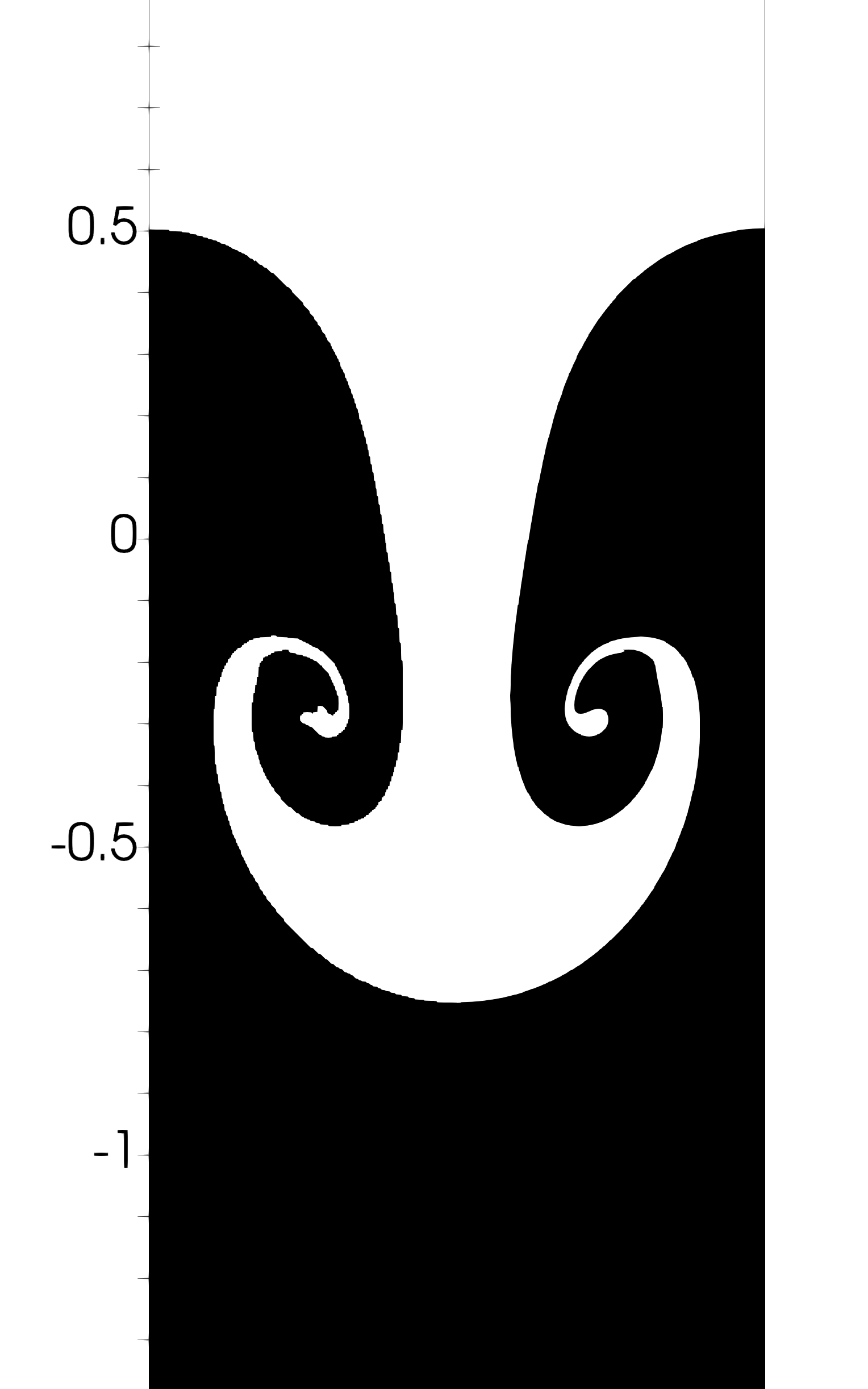} & 
\includegraphics[trim=50 0 100 0,clip,width=0.146\textwidth]{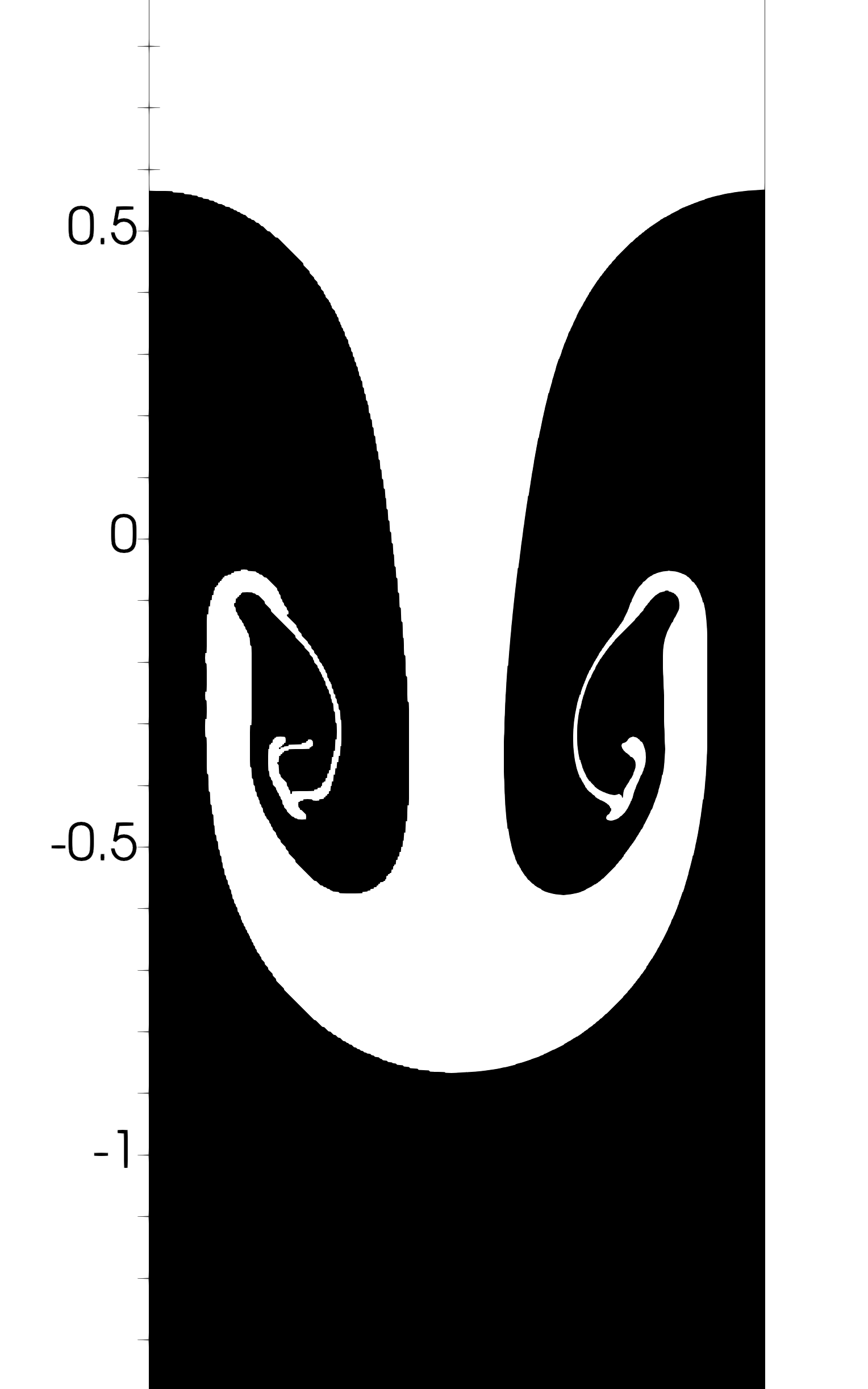} & 
\includegraphics[trim=50 0 100 0,clip,width=0.146\textwidth]{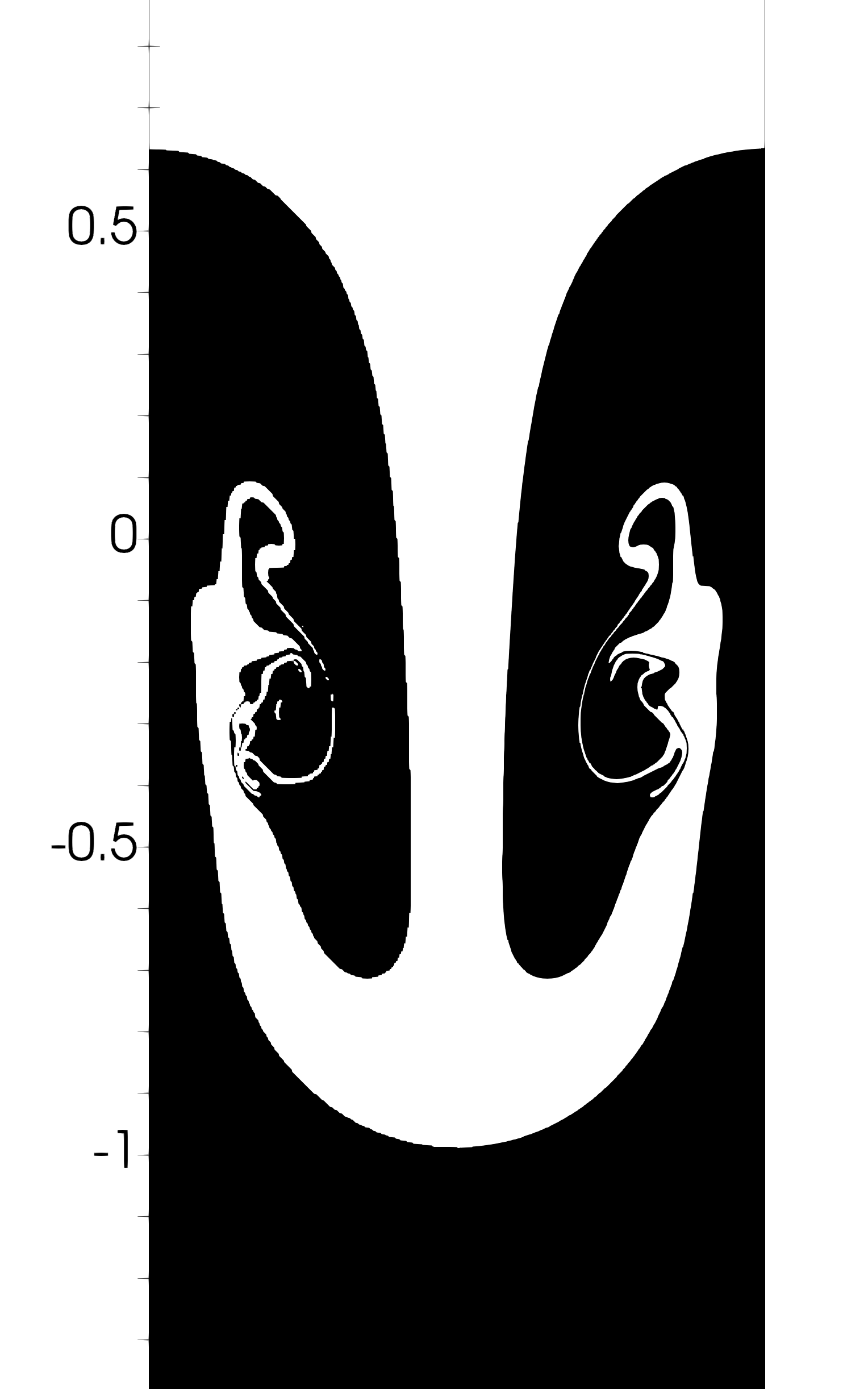} & 
\includegraphics[trim=50 0 100 0,clip,width=0.146\textwidth]{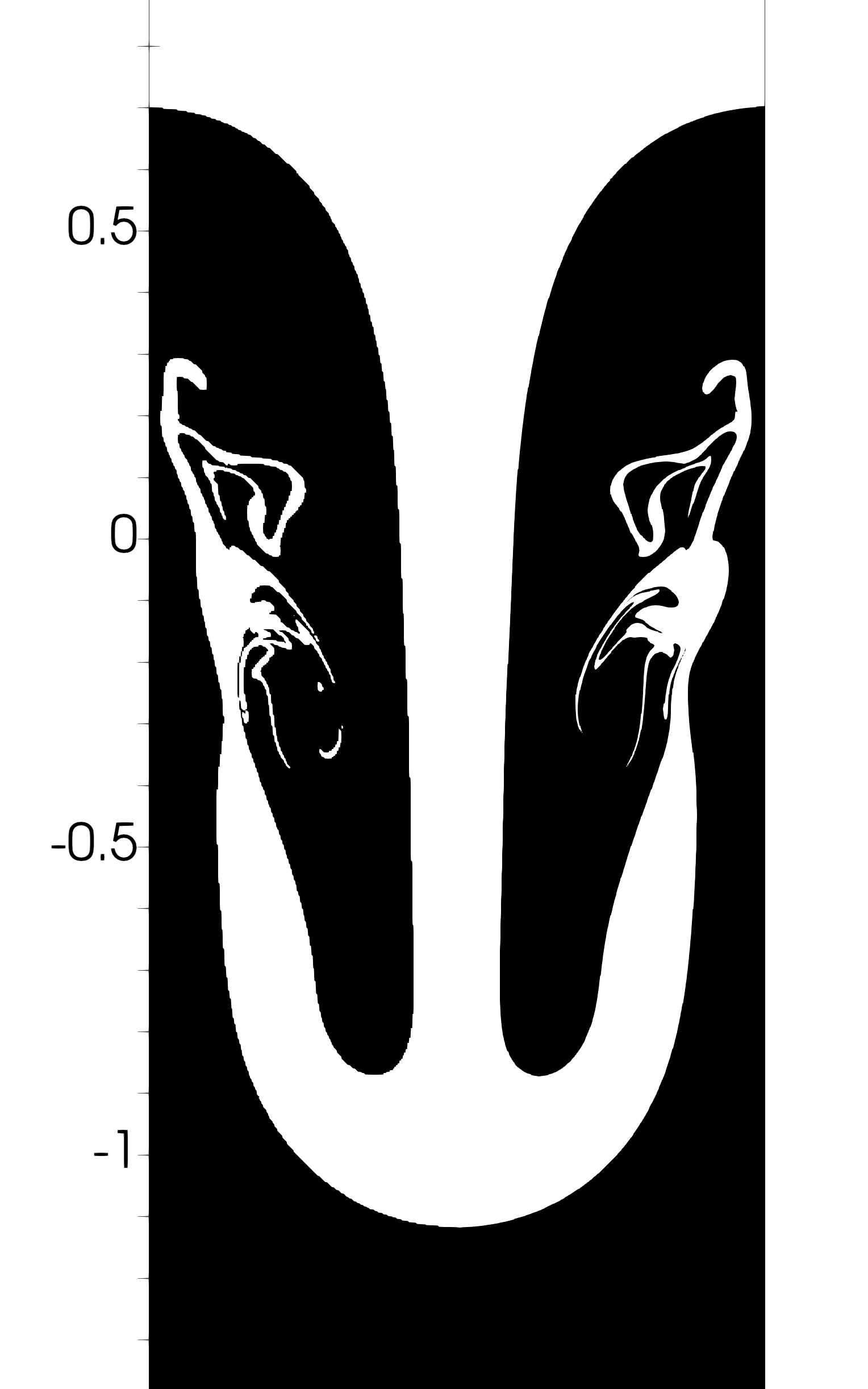} & 
\includegraphics[trim=50 0 100 0,clip,width=0.146\textwidth]{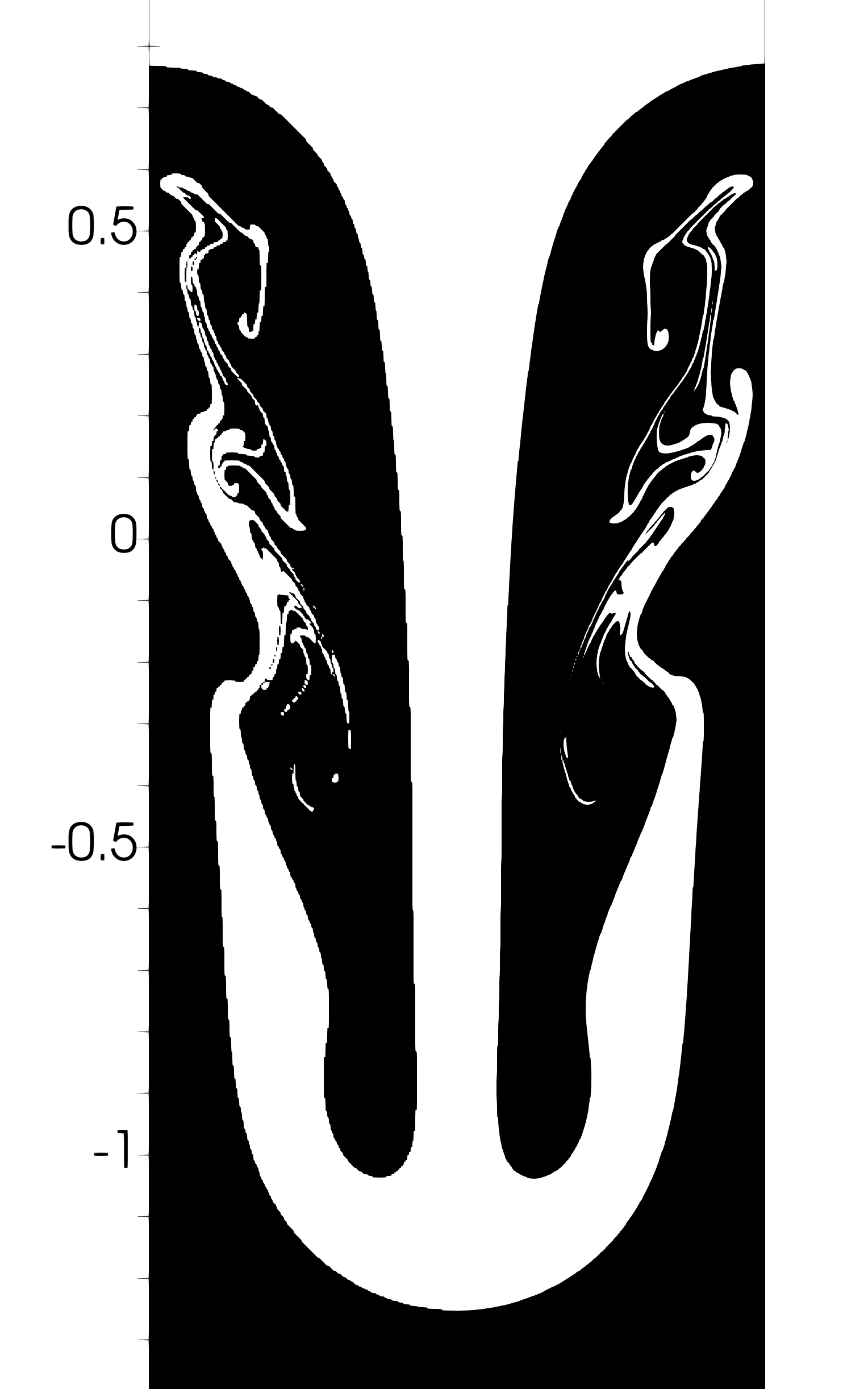} \\
\includegraphics[trim=50 0 100 0,clip,width=0.146\textwidth]{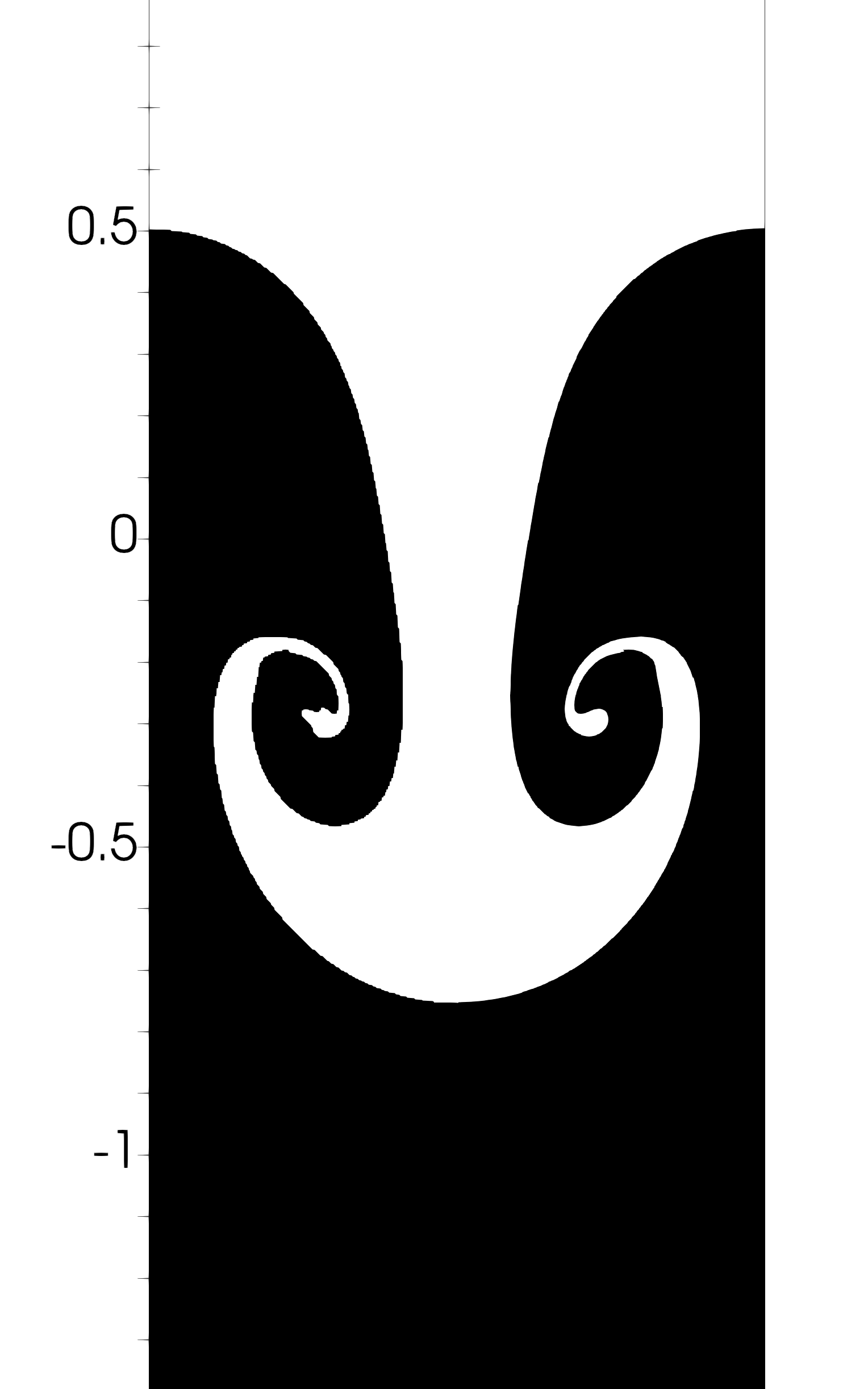} & 
\includegraphics[trim=50 0 100 0,clip,width=0.146\textwidth]{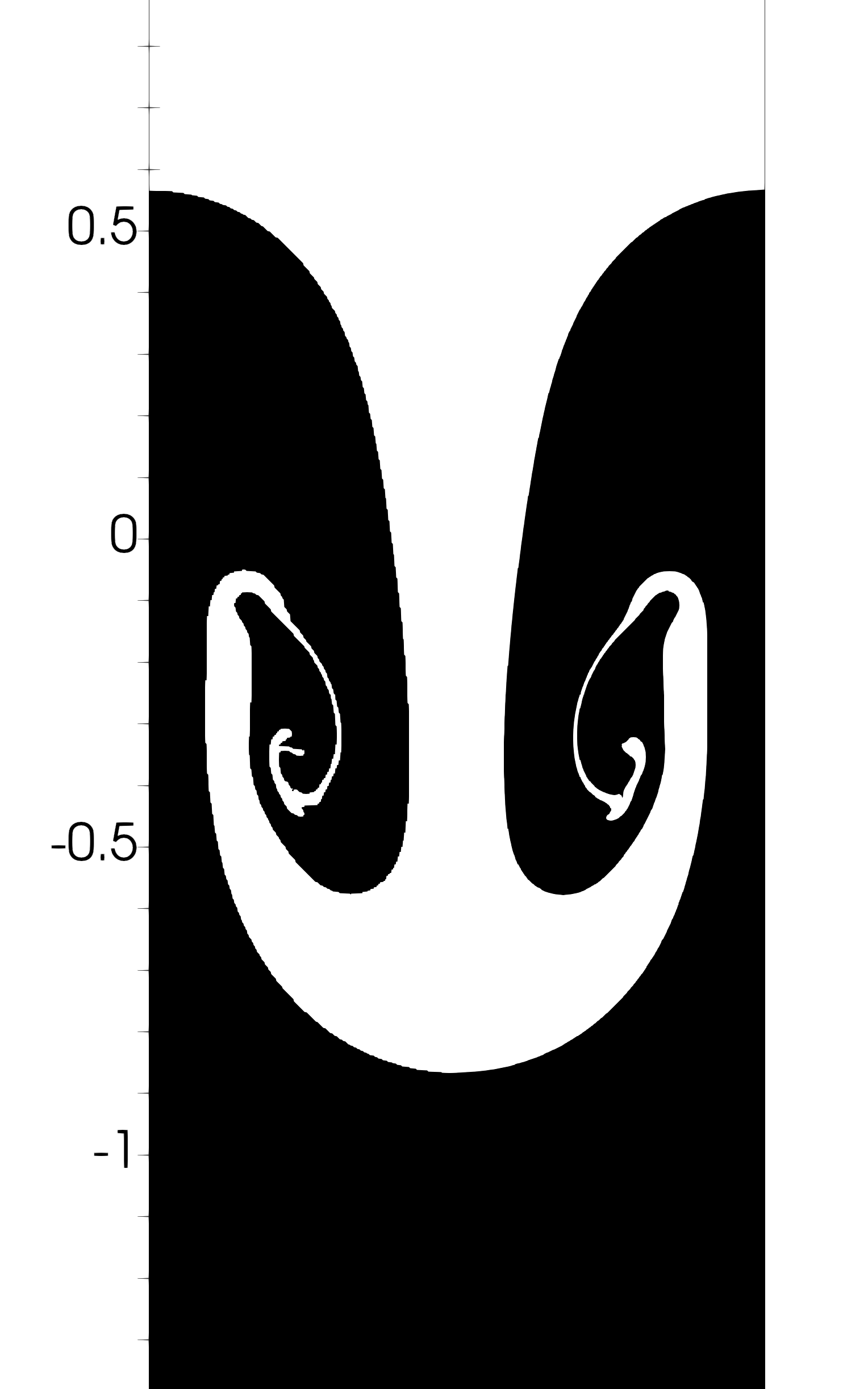} & 
\includegraphics[trim=50 0 100 0,clip,width=0.146\textwidth]{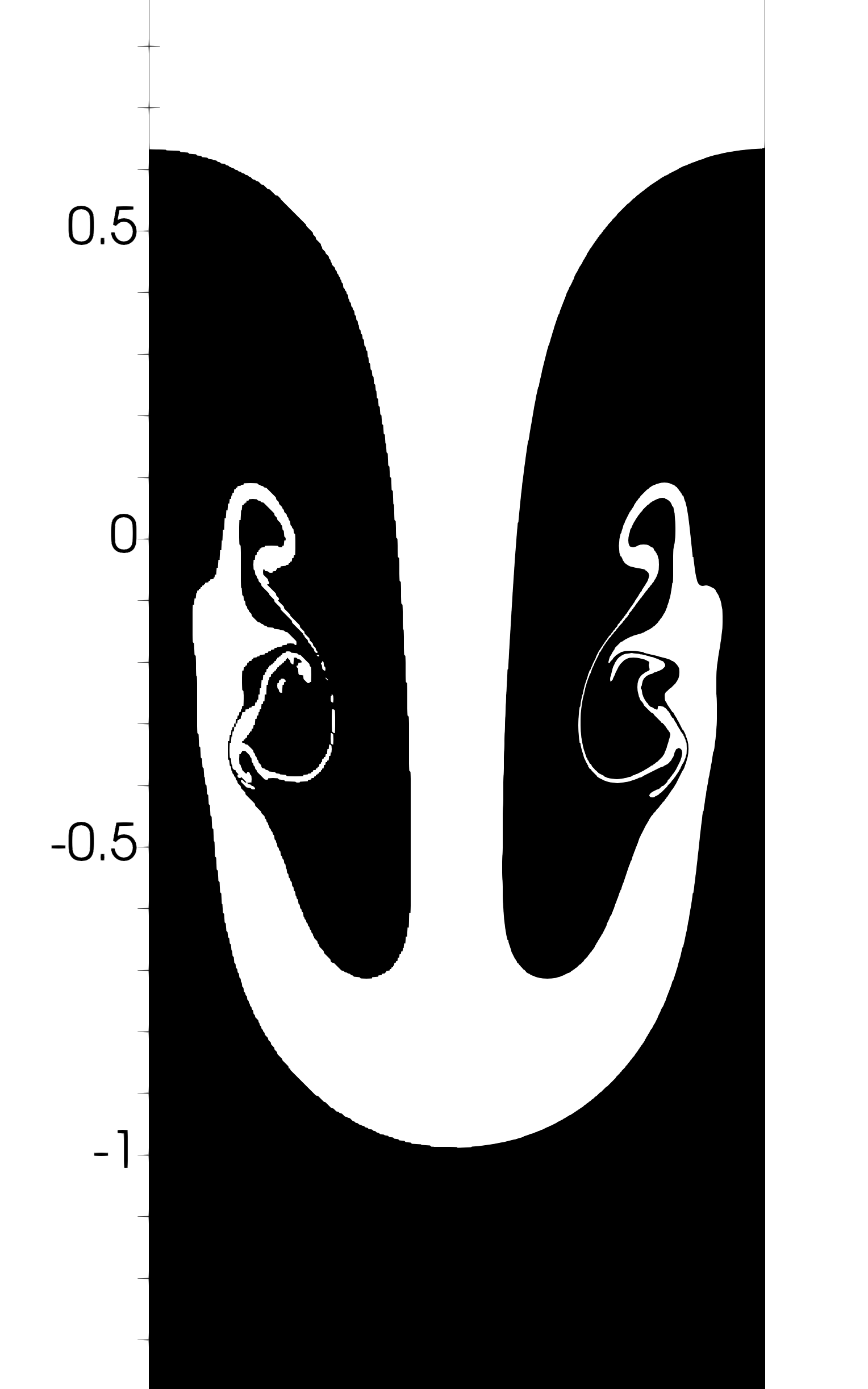} & 
\includegraphics[trim=50 0 100 0,clip,width=0.146\textwidth]{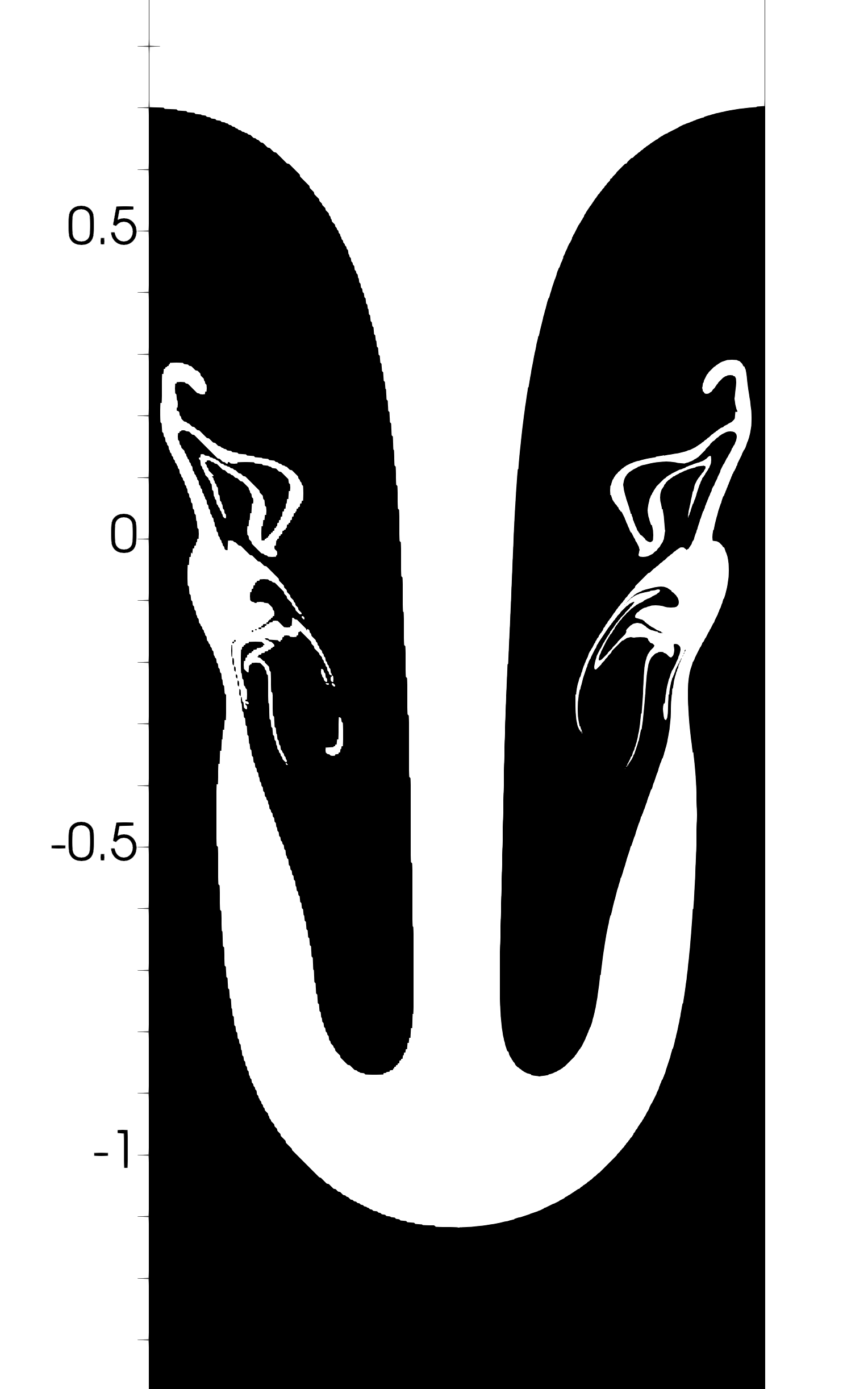} & 
\includegraphics[trim=50 0 100 0,clip,width=0.146\textwidth]{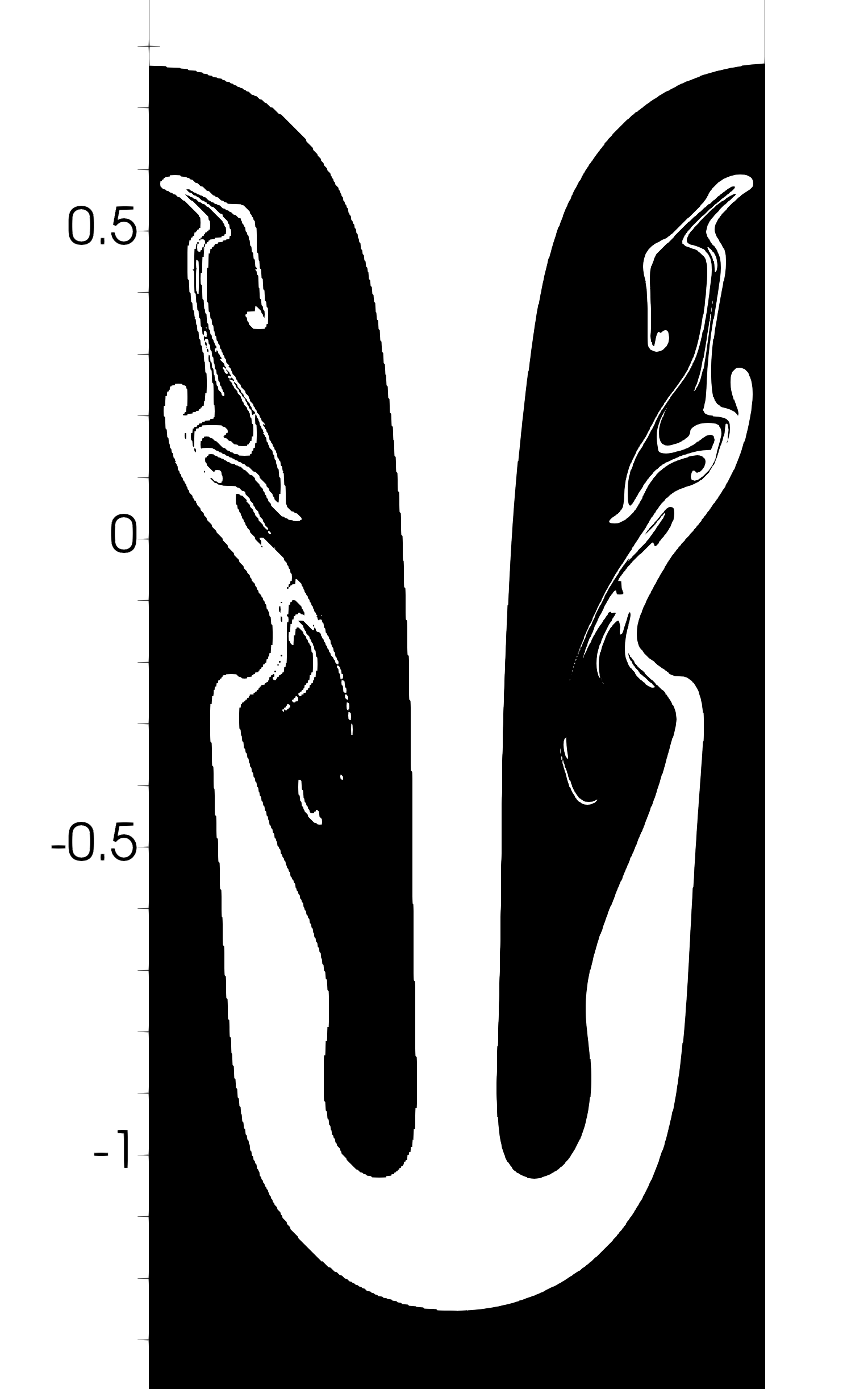} \\
\end{tabular}
\caption{RTI plume close-up view at $\Atwood{=}0.5$ and $\Reynolds{=}1~000$ with HHO-$\pi^k$ (\emph{top row}) and HHO-$\pi^{k+1}$ (\emph{bottom row}). 
\emph{From left to right}, evolution of the interface at selected time points $t_{\mathrm{Trg}} = t \sqrt{\Atwood} \simeq 1.75, 2, 2.25, 2.5, 2.75$. 
In each frame, results obtained with $k{=}6$ over the coarse grid and $k{=}1$ over the fine grid results stands on the left and the right, 
respectively, with respect to a vertical center-line.
         \label{fig:RTI1000comp}}
\end{figure}

\begin{figure}[!htb]
\centering
\begin{tabular}{ccccccc}
\includegraphics[trim=50 0 100 0,clip,width=0.146\textwidth]{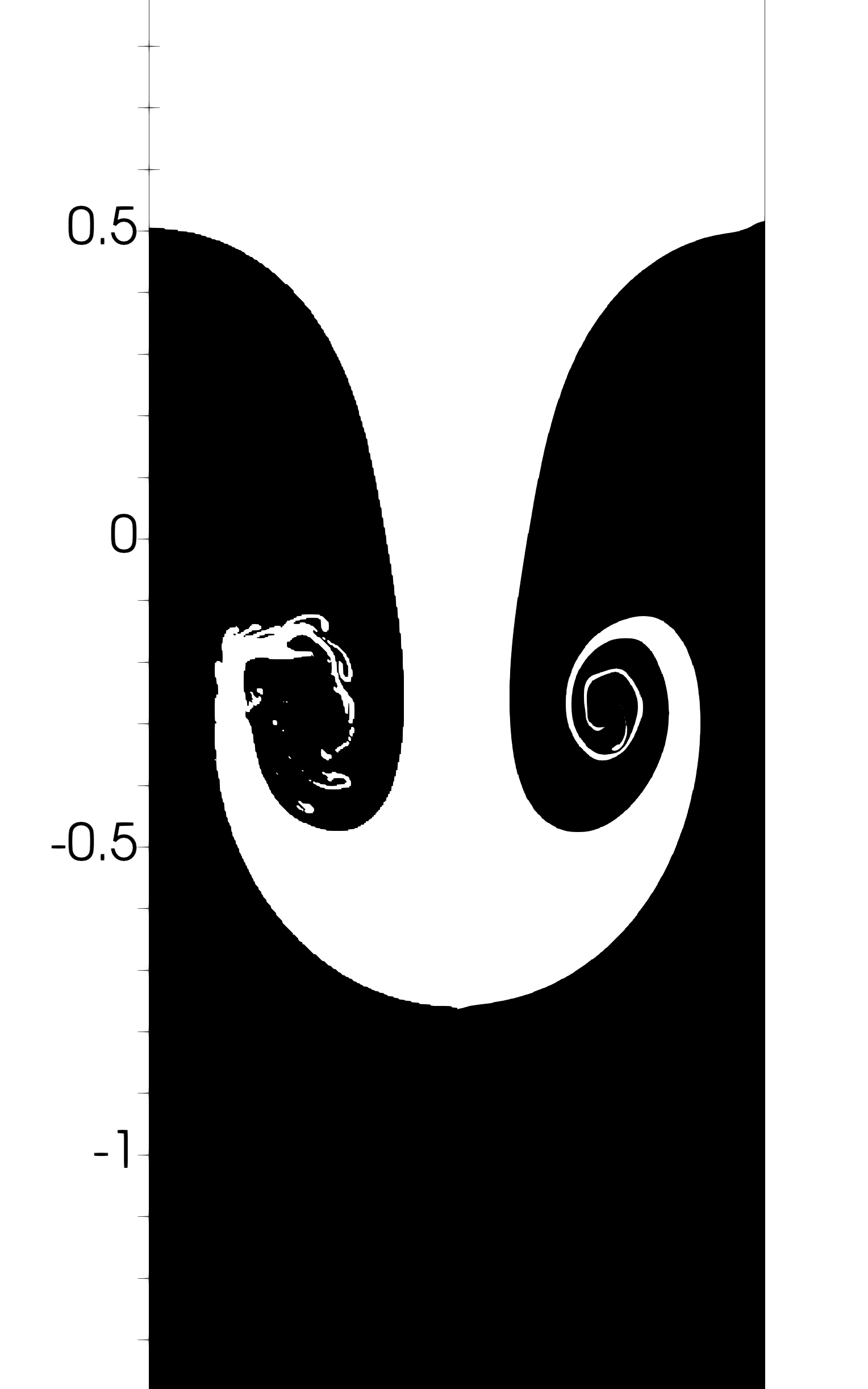} & 
\includegraphics[trim=50 0 100 0,clip,width=0.146\textwidth]{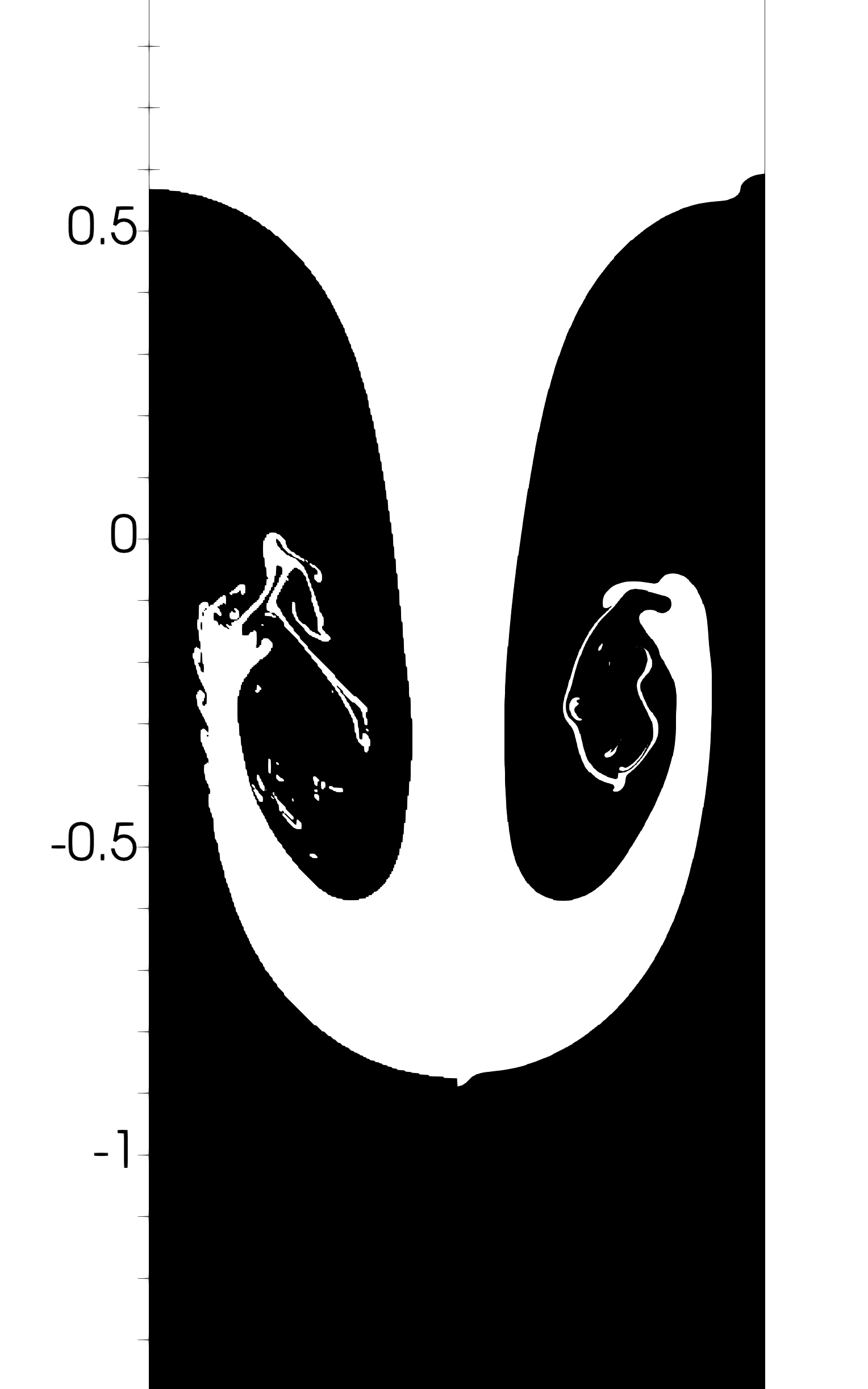} & 
\includegraphics[trim=50 0 100 0,clip,width=0.146\textwidth]{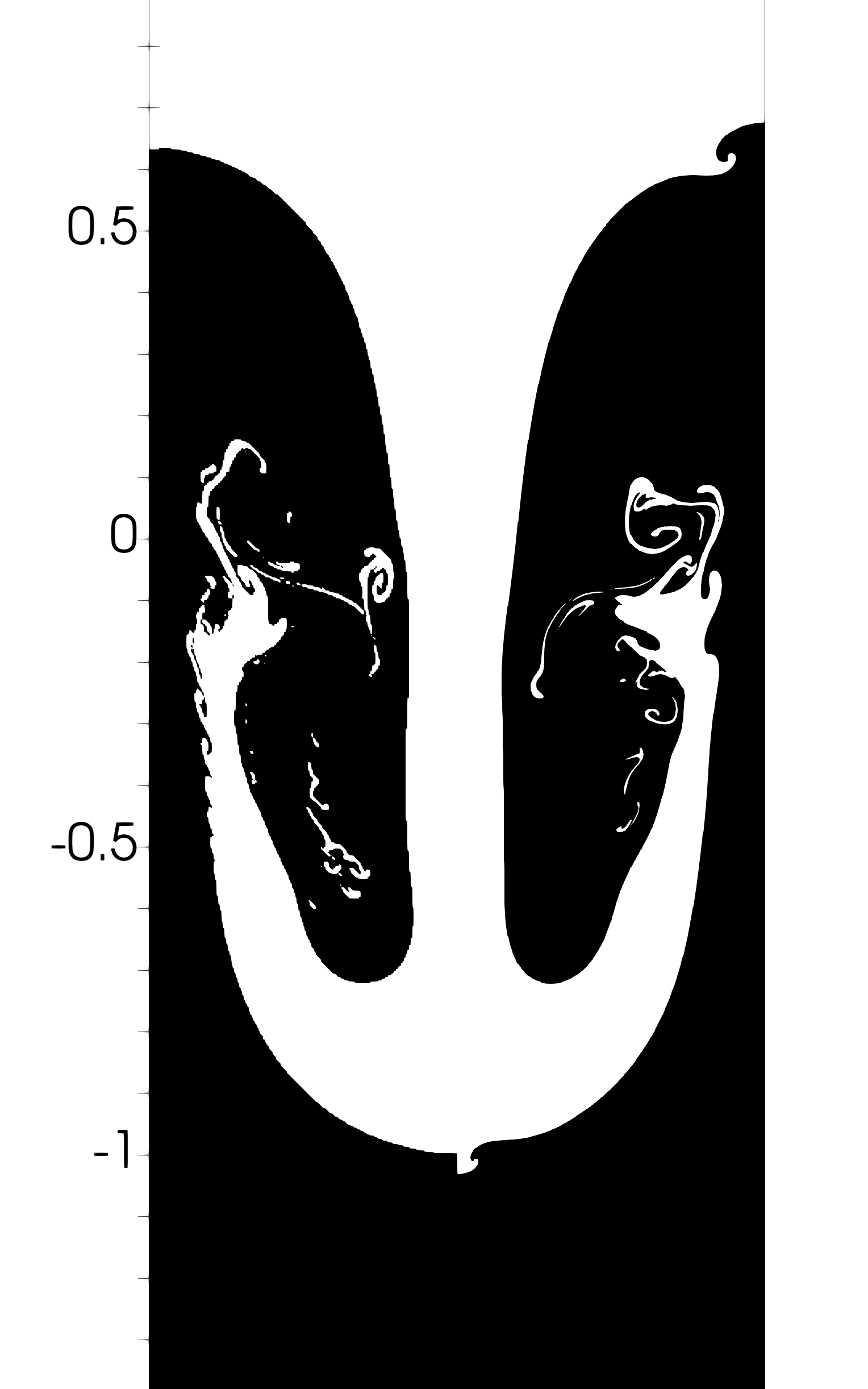} & 
\includegraphics[trim=50 0 100 0,clip,width=0.146\textwidth]{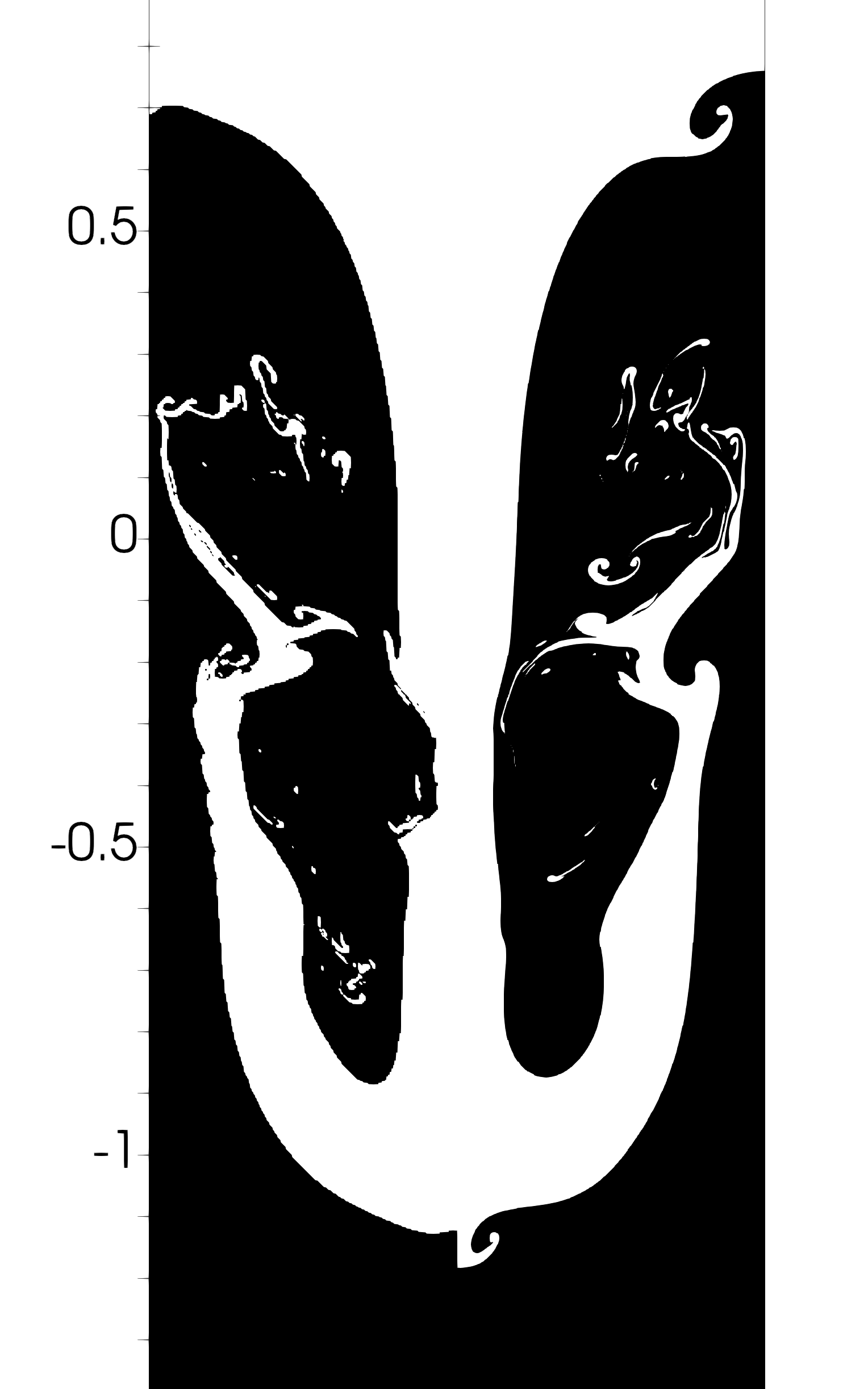} & 
\includegraphics[trim=50 0 100 0,clip,width=0.146\textwidth]{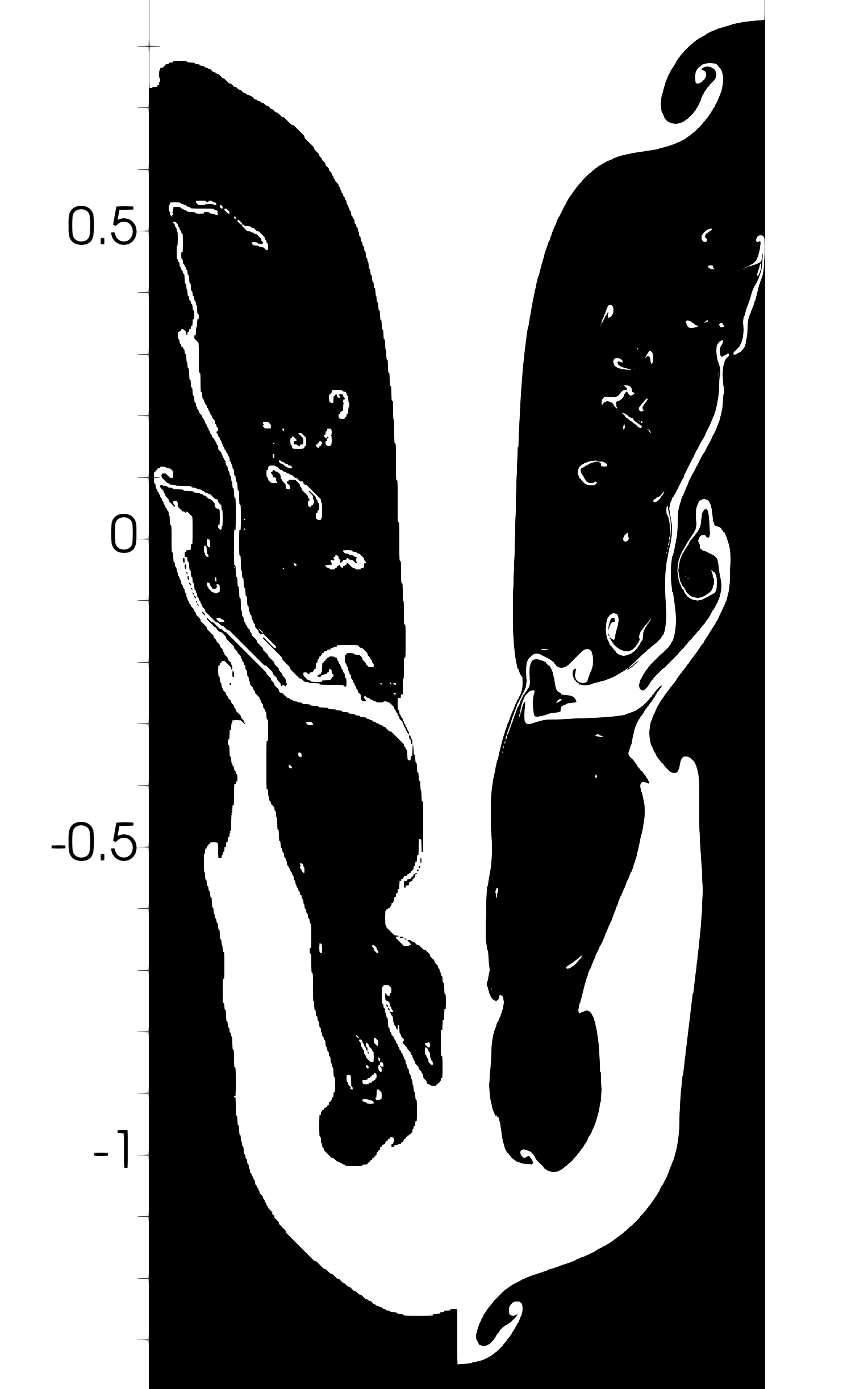} \\
\includegraphics[trim=50 0 100 0,clip,width=0.146\textwidth]{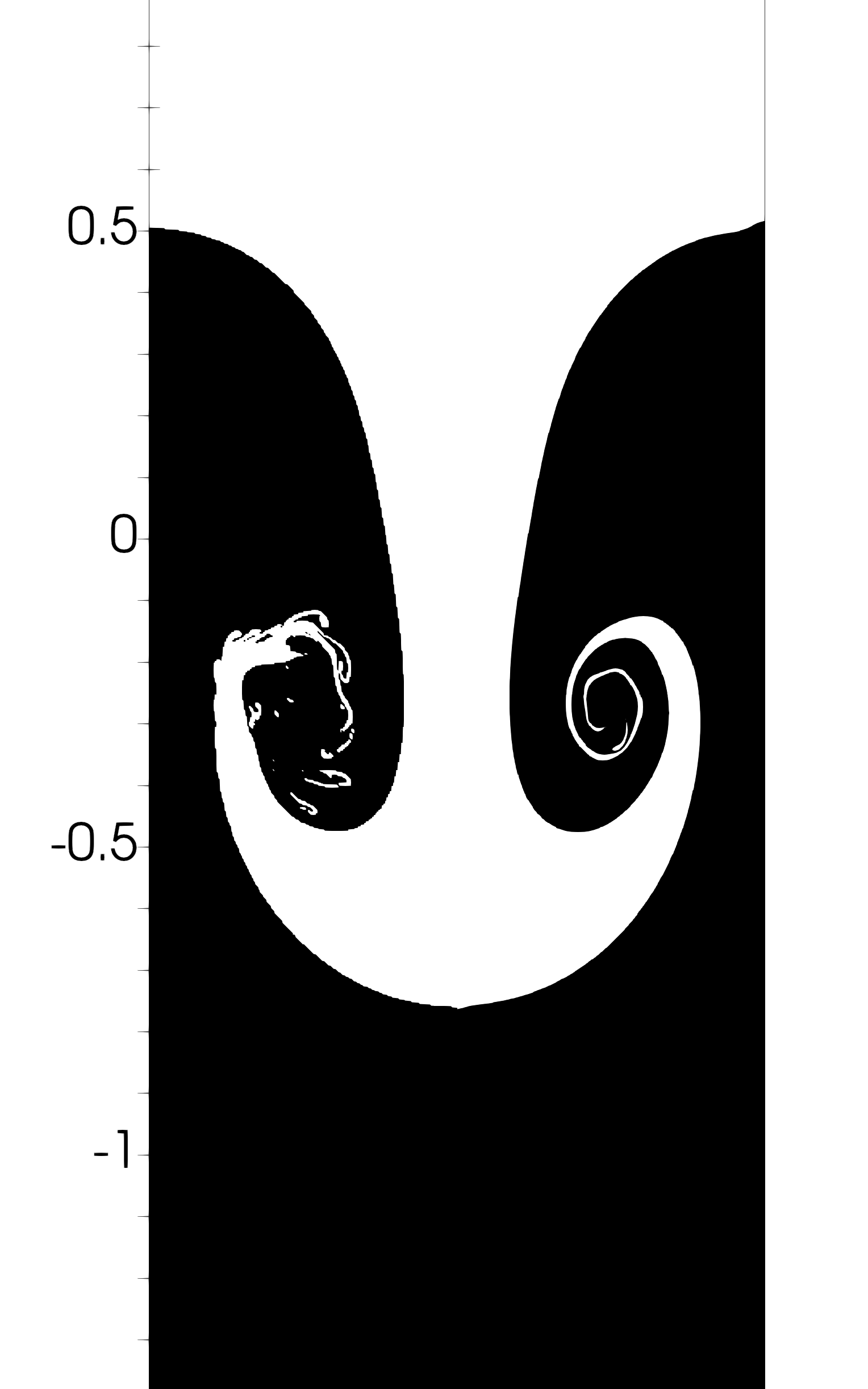} & 
\includegraphics[trim=50 0 100 0,clip,width=0.146\textwidth]{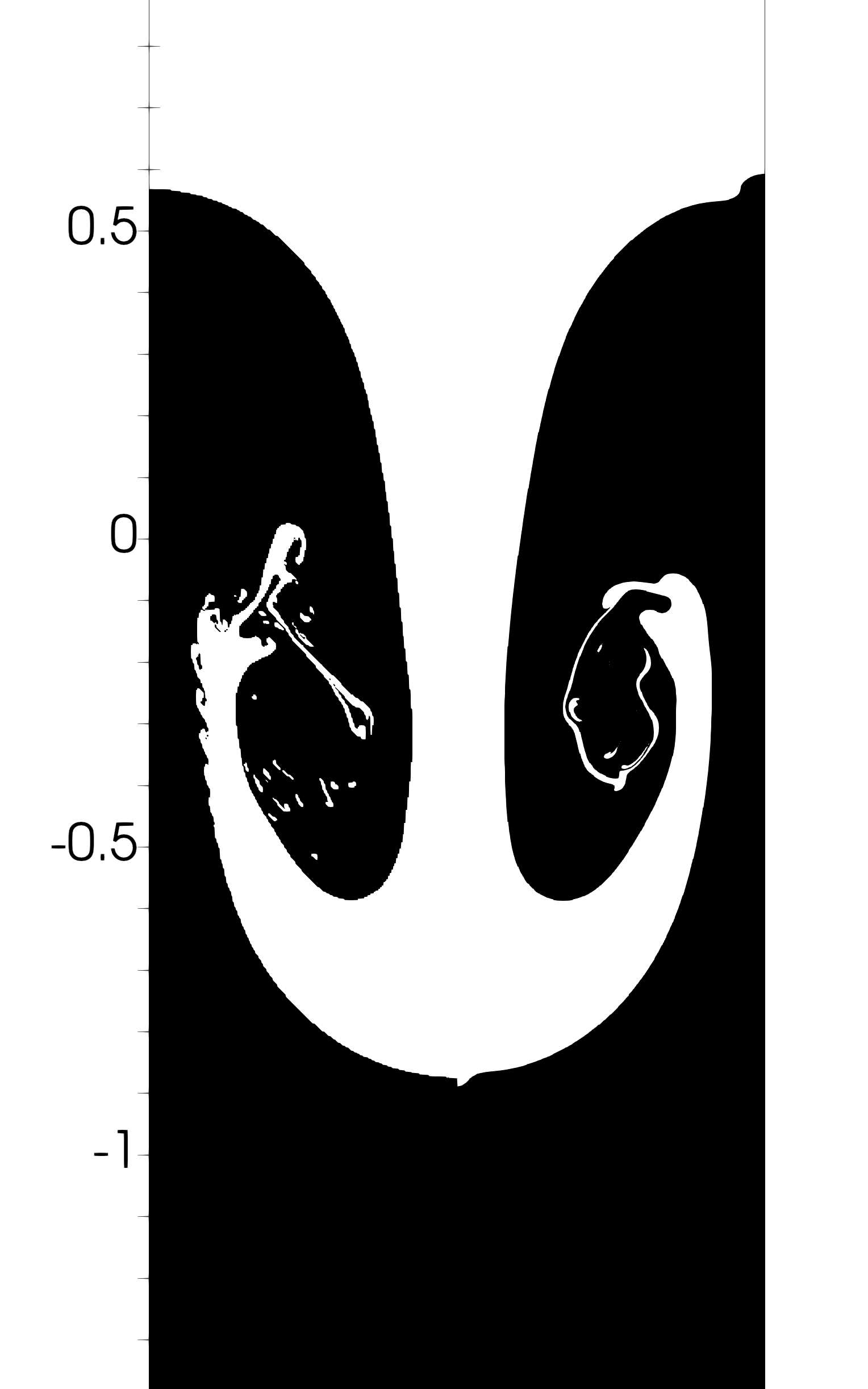} & 
\includegraphics[trim=50 0 100 0,clip,width=0.146\textwidth]{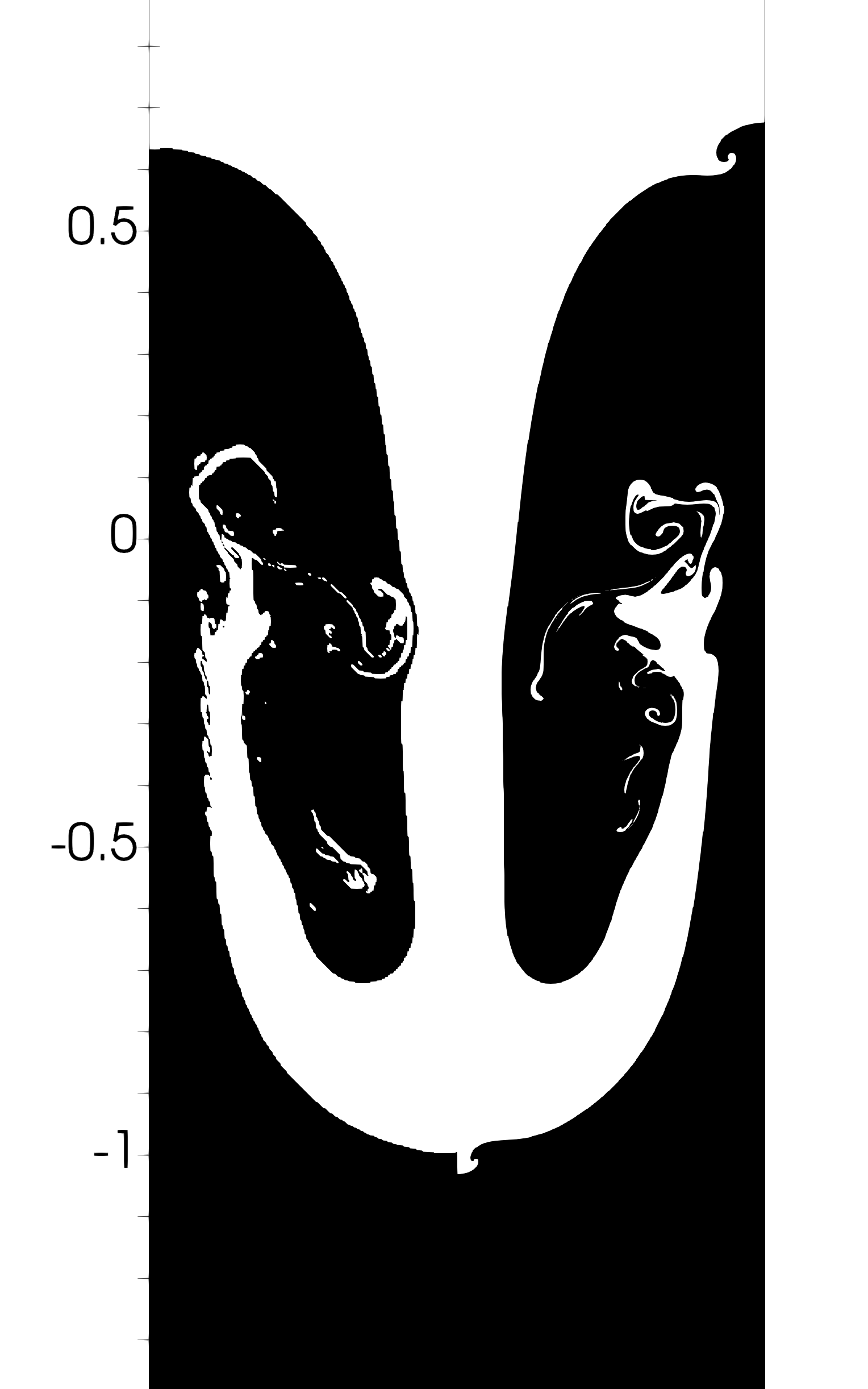} & 
\includegraphics[trim=50 0 100 0,clip,width=0.146\textwidth]{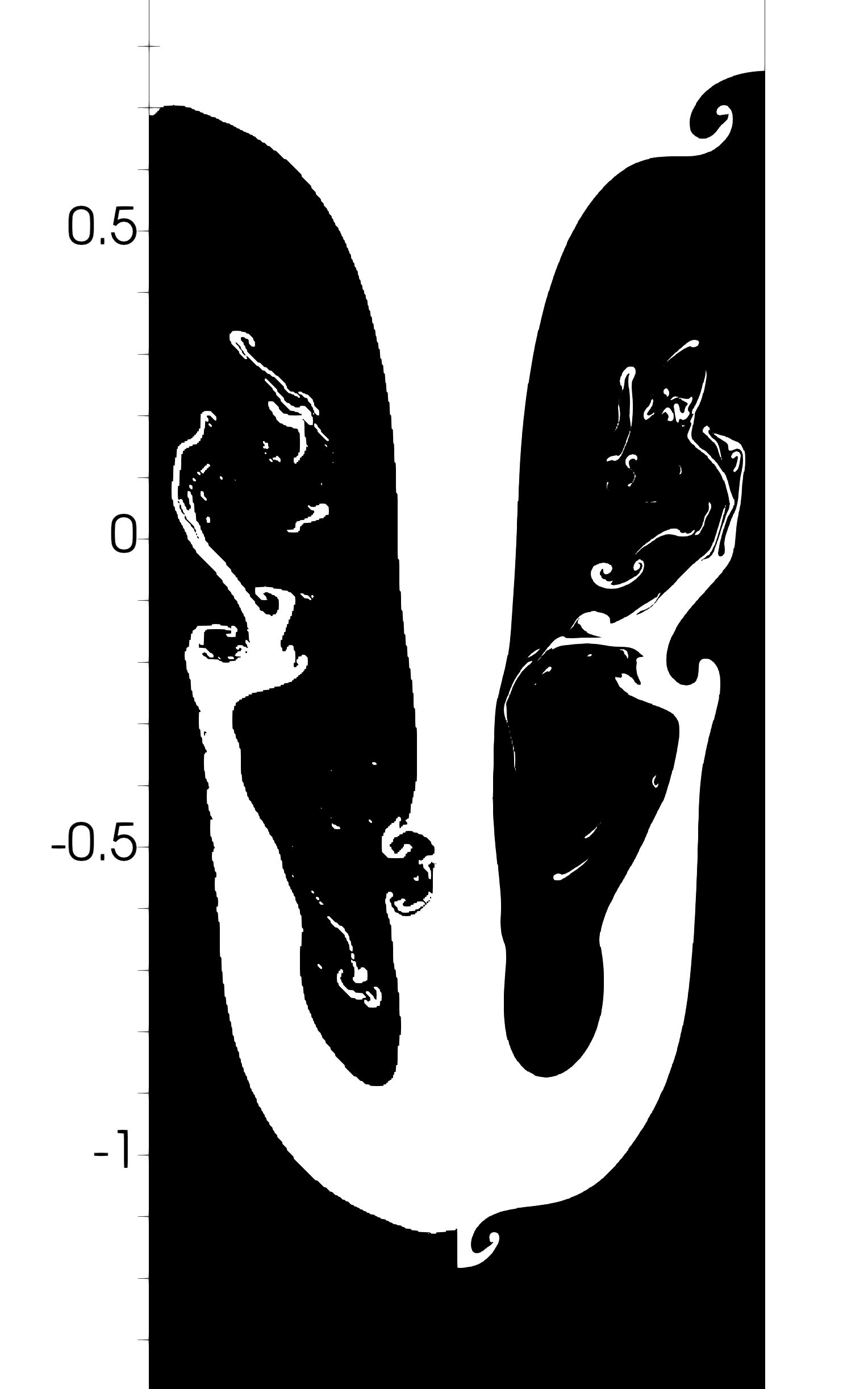} & 
\includegraphics[trim=50 0 100 0,clip,width=0.146\textwidth]{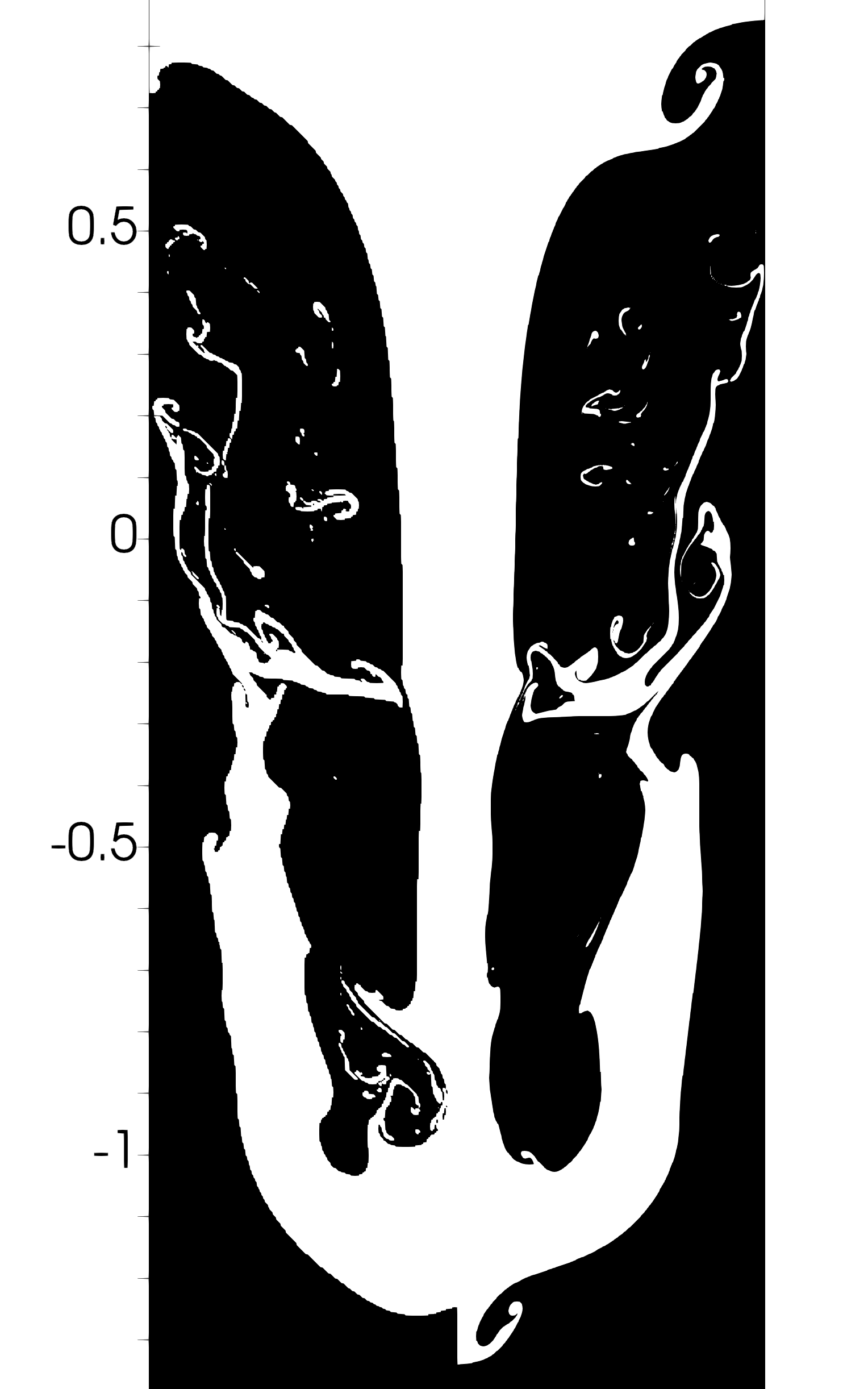} \\
\end{tabular}
\caption{RTI plume close-up view at $\Atwood{=}0.5$ and $\Reynolds{=}5~000$ with HHO-$\pi^k$ (\emph{top row}) and HHO-$\pi^{k+1}$ (\emph{bottom row}). 
\emph{From left to right}, evolution of the interface at selected time points $t_{\mathrm{Trg}} = t \sqrt{\Atwood} \simeq 1.75, 2, 2.25, 2.5, 2.75$. 
In each frame, results obtained with $k{=}6$ over the coarse grid and $k{=}1$ over the fine grid results stands on the left and the right, 
respectively, with respect to a vertical center-line.
         \label{fig:RTI5000comp}}
\end{figure}

A comparison between HHO-$\pi^k$ and HHO-$\pi^{k+1}$ is reported in Figure \ref{fig:RTI1000comp} and Figure \ref{fig:RTI5000comp}, 
focusing on the position of the interface between the two fluids. 
Accordingly, by setting the threshold to the average density ($\rho{=}2$), 
we are able to distinguish between the heavy fluid, depicted in white, 
and the light fluid, depicted in black.
To highlight the differences between high-order and low-order accurate results, 
we depict the whole computational domain $\Omega = (-\frac{d}{2},\frac{d}{2})\times(-2d,2d)$
by gluing together $k{=}6$ computations rotated by 180 degrees, as if $\Omega^{k=6} = (-\frac{d}{2},0)\times(-2d,2d)$ was the computational domain, 
with $k{=}1$ computations over $\Omega^{k=1} = (0,\frac{d}{2})\times(-2d,2d)$, \ie $\Omega =\Omega^{k=6} \bigcup \Omega^{k=1}$.

At both $Re{=}1~000$ and $Re{=}5~000$, HHO-$\pi^k$ and HHO-$\pi^{k+1}$ computations are in very good agreement when $k=1$, while some differences  
can be appreciated comparing $k=6$ results at $Re{=}5~000$. 
As previously mentioned, we can conclude that spurious oscillations influence the most the flow field evolution 
when the interface is sharp and high-degree polynomials are considered.
This confirms the statement that an accurate and detailed prediction of the RTI flow for $t\geq 1.5$ and
and $\Reynolds \geq 5000$ is a difficult task, as observed by \C{Guermond \ea~}\cite{Guermond2000}.


\subsubsection{High Atwood number results}
\begin{figure}[!htb]
\centering
\begin{tabular}{ccccccc}
\includegraphics[trim=40 0 80 0,clip,width=0.108\textwidth]{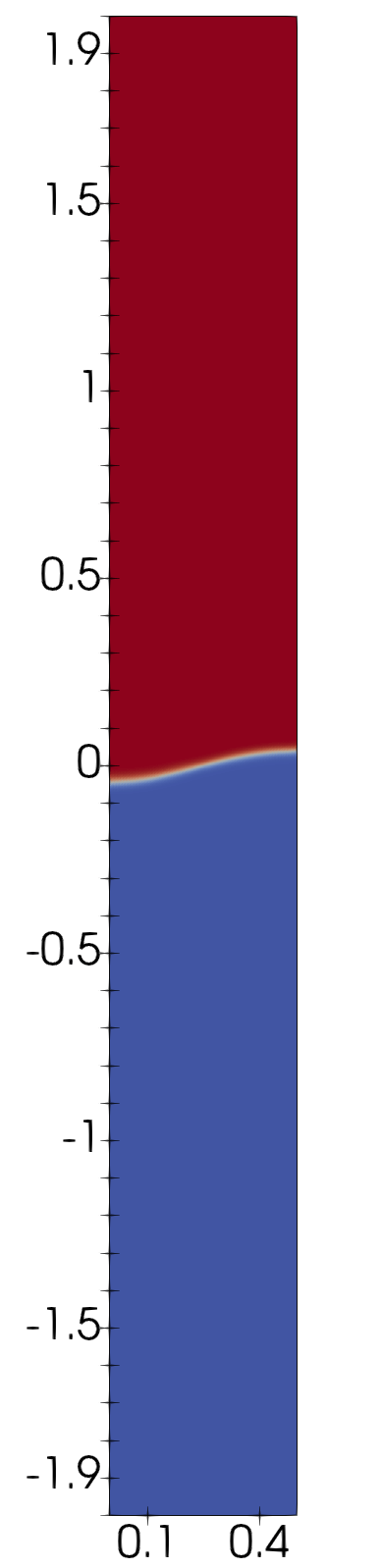} & 
\includegraphics[trim=40 0 80 0,clip,width=0.108\textwidth]{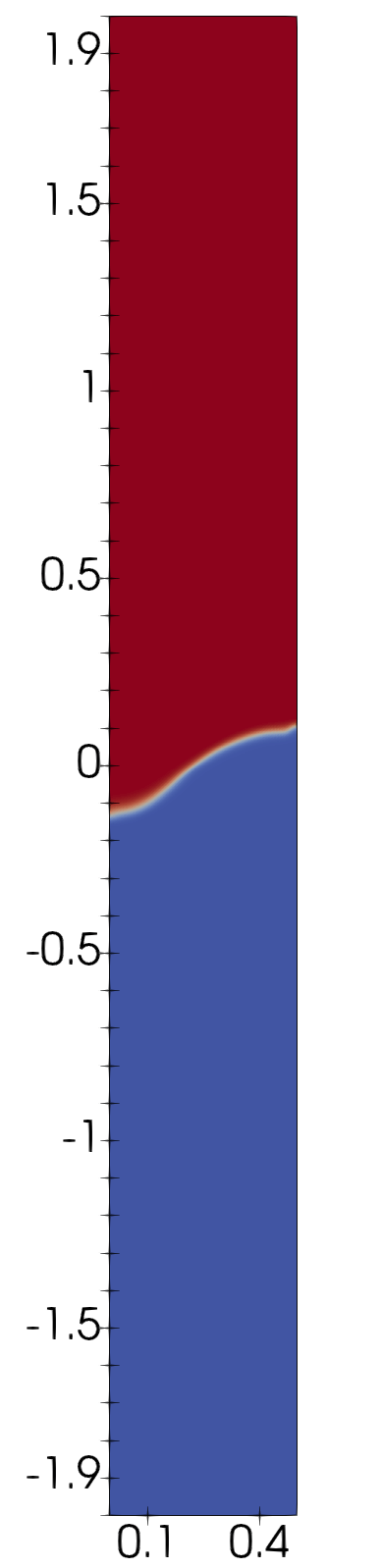} & 
\includegraphics[trim=40 0 80 0,clip,width=0.108\textwidth]{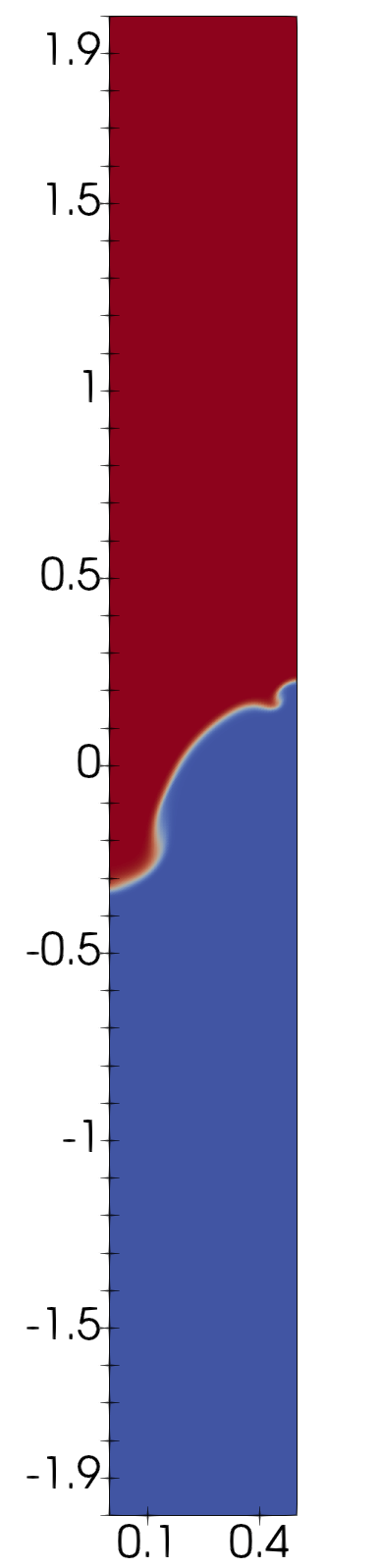} & 
\includegraphics[trim=40 0 80 0,clip,width=0.108\textwidth]{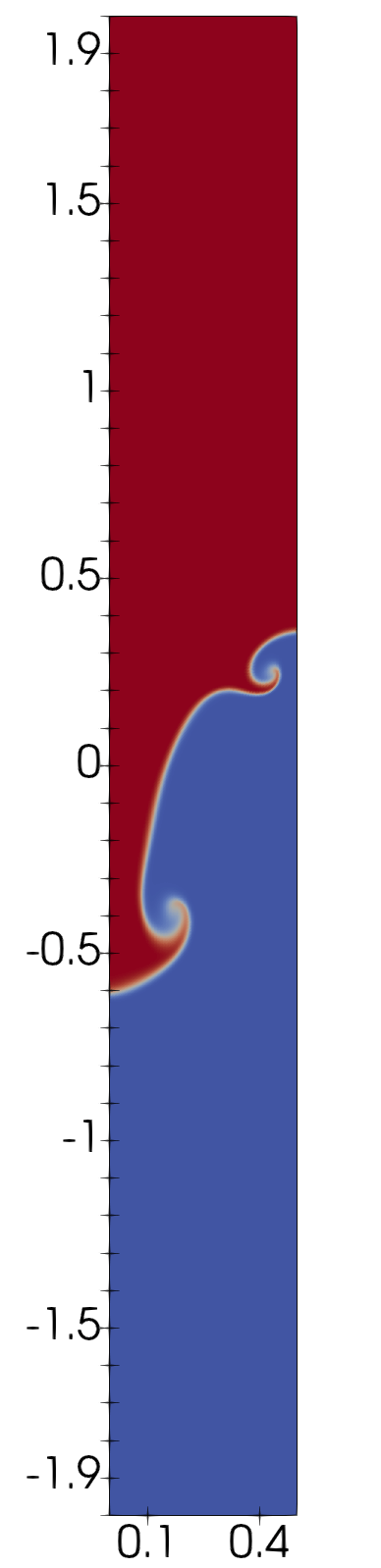} & 
\includegraphics[trim=40 0 80 0,clip,width=0.108\textwidth]{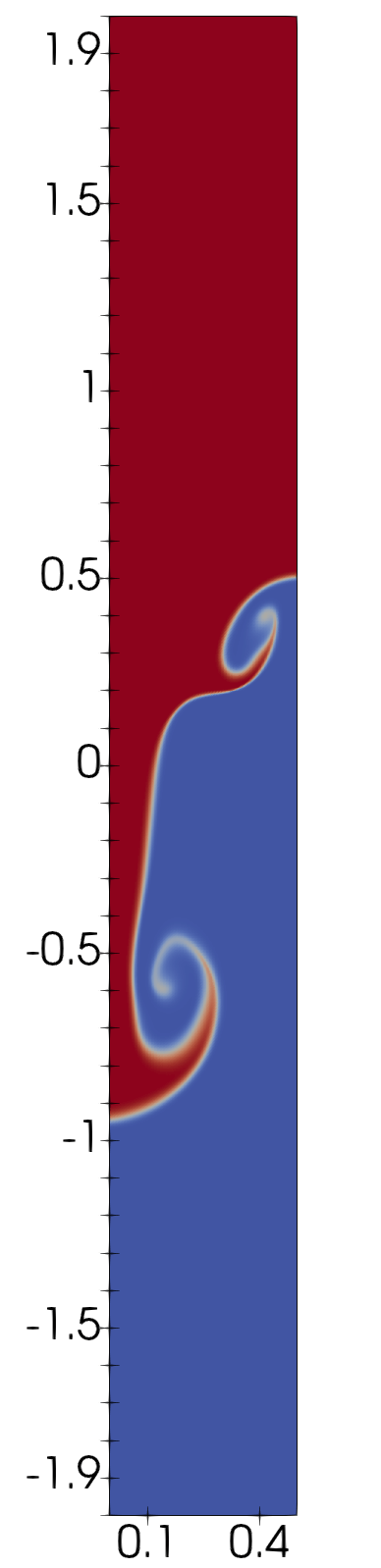} &
\includegraphics[trim=40 0 80 0,clip,width=0.108\textwidth]{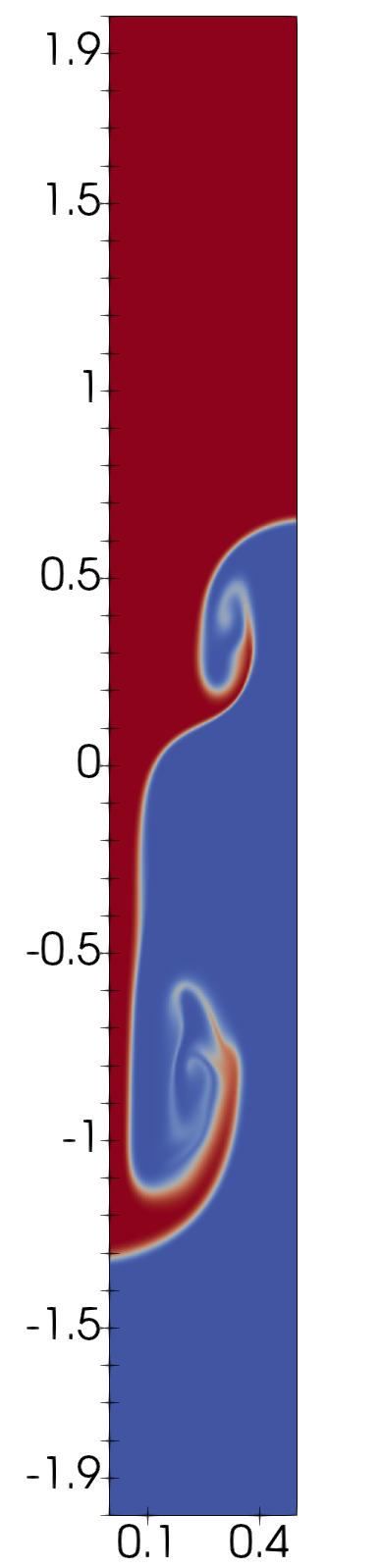} &
\includegraphics[trim=40 0 80 0,clip,width=0.108\textwidth]{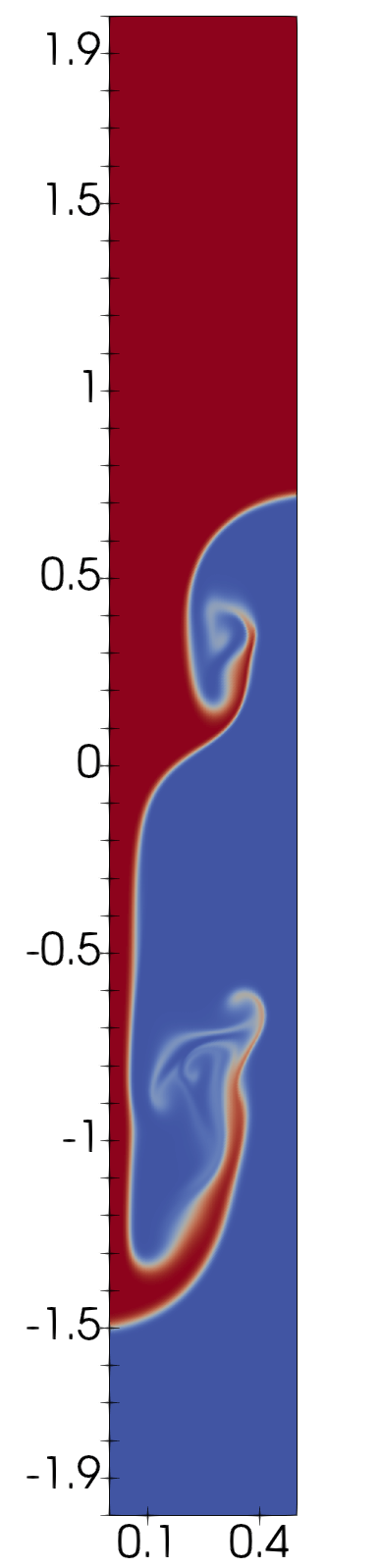} \\
\includegraphics[trim=40 0 80 0,clip,width=0.108\textwidth]{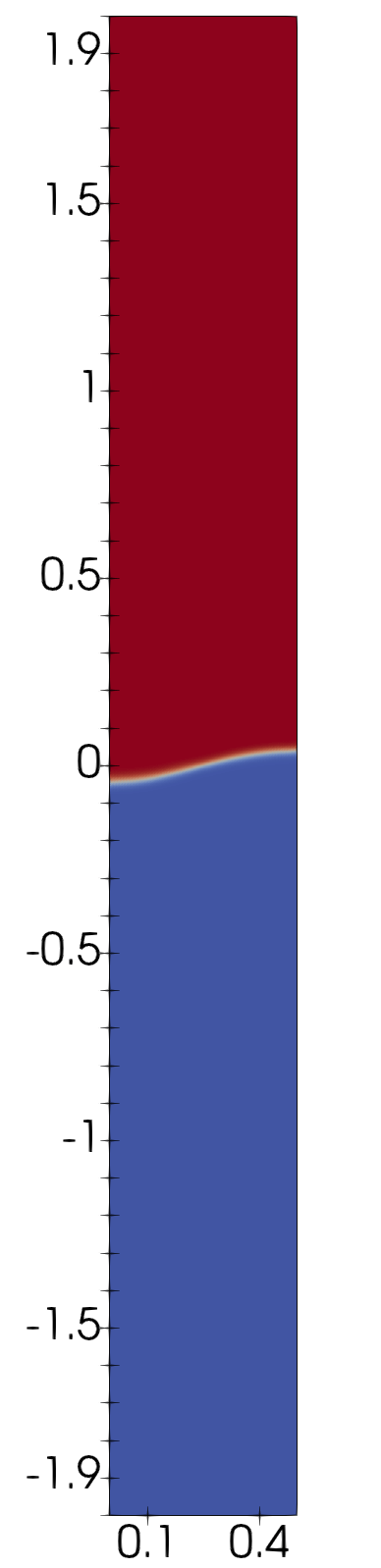} & 
\includegraphics[trim=40 0 80 0,clip,width=0.108\textwidth]{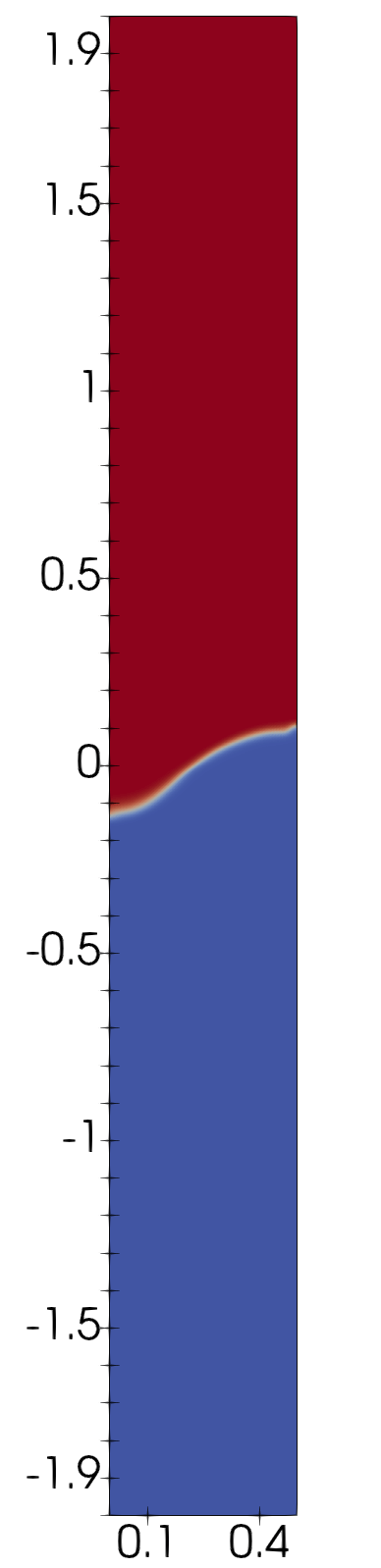} & 
\includegraphics[trim=40 0 80 0,clip,width=0.108\textwidth]{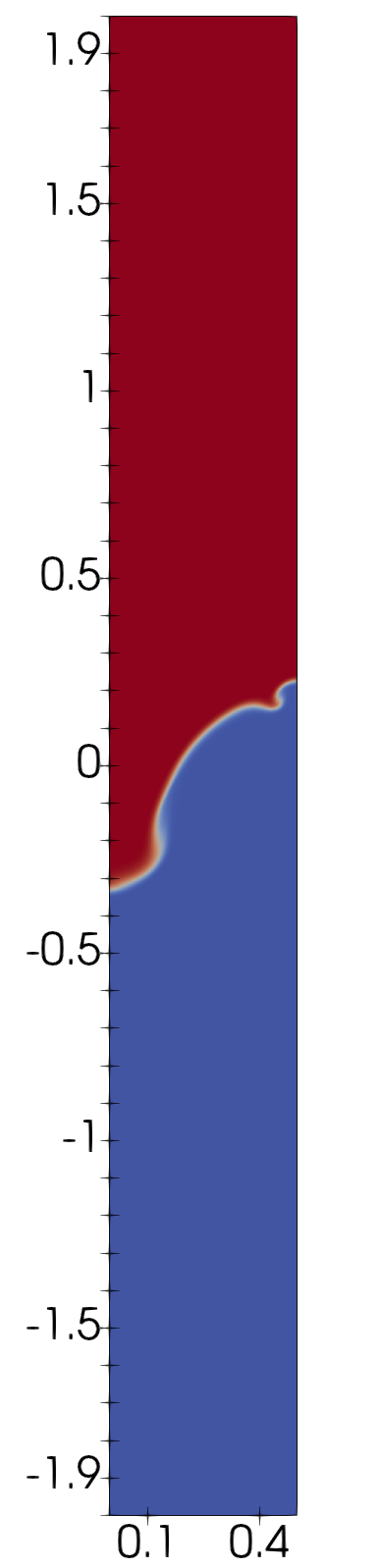} & 
\includegraphics[trim=40 0 80 0,clip,width=0.108\textwidth]{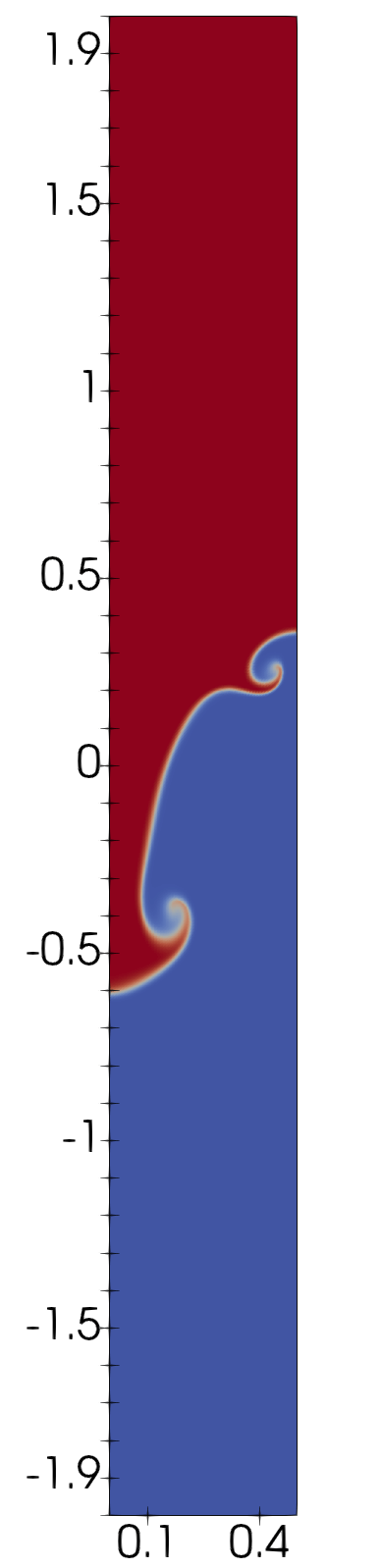} & 
\includegraphics[trim=40 0 80 0,clip,width=0.108\textwidth]{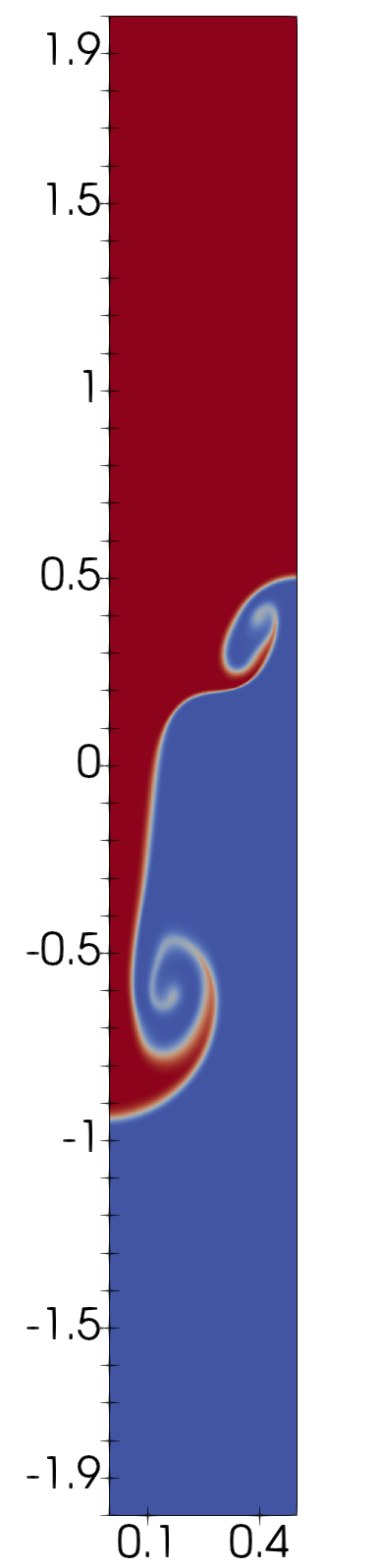} &
\includegraphics[trim=40 0 80 0,clip,width=0.108\textwidth]{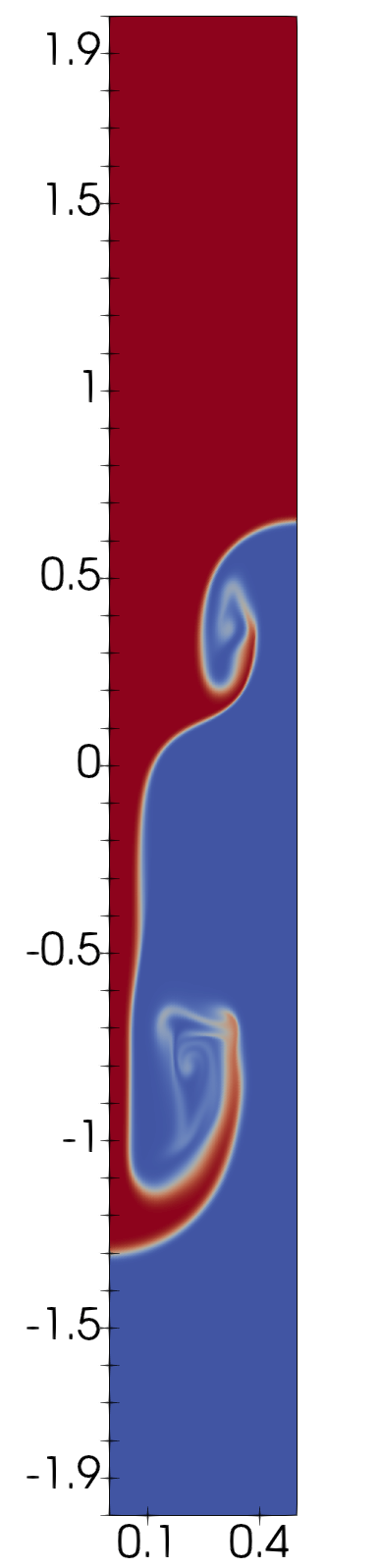} &
\includegraphics[trim=40 0 80 0,clip,width=0.108\textwidth]{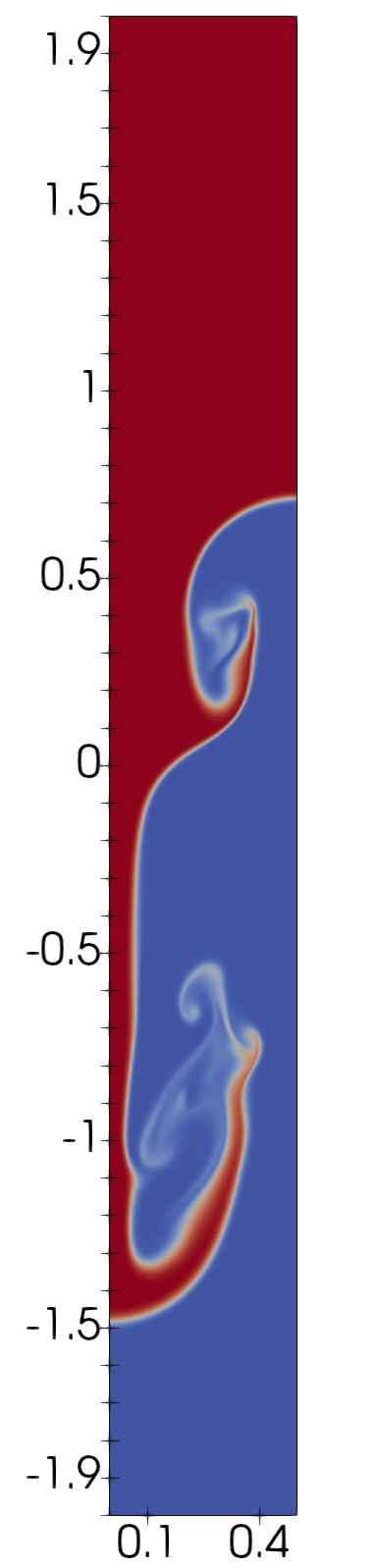} 
\end{tabular}
\caption{RTI at $\Atwood{=}0.75$ and $\Reynolds{=}1~000$. 
\emph{Top and bottom row}, $k{=}0$ HHO-$\pi^k$ and HHO-$\pi^{k+1}$ computations, respectively. 
\emph{From left to right}, evolution of the density field at selected time points 
$t \simeq 1, 1.5, 2, 2.5, 3, 3.5, 3.75$. 
         \label{fig:RTI1000hAt}}
\end{figure}

The high density ratio configuration is challenging from the stability viewpoint,
indeed, when high-order accurate formulations are employed, spurious oscillation might lead to negative densities and, eventually, singular mass matrices.
Accordingly, we consider $k{=}0$ HHO-$\pi^k$ and HHO-$\pi^{k+1}$ discretizations over the fine one million elements grid.
The Reynolds number, defined as $\Reynolds = \frac{\rho_{\mathrm{b}} d^{\frac{3}{2}} g^{\frac{1}{2}}  }{\mu}$, is $\Reynolds {=} 1~000$ and we set $\mu=\frac{1}{\Reynolds}$.
The time evolution of the density field is shown in Figure \ref{fig:RTI1000hAt} at times $t \simeq 1, 1.5, 2, 2.5, 3, 3.5, 3.75$.
In practice, for a given a value of $t$, we select the closest available time point among the $3~200$ available 
by rounding off to the closest integer the quantity $\frac{t}{\delta t}$, where $\delta t = \frac{t_F}{3~200}$.

We remark that both HHO-$\pi^k$ and HHO-$\pi^{k+1}$ are bounds preserving at $k{=}0$, meaning that the maximum and the minimum density values observed 
all along the density field evolution correspond to the density of the heavy ($\rho=7$) and the light ($\rho=1$) fluid. 
Comparing with the reference results of \cite{Guermond2009}, it is clear
that the spatial accuracy is not sufficient to properly capture the evolution of the sharp interface between the two fluids.
We remark that our computations are in better agreement with the earlier results of \cite{Guermond2000} 
where a coarser computational mesh was employed as compared with \cite{Guermond2009}. 
Even if the tiniest flow features are missing or misplaced, the position and shape of the main plume is properly replicated.
Note that HHO-$\pi^{k+1}$ seems to introduce less numerical dissipation as compared with HHO-$\pi^{k}$, confirming the better convergence
rates observed in Section~\ref{test:kova}.

\section{Conclusions}
\label{sec:conclusion}
We devised and numerically validated 
an ESDIRK-HHO formulation of the incompressible Navier--Stokes equations with variable density 
that is capable of simulating mixtures of immiscible fluids reaching high-orders of accuracy in both space and time.
Interestingly, the role of the pressure at the continuous level is closely replicated at the discrete level. 
Indeed, in the one hand pressure is the Lagrange multiplier of the incompressibilty constraint and,
on the other hand, the hybrid pressure space enforces $\boldsymbol{H}(\operatorname{div})$-conformity starting from non-conforming polynomial spaces.
The variant of the method named HHO-$\pi^k$ is provably stable thanks to the use of \emph{skew-symmetric} forms for the transport of mass and momentum.
The distinguishing features, that is pressure-robustness, exact incompressibility and pure advection of the density, 
have been numerically demonstrated. 

Since the spatial discretization requires primitive variables in order to achieve $\boldsymbol{H}(\operatorname{div})$-conformity,
while the time derivatives are formulated in terms of conservative variables, 
the ESDIRK time marching strategy has been carefully designed. 
The resulting scheme is fully implicit and fully hybrid, and static condensation of element unknowns
has been employed to alleviate the memory footprint of the Jacobian matrix.
To further improve the performance of iterative linear solvers a $p$-multigrid preconditioner has been fruitfully exploited when $k>0$. 
Note that, thanks to the use of non-conforming polynomial spaces, inherited coarse grid operators are trivial to construct.

The ability of simulating realistic flow configurations is demonstrated tackling the Rayleigh-Taylor instability at different Atwood and Reynolds numbers.
In the low-Atwood number number setup, the evolution of the density field is obtained with both high-order accurate ($k=6$) formulations 
on coarse meshes and low-polynomial degree ($k=1$) expansions on fine meshes.
Notably, for $\Atwood=0.5$ and $\Reynolds=1000$, the numerical solutions are in very good agreement despite the major difference in degrees-of-freedom count,
resulting in significant advantages in terms of computational expense for high-order accurate computations.
High-Atwood number computations demonstrate the HHO-ESDIRK formulation is bounds preserving when considering the lowest polynomial degree ($k=0$),
which implies that high density ratios can be tackled without compromising stability.

Further research efforts are needed to address physically relevant multi-phase flows,
in particular, future works will consider extending the present HHO-ESDIRK formulation to the Cahn-Hilliard-Navier-Stokes (CHNS) system.
HHO formulations that include heterogenous viscosity coefficients and strain-rate related viscous stresses are already available in literature, 
see, \eg~\cite{Botti.Di-Pietro.ea:18,Mbotti2021}, alongside, HHO methods have already been succesfully applied to the Cahn-Hilliard problem, see \eg~\cite{Chave.Di-Pietro.ea:16}. 
We plan to investigate up to which extent the intrinsic interface thickness control provided by the Cahn-Hilliard model 
will help in mitigating the onset of spurious oscillations when higher-order ($k>0$) formulations are considered.

  \bibliographystyle{plain} 

\begin{thebibliography}{10}
\bibitem{petscTR}
Satish Balay, Shrirang Abhyankar, Mark~F. Adams, Jed Brown, Peter Brune, Kris
  Buschelman, Lisandro Dalcin, Victor Eijkhout, William~D. Gropp, Dinesh
  Kaushik, Matthew~G. Knepley, Lois~Curfman McInnes, Karl Rupp, Barry~F. Smith,
  Stefano Zampini, and Hong Zhang.
\newblock {PETS}c users manual.
\newblock Technical Report ANL-95/11 - Revision 3.6, Argonne National
  Laboratory, 2015.

\bibitem{Bassi.Botti.ea:12}
F.~Bassi, L.~Botti, A.~Colombo, D.~A. Di~Pietro, and P.~Tesini.
\newblock On the flexibility of agglomeration based physical space
  discontinuous {Galerkin} discretizations.
\newblock {\em J. Comput. Phys.}, 231(1):45--65, 2012.

\bibitem{BassiVDAss22}
Francesco Bassi, Lorenzo Botti, Alessandro Colombo, and Francesco~Carlo Massa.
\newblock Assessment of an implicit discontinuous {G}alerkin solver for
  incompressible flow problems with variable density.
\newblock {\em Applied Sciences}, 12(21), 2022.

\bibitem{Bassi.Massa.ea:18}
Francesco Bassi, {Francesco Carlo} Massa, {Lorenzo} Botti, and Alessandro
  Colombo.
\newblock Artificial compressibility {G}odunov fluxes for variable density
  incompressible flows.
\newblock {\em Comput. Fluids}, 169:186--200, 2018.

\bibitem{Beirao-da-Veiga.Di-Pietro.ea:25}
L.~Beir\~{a}o~da Veiga, D.~A. Di~Pietro, J.~Droniou, K.~B. Haile, and T.~J.
  Radley.
\newblock A {R}eynolds-semi-robust method with hybrid velocity and pressure for
  the unsteady incompressible {Navier--Stokes} equations.
\newblock {\em {SIAM} Journal on Numerical Analysis}, 63(6):2317--2342, 2025.

\bibitem{Botti:12}
L.~Botti.
\newblock Influence of reference-to-physical frame mappings on approximation
  properties of discontinuous piecewise polynomial spaces.
\newblock {\em J. Sci. Comput.}, 52(3):675--703, 2012.

\bibitem{Botti.Di-Pietro:18}
L.~Botti and D.~A. Di~Pietro.
\newblock Numerical assessment of {Hybrid High-Order} methods on curved meshes
  and comparison with discontinuous {Galerkin} methods.
\newblock {\em J. Comput. Phys.}, 370:58--84, 2018.

\bibitem{BottiDiPietroHHOpMG2021}
Lorenzo Botti and Daniele~A. Di~Pietro.
\newblock p-{M}ultilevel preconditioners for {HHO} discretizations of the
  {S}tokes equations with static condensation.
\newblock {\em Commun. Appl. Math. Comput.}, 4:783--822, 2022.

\bibitem{Botti.Di-Pietro.ea:18}
Lorenzo Botti, Daniele~A. Di~Pietro, and J\'{e}r\^{o}me Droniou.
\newblock A {H}ybrid {H}igh-{O}rder discretisation of the {B}rinkman problem
  robust in the {D}arcy and {S}tokes limits.
\newblock {\em Comput. Methods Appl. Mech. Engrg.}, 341:278--310, 2018.

\bibitem{BottiDiPietroMassa25}
Lorenzo Botti, Daniele~A. {Di Pietro}, and Francesco~Carlo Massa.
\newblock Hybrid high-order formulations with turbulence modelling capabilities
  for incompressible flow problems.
\newblock {\em Computers \& Fluids}, 305:106915, 2026.

\bibitem{BottiMassa22}
Lorenzo Botti and Francesco~Carlo Massa.
\newblock {HHO} methods for the incompressible {N}avier-{S}tokes and the
  incompressible {E}uler equations.
\newblock {\em Journal of Scientific Computing}, 92(28):397--434, 06 2022.

\bibitem{Mbotti2021}
Michele Botti, Daniel Castanon~Quiroz, Daniele~A. Di~Pietro, and Andr\'e
  Harnist.
\newblock A {H}ybrid {H}igh-{O}rder method for creeping flows of
  {non-Newtonian} fluids.
\newblock {\em ESAIM: Mathematical Modelling and Numerical Analysis},
  55(5):2045--2073, 2021.

\bibitem{Cai2025}
Yunzhu Cai, Jiawei Wan, and Ahsan Kareem.
\newblock On convergence of implicit {R}unge-{K}utta methods for the
  incompressible {N}avier-{S}tokes equations with unsteady inflow.
\newblock {\em Journal of Computational Physics}, 523:113627, 2025.

\bibitem{Chave.Di-Pietro.ea:16}
F.~Chave, D.~A. Di~Pietro, F.~Marche, and F.~Pigeonneau.
\newblock A hybrid high-order method for the {Cahn--Hilliard} problem in mixed
  form.
\newblock {\em SIAM J. Numer. Anal.}, 54(3):1873--1898, 2016.

\bibitem{Cockburn.Di-Pietro.ea:16}
B.~Cockburn, D.~A. Di~Pietro, and A.~Ern.
\newblock Bridging the {Hybrid High-Order} and {Hybridizable Discontinuous
  Galerkin} methods.
\newblock {\em ESAIM: M2AN}, 50(3):635--650, 2016.

\bibitem{dauphin2026}
Mathias Dauphin, Daniele A.~Di Pietro, Jérôme Droniou, and Alexandros
  Skouras.
\newblock A low-order hybrid method for the variable-density incompressible
  {N}avier-{S}tokes equations, 2026.

\bibitem{Di-Pietro.Droniou:20}
D.~A. Di~Pietro and J.~Droniou.
\newblock {\em The {H}ybrid {H}igh-{O}rder method for polytopal meshes}.
\newblock Number~19 in Modeling, Simulation and Application. Springer, Cham,
  2020.

\bibitem{Di-Pietro.Droniou.ea:15}
D.~A. Di~Pietro, J.~Droniou, and A.~Ern.
\newblock A discontinuous-skeletal method for advection-diffusion-reaction on
  general meshes.
\newblock {\em SIAM J. Numer. Anal.}, 53(5):2135--2157, 2015.

\bibitem{FU2020}
Guosheng Fu.
\newblock A divergence-free {HDG} scheme for the {C}ahn-{H}illiard phase-field
  model for two-phase incompressible flow.
\newblock {\em Journal of Computational Physics}, 419:109671, 2020.

\bibitem{GoudonKrell14}
Thierry Goudon and Stella Krell.
\newblock A {DDFV} scheme for incompressible {N}avier-{S}tokes equations with
  variable density.
\newblock In {\em Finite volumes for complex applications {VII}. {E}lliptic,
  parabolic and hyperbolic problems}, volume~78 of {\em Springer Proc. Math.
  Stat.}, pages 627--635. Springer, Cham, 2014.

\bibitem{Guermond2000}
J.-L. Guermond and L.~Quartapelle.
\newblock A projection {FEM} for variable density incompressible flows.
\newblock {\em Journal of Computational Physics}, 165(1):167--188, 2000.

\bibitem{Guermond2009}
J.-L. Guermond and Abner Salgado.
\newblock A splitting method for incompressible flows with variable density
  based on a pressure {P}oisson equation.
\newblock {\em Journal of Computational Physics}, 228(8):2834--2846, 2009.

\bibitem{Kovasznay:48}
Leslie I.~George Kovasznay.
\newblock Laminar flow behind a two-dimensional grid.
\newblock {\em Math. Proc. Cambridge}, 44(1):58--62, 1948.

\bibitem{LANDET2020}
Tormod Landet, Kent-Andre Mardal, and Mikael Mortensen.
\newblock Slope limiting the velocity field in a discontinuous {G}alerkin
  divergence-free two-phase flow solver.
\newblock {\em Computers \& Fluids}, 196:104322, 2020.

\bibitem{Lehrenfeld:10}
C.~Lehrenfeld.
\newblock {\em Hybrid Discontinuous Galerkin methods for solving incompressible
  flow problems}.
\newblock PhD thesis, Rheinisch-Westf\"alischen Technischen Hochschule Aachen,
  2010.

\bibitem{LiQiuYang22}
Buyang Li, Weifeng Qiu, and Zongze Yang.
\newblock A convergent post-processed {D}iscontinuous {G}alerkin method for
  incompressible flow with variable density.
\newblock {\em Journal of Scientific Computing}, 91(2), 2022.

\bibitem{Manzanero.Rubio.ea:20}
Juan Manzanero, Gonzalo Rubio, David~A. Kopriva, Esteban Ferrer, and Eusebio
  Valero.
\newblock An entropy–stable discontinuous {G}alerkin approximation for the
  incompressible {N}avier–{S}tokes equations with variable density and
  artificial compressibility.
\newblock {\em Journal of Computational Physics}, 408:109241, 2020.

\bibitem{NORDSTROM25}
Jan Nordström and Arnaud~G. Malan.
\newblock {A}n energy stable incompressible multi-phase flow formulation.
\newblock {\em Journal of Computational Physics}, 523:113685, 2025.

\bibitem{Skvorstov2010}
L.~M. Skvortsov.
\newblock Diagonally implicit {R}unge-{K}utta methods for differential
  algebraic equations of indices two and three.
\newblock {\em Computational Mathematics and Mathematical Physics},
  50:993--1005, 2010.

\bibitem{Skvorstov2022}
L.~M. Skvortsov.
\newblock Third- and fourth-order {ESDIRK} methods for stiff and
  differential-algebraic problems.
\newblock {\em Computational Mathematics and Mathematical Physics},
  62:766--783, 2022.

\bibitem{Tryggvason1988}
Grétar Tryggvason.
\newblock Numerical simulations of the {R}ayleigh-{T}aylor instability.
\newblock {\em Journal of Computational Physics}, 75(2):253--282, 1988.

\end{thebibliography}

\end{document}